\documentclass{aa}  
\usepackage{graphicx}
\graphicspath{ {./Images/} }
\usepackage[font=small]{caption}
\usepackage{txfonts}
\usepackage[dvipsnames]{xcolor}
\usepackage{subcaption} 
\newcommand\Tstrut{\rule{0pt}{2.3ex}}   
\usepackage{float}
\usepackage{upgreek}
\usepackage{fontawesome5}

\newcommand{\orcid}[1]{\href{https://orcid.org/#1}{\textcolor[HTML]{A6CE39}{\faOrcid}}}

\usepackage[colorlinks=true,linkcolor=blue,citecolor=blue]{hyperref}
\begin{document}

   \title{A benchmark for extreme conditions of the multiphase interstellar medium in the most luminous hot dust-obscured galaxy at $z=4.6$}

   \author{Román Fernández Aranda\inst{1,2,3}\orcid{0000-0002-7714-688X}
   \and Tanio Díaz Santos\inst{2,4}\orcid{0000-0003-0699-6083}
   \and Evanthia Hatziminaoglou\inst{3}\orcid{0000-0003-0917-9636} 
   \and Roberto J. Assef\inst{5}\orcid{0000-0002-9508-3667}
   \and Manuel Aravena\inst{5}\orcid{0000-0002-6290-3198}
   \and Peter R. M. Eisenhardt\inst{6}
   \and Carl Ferkinhoff\inst{7}
   \and Antonio Pensabene\inst{8}\orcid{0000-0001-9815-4953}
   \and Thomas Nikola\inst{9}
   \and Paola Andreani\inst{3}
   \and Amit Vishwas\inst{10}\orcid{0000-0002-4444-8929}
   \and Gordon J. Stacey\inst{10}\orcid{0000-0003-1260-5448}
   \and Roberto Decarli\inst{11}
   \and Andrew W. Blain\inst{12}\orcid{0000-0001-7489-5167}
   \and Drew Brisbin\inst{13} 
   \and Vassilis Charmandaris\inst{1,2,4}\orcid{0000-0002-2688-1956}
   \and Hyunsung D. Jun\inst{14}\orcid{0000-0003-1470-5901}
   \and Guodong Li\inst{15,16}
   \and Mai Liao\inst{5,15,17}\orcid{0000-0002-9137-7019}
   \and Lee R. Martin\inst{12}\orcid{0000-0002-4950-7940}
   \and Daniel Stern\inst{6}\orcid{0000-0003-2686-9241}
   \and Chao-Wei Tsai\inst{15,16,18}\orcid{0000-0002-9390-9672}
   \and Jingwen Wu\inst{15,16}
   \and Dejene Zewdie\inst{5}\orcid{0000-0003-4293-7507}
    }
    
   \institute{Department of Physics, University of Crete, 70013, Heraklion, Greece
         \and Institute of Astrophysics, Foundation for Research and Technology - Hellas (FORTH), Voutes, 70013 Heraklion, Greece
        \and European Southern Observatory (ESO), Karl-Schwarzschild-Strasse 2, D-85748 Garching b. München, Germany
        \and School of Sciences, European University Cyprus, Diogenes street, Engomi, 1516 Nicosia, Cyprus
        \and Instituto de Estudios Astrofísicos, Facultad de Ingeniería y Ciencias, Universidad Diego Portales, Av. Ejército Libertador 441, Santiago, Chile
        \and Jet Propulsion Laboratory, California Institute of Technology, 4800 Oak Grove Drive, Pasadena, CA 91109, USA
        \and Department of Physics, Winona State University, Winona, MN 55987, USA
        \and Dipartimento di Fisica “G. Occhialini”, Università degli Studi di Milano-Bicocca, Piazza della Scienza 3, 20126 Milano, Italy  
        \and Cornell Center for Astrophysics and Planetary Science, Cornell University, Ithaca, NY 14853, USA
        \and Department of Astronomy, Cornell University, Ithaca, NY 14853, USA    
        \and INAF – Osservatorio di Astrofisica e Scienza dello Spazio di Bologna, Via Gobetti 93/3, 40129 Bologna, Italy
        \and Physics \& Astronomy, University of Leicester, 1 University Road, Leicester LE1 7RH, UK
        \and Joint ALMA Observatory, Alonso de Córdova 3101, Santiago, 7630410, Chile
        \and Department of Physics, Northwestern College, 101 7th St SW, Orange City, IA 51041, USA        
        \and National Astronomical Observatories, Chinese Academy of Sciences, 20A Datun Road, Beijing 100101, People’s Republic of China
        \and University of Chinese Academy of Sciences, Beijing 100049, People’s Republic of China
        \and Chinese Academy of Sciences South America Center for Astronomy, National Astronomical Observatories, CAS, Beijing, 100101, China
        \and Institute for Frontiers in Astronomy and Astrophysics, Beijing Normal University, Beijing 102206, People’s Republic of China
             }

   \date{Received ...; accepted ...}

 
  \abstract
    {WISE J224607.6--052634.9 (W2246--0526) is a hot dust-obscured galaxy at $z$ = 4.601, and the most luminous obscured quasar known to date. W2246--0526 harbors a heavily obscured supermassive black hole that is most likely accreting above the Eddington limit. We present observations with the Atacama Large Millimeter/submillimeter Array (ALMA) in seven bands, including band 10, of the brightest far-infrared (FIR) fine-structure emission lines of this galaxy: [OI]$_{63\upmu \mathrm{m}}$, [OIII]$_{88\upmu \mathrm{m}}$, [NII]$_{122\upmu \mathrm{m}}$, [OI]$_{145\upmu \mathrm{m}}$, [CII]$_{158\upmu \mathrm{m}}$, [NII]$_{205\upmu \mathrm{m}}$, [CI]$_{370\upmu \mathrm{m}}$, and [CI]$_{609\upmu \mathrm{m}}$. A comparison of the data to a large grid of \textsc{Cloudy} radiative transfer models reveals that a high hydrogen density ($n_{\mathrm{H}}\sim3\times10^3$ cm$^{-3}$) and extinction ($A_{\mathrm{V}}\sim300$ mag), together with extreme ionization ($log(U)=-0.5$) and a high X-ray to UV ratio ($\alpha_{\mathrm{ox}}\geq-0.8$) are required to reproduce the observed nuclear line ratios. The values of $\alpha_{\mathrm{ox}}$ and $U$ are among the largest found in the literature and imply the existence of an X-ray-dominated region (XDR). In fact, this component explains the a priori very surprising non-detection of the [OIII]$_{88\upmu \mathrm{m}}$ emission line, which is actually suppressed, instead of boosted, in XDR environments. Interestingly, the best-fitted model implies higher X-ray emission and lower CO content than what is detected observationally, suggesting the presence of a molecular gas component that should be further obscuring the X-ray emission over larger spatial scales than the central region that is being modeled. These results highlight the need for multiline infrared observations to characterize the multiphase gas in high redshift quasars and, in particular, W2246--0526 serves as an extreme benchmark for comparisons of interstellar medium conditions with other quasar populations at cosmic noon and beyond.}

   \keywords{galaxies: ISM – galaxies: nuclei – galaxies: active - galaxies: individual (WISE J2246-0526) –   quasars: emission lines}
    \titlerunning{A benchmark for extreme conditions of the ISM in the most luminous Hot DOG}
    \authorrunning{R. Fernández Aranda et al.}
   \maketitle
%

\section{Introduction}
\label{sec:intro}

"Cosmic noon" ($z\sim1-3$) and the epoch leading to it ($z\sim3-6$) are the two most important periods of cosmic history in terms of galaxy growth and supermassive black hole (SMBH) activity. It is during these periods that SMBHs and their hosts assembled most of their mass, as attested by the tight relationship between the stellar mass of galaxies and their central SMBH mass across cosmic time \citep[e.g.,][]{magorrian1998,kormendy2013}. Understanding the properties of the interstellar medium (ISM) of actively star-forming galaxies with an actively accreting SMBH provides important insights into the co-evolution of a SMBH and their host galaxies. Additionally, observations of such extreme systems allow meaningful constraints to be placed on cosmological simulations that aim to reproduce correlations between galaxy properties on spatial scales ranging from approximately parsecs (SMBH gas accretion) to tens of kiloparsecs (evolution of the host galaxy).

In particular, quasars, powered by accreting SMBHs at their centers, exert a profound influence on the ISM that surrounds them. The intense radiation and high-energy photons emitted by active galactic nuclei (AGNs) can significantly impact the physical and chemical conditions of the ISM, playing a crucial role in shaping the evolution of galaxies and regulating the growth of their central SMBHs \citep[e.g.,][]{dimatteo2005}. Nevertheless, most of the accretion onto SMBHs is expected to be heavily obscured by dust and gas \citep{hickox2018}, making the study of heavily obscured quasars key to understand the interaction between AGN and their surrounding gas. 

Among the large variety of dust enshrouded quasars, hot, dust-obscured galaxies \citep[Hot DOGs;][]{eisenhardt2012,wu2012} are a population of hyperluminous quasars ($L_{bol}\gtrsim 10^{13} \mathrm{L_{\odot}}$), firstly discovered by NASA’s \textit{Wide-field Infrared Survey Explorer} \citep[WISE;][]{wright2010}. Hot DOGs are characterized by very high bolometric luminosities ($L_{bol}\gtrsim 10^{13} \ \mathrm{L_{\odot}}$) and high dust temperatures \citep[$>60$ K,][]{wu2012}, powered by highly obscured AGNs that dominate the spectral energy distributions (SEDs) beyond $\lambda >1 \ \upmu \mathrm{m}$ \citep{tsai2015}, and Eddington-limited accretion \citep{wu2018}. Signs of quasar feedback have been observed in the Hot DOG population, including the highly turbulent [CII]$_{158\upmu \mathrm{m}}$ gas \citep{2021A&A...654A..37D} and the high-velocity ionized gas outflows detected in the rest-frame optical  \citep{2020ApJ...905...16F,2020ApJ...888..110J}. Individual and statistical studies also suggest that Hot DOGs live in over-dense environments \citep{jones2014,assef2015,2022NatCo..13.4574G,2022ApJ...935...80L,zewdie2023} and their activity is likely merger-driven \citep{2016ApJ...822L..32F,diaz2018}. Moreover, these objects could be experiencing several recurrent episodes of extreme accretion and obscuration \citep{2021A&A...654A..37D}. This scenario is in agreement with recent cosmological simulations of high redshift quasars \citep{2021ApJ...917...53A}, but it is in tension with the paradigm for the formation of luminous quasars in the local Universe, that is, involving a single encounter of two massive, gas-rich spiral galaxies \citep[e.g.,][]{sanders1988,hopkins2008,treister2012}. 

WISE J224607.6--052634.9 (hereafter W2246--0526) is a Hot DOG at $z_{\mathrm{[CII]}} = 4.601$ \citep{2016ApJ...816L...6D} with a bolometric luminosity of $3.6\times10^{14} \ \mathrm{L_{\odot}}$, making it one of the most luminous galaxies known to date \citep{tsai2015,2018ApJ...868...15T}. It is part of a multiple merger \citep{diaz2018}, where W2246--0526 is connected to at least three galaxy companions by dusty tidal streamers. W2246--0526 hosts an AGN with a SMBH mass of $M_{\bullet} = 4.0\times10^{9} \ \mathrm{M_{\odot}}$, and it radiates well above the Eddington limit, with an Eddington ratio of $\lambda_{Edd}=2.8$ \citep[][assuming that the MgII $2799 \ Å$ emission line is being broadened by the SMBH gravity]{2018ApJ...868...15T}. W2246--0526 is a radio-quiet quasar \citep[according to the classification of ][]{1989AJ.....98.1195K}, but the central SMBH powers parsec-scale radio activity \citep{2020ApJ...905L..32F}. Given the extreme conditions, W2246--0526 provides a key laboratory to study the effect of the AGN on its surrounding ISM. Moreover, because of the extreme luminosity of W2246--0526, one might expect bright far-infrared (FIR) spectral lines that trace the multiphase ISM, and that are straightforward to detect with the Atacama Large Millimeter/submillimeter Array (ALMA). 

ALMA covers a wavelength range where, at $z\sim4.6$, fine-structure lines from [OI]$_{63\upmu \mathrm{m}}$ to [CI]$_{609\upmu \mathrm{m}}$ can be observed. Previous observations of [CII]$_{158\upmu \mathrm{m}}$ and CO$(J=2-1)$ \citep{2016ApJ...816L...6D,diaz2018} suggest that the ionized, neutral, and molecular phases of the ISM are very turbulent, likely due to the feedback from the central AGN, out to scales of at least a few kiloparsecs. Therefore, we can expect all the fine-structure lines from all regions to be strongly affected by the central energy source. 

\begin{figure}[h!]
    \centering
        \subfloat{\includegraphics[width=78mm]{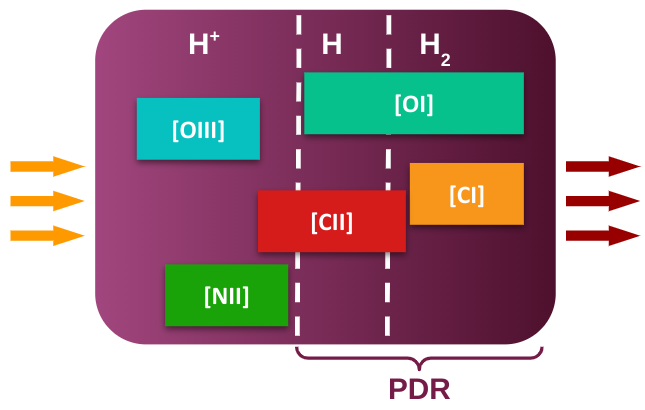}}
        \par\vspace{0.0mm}        \subfloat{\includegraphics[width=78mm]{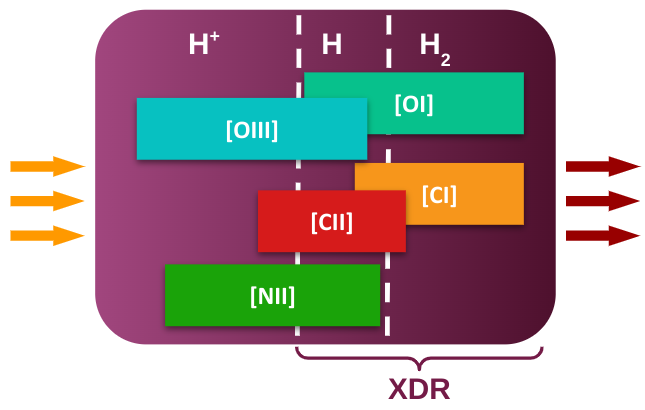}}
        \caption{Sketches of the emission of the targeted lines in the different phases of the ISM, for a PDR on the top and an XDR on the bottom. The incident photons, far-UV for PDRs and X-rays for XDRs, enter from the left. Thick dashed lines separate the different hydrogen phases: ionized ($H^+$), neutral ($H$), and molecular ($H_2$). We do not show a scale regarding the depth of the cloud because it strongly depends on the ionization source and cloud properties. We note that in the XDR case, ionized transitions arise also where the gas is mostly neutral, and neutral carbon emission is distributed much further into the cloud since X-rays can penetrate much more deeply than far-UV photons.}
        \label{fig:PDR}
\end{figure}

From the photodissociation regions (PDRs) paradigm, we expect many of these lines (e.g., [CII], [OI], and [CI]) to arise from the surface of molecular clouds that are exposed to UV radiation. In the ionized gas phase outside the PDR, we find ionized nitrogen, and doubly ionized oxygen, which originates in highly ionized HII regions requiring radiation from early-type O stars or an AGN. [NII] and [OIII] lines will therefore trace this region. Tracking from the UV-exposed surface of the PDR, carbon is initially ionized, and oxygen is in atomic form.  This region is therefore a strong emitter of both [CII] and [OI].  Beyond, the far-UV field is extenuated to the extent that carbon is in atomic form. Deeper into the cloud, carbon is found in CO. The neutral carbon region, traced in its [CI] line emission, is relatively much thinner and cooler than the ionized carbon region. However, some AGNs can also generate X-ray-dominated regions \citep[XDRs, e.g.,][]{maloney1996}, which significantly changes both the ionization structure and the heating of the molecular cloud, and therefore the regions from where fine-structure FIR lines arise (see Fig.~\ref{fig:PDR}). Recent studies targeting these and other FIR lines, in combination with ISM models, have proven to be very successful in characterizing the properties and phases of the ISM of high-redshift quasars \citep{novak2019,pensabene2021,meyer2022,decarli2023}. 

In this paper, we analyze ALMA observations of the [OI]$_{63\upmu \mathrm{m}}$, [OIII]$_{88\upmu \mathrm{m}}$, [NII]$_{122\upmu \mathrm{m}}$, [OI]$_{145\upmu \mathrm{m}}$, [CII]$_{158\upmu \mathrm{m}}$, [NII]$_{205\upmu \mathrm{m}}$, [CI]$_{370\upmu \mathrm{m}}$, and [CI]$_{609\upmu \mathrm{m}}$ emission lines in W2246--0526. Modeling the AGN and ISM with the brightest FIR lines observed with ALMA provides an understanding of the role of quasar feedback in galaxy evolution.

The paper is structured as follows: Section~\ref{sec:obs} describes the observational setup, data reduction, and line measurements of the ALMA observations. In Sect.~\ref{sec:results} we analyze individual line ratios that provide diagnostics of the ISM, describe the setup of the ISM modeling, and fit all the lines with the models. In Sect.~\ref{sec:discussion} we discuss the results and place them into a wider context. Finally, in Sect.~\ref{sec:conclusions} we summarize and draw our conclusions.  

Throughout the paper, we assume a flat $\Lambda$CDM cosmology with $H_0 = 70$ km s$^{-1}$ Mpc$^{-1}$ and $\Omega_{\mathrm{M}}=0.3$. For reference, in this cosmology, $1\arcsec$ is equivalent to 6.54 kpc at $z=4.601$.


\section{Observations, data reduction, and emission line measurements}
\label{sec:obs}

This work uses data from six different ALMA projects observed during Cycles 3 through 9, with a total on-source observation time of 19 hours with the ALMA main 12-m array. These observations were conducted between June 2016 and October 2022, and targeted the following FIR emission lines: [OI]$_{63\upmu \mathrm{m}}$ (band 10, one of the few detections in the literature to date), [OIII]$_{88\upmu \mathrm{m}}$ (band 9), [NII]$_{122\upmu \mathrm{m}}$ (band 8), [OI]$_{145\upmu \mathrm{m}}$ (band 7), [CII]$_{158\upmu \mathrm{m}}$ (band 7), [NII]$_{205\upmu \mathrm{m}}$ (band 6), [CI]$_{370\upmu \mathrm{m}}$ (band 4), and [CI]$_{609\upmu \mathrm{m}}$ (band 3). The properties of each transition are listed in Table~\ref{tab1}, and the details of the observations are listed in Table~\ref{tab2}.

\begin{table*}
    \centering
    \caption{Fundamental physical parameters for the targeted lines}
    \begin{tabular}{*{9}{c}}
    \hline
     Line & Transition & $\lambda$ \newline  & $\nu$ & F.P. & $n_{\mathrm{cr,e}}$ & $n_{\mathrm{cr,H}}$ &$E_{\mathrm{ul}}$ \\
      & & [$\upmu$m] & [GHz] & [eV] & [cm$^{-3}$] & [cm$^{-3}$] & [K] \\
     \hline
     {[OI]$_{63\upmu \mathrm{m}}$} & $^{3}\textrm{P}_1 - ^{3}\textrm{P}_2$ & 63.18 & 4744.8 & ...  & ... & $2.5\times10^5$ & 228  \\      
     {[OIII]$_{88\upmu \mathrm{m}}$} & $^{3}\textrm{P}_1 - ^{3}\textrm{P}_0$ & 88.36 & 3393.0 & 35.12 & $1.7\times10^3$ & ... & 163 \\     
     {[NII]$_{122\upmu \mathrm{m}}$} & $^{3}\textrm{P}_2 - ^{3}\textrm{P}_1$ & 121.90 & 2459.4 & 14.53 & 299 & ... & 118  \\ 
     {[OI]$_{145\upmu \mathrm{m}}$} & $^{3}\textrm{P}_0 - ^{3}\textrm{P}_1$ & 145.53 & 2060.1 & ... & ... & $2.3\times10^4$ & 99  \\ 
     {[CII]$_{158\upmu \mathrm{m}}$} & $^{3}\textrm{P}_{3/2} - ^{3}\textrm{P}_{1/2}$ & 157.74 & 1900.5 & 11.26 & 52 & $2.0\times10^3$ & 92  \\ 
     {[NII]$_{205\upmu \mathrm{m}}$} & $^{3}\textrm{P}_1 - ^{3}\textrm{P}_0$ & 205.18 & 1461.1 & 14.53 & 190 & ... & 70  \\
     {[CI]$_{370\upmu \mathrm{m}}$} & $^{3}\textrm{P}_2 - ^{3}\textrm{P}_1$ & 370.42 & 809.34 & ... & ... & 720 & 39  \\  
     {[CI]$_{609\upmu \mathrm{m}}$} & $^{3}\textrm{P}_1 - ^{3}\textrm{P}_0$ & 609.14 & 492.16 & ... & ... & 620 & 24 \\ 
     \hline
     \\
    \end{tabular}
    \caption*{Note: Columns correspond to the emission line, transition, central wavelength, central frequency, formation potential (FP), electron ($n_{\mathrm{cr,e}}$), and Hydrogen ($n_{\mathrm{cr,H}}$) critical densities, and energy difference between upper and lower level ($E_{\mathrm{ul}}$). We note that $n_{\mathrm{cr,e}}$ were extracted from the \textsc{Cloudy} database, computed for a gas temperature of 10000 K. The last two columns are taken from \cite{draine2011}, with the $n_{\mathrm{cr,H}}$ computed for a gas temperature of 100 K.
    }
    \label{tab1}
\end{table*}

\begin{table*}
    \centering
    \caption{Details of the ALMA observations}
    \begin{tabular}{*{10}{c}}
    \hline
     Line & Date & Time  & Antennas & Baselines & ALMA & Phase & Flux & Project code \\
     & & on-source & & & Band & calibrator & calibrator & \\
      & [yyyy/mm/dd] & [s] & & [m] \\
     \hline
     {[OI]$_{63\upmu \mathrm{m}}$} & 2018-05-23 & 1816 & 45 & 15-314 & 10 & J2229-0832 & J2253+1608 & 2017.1.00899.S  \\     
     {[OIII]$_{88\upmu \mathrm{m}}$} & 2022-04-28 & 2978 & 45 & 15-500 & 9 & J2301-0158 & J1924-2914 & 2021.1.00726.S  \\
      & 2022-09-24 & 1659 & 44 & 15-500 & 9 & J2229-0832 & J1924-2914 & 2021.1.00726.S \\
      & 2022-09-30 & 2980 & 23 & 15-500 & 9 & J2301-0158 & J2253+1608 & 2021.1.00726.S \\
      & 2022-10-08 & 2979 & 41 & 15-500 & 9 & J2229-0832 & J1924-2914 & 2021.1.00726.S \\
     {[NII]$_{122\upmu \mathrm{m}}$} & 2017-04-28 & 1820 & 38 & 15-460 & 8 & J2229-0832 & J2253+1608 & 2016.1.00668.S  \\   
     {[OI]$_{145\upmu \mathrm{m}}$} & 2018-09-04 & 2541 & 45 & 15-784 & 7 & J2229-0832 & J2253+1608 & 2017.1.00899.S  \\
       & 2018-09-04 & 2542 & 45 & 15-784 & 7 & J2229-0832 & J2253+1608 & 2017.1.00899.S\\
       & 2018-09-05 & 2525 & 46 & 15-784 & 7 & J2229-0832 & J2253+1608 & 2017.1.00899.S\\
     {[CII]$_{158\upmu \mathrm{m}}$} & 2014-06-29 & 1050 & 32 & 21-558 & 7 & J2225-0457 & J2232+117 & 2013.1.00576.S  \\  
     {[NII]$_{205\upmu \mathrm{m}}$} & 2016-06-20 & 3045 & 38 & 15-704 & 6 & J2243-0609 & J2148+0657 & 2015.1.00883.S  \\
     & 2016-06-20 & 3038 & 41 & 15-704 & 6 & J2243-0609 & Pallas & 2015.1.00883.S \\
     & 2016-07-13 & 3046 & 39 & 15-867 & 6 & J2243-0609 & Pallas & 2015.1.00883.S \\
     {[CI]$_{370\upmu \mathrm{m}}$} & 2018-10-16 & 1519 & 45 & 15-2517 & 4 & J2301-0158 & J2148+0657 & 2018.1.00119.S \\
     {[CI]$_{609\upmu \mathrm{m}}$} & 2019-08-12 & 1869 & 48 & 41-3638 & 3 & J2236-0406 & J0006-0623 & 2018.1.00119.S \\   
     \hline
     \\
    \end{tabular}
    \label{tab2}
\end{table*}

We reduced and calibrated the data using the Common Astronomy Software Applications \citep[CASA;][]{2007ASPC..376..127M}. We processed the data with the \texttt{ScriptForPI.py} provided through the ALMA science archive to generate the measurement sets, using the version of the pipeline used by the observatory to reduce the data from each Cycle. We note that the automatic pipeline was run without the need for manual intervention in all bands. We imaged the measurement sets to create spectral cubes for the emission lines using the \texttt{tclean} task of CASA v6.4.3. We adopted a uniform pixel scale of 0.05$\arcsec$ to sample the synthesized beams (ranging from $\sim0.3\arcsec$ to $0.5\arcsec$ in the different observations). We run the Hogbom cleaning algorithm \citep{1974A&AS...15..417H} to a flux density threshold of two times the root mean square of each cube with the Briggs weighting mode set to a robustness = 2.0 (close to natural weighting), and using a mask with radius $1.25\arcsec$ centered at the peak luminosity pixel, corresponding to $\sim8.2$ kpc at $z=4.601$, enough to cover the host galaxy. We created the line cubes by combining appropriate spectral windows to select the continuum from neighboring line-free channels and subtracting it using the CASA task \texttt{uvcontsub}. The intensity maps of all the lines as well as each of the corresponding spectra are presented in Fig.~\ref{fig:observations}.

\begin{figure*}[!htbp]
    \centering
        \subfloat{\includegraphics[width=84mm]{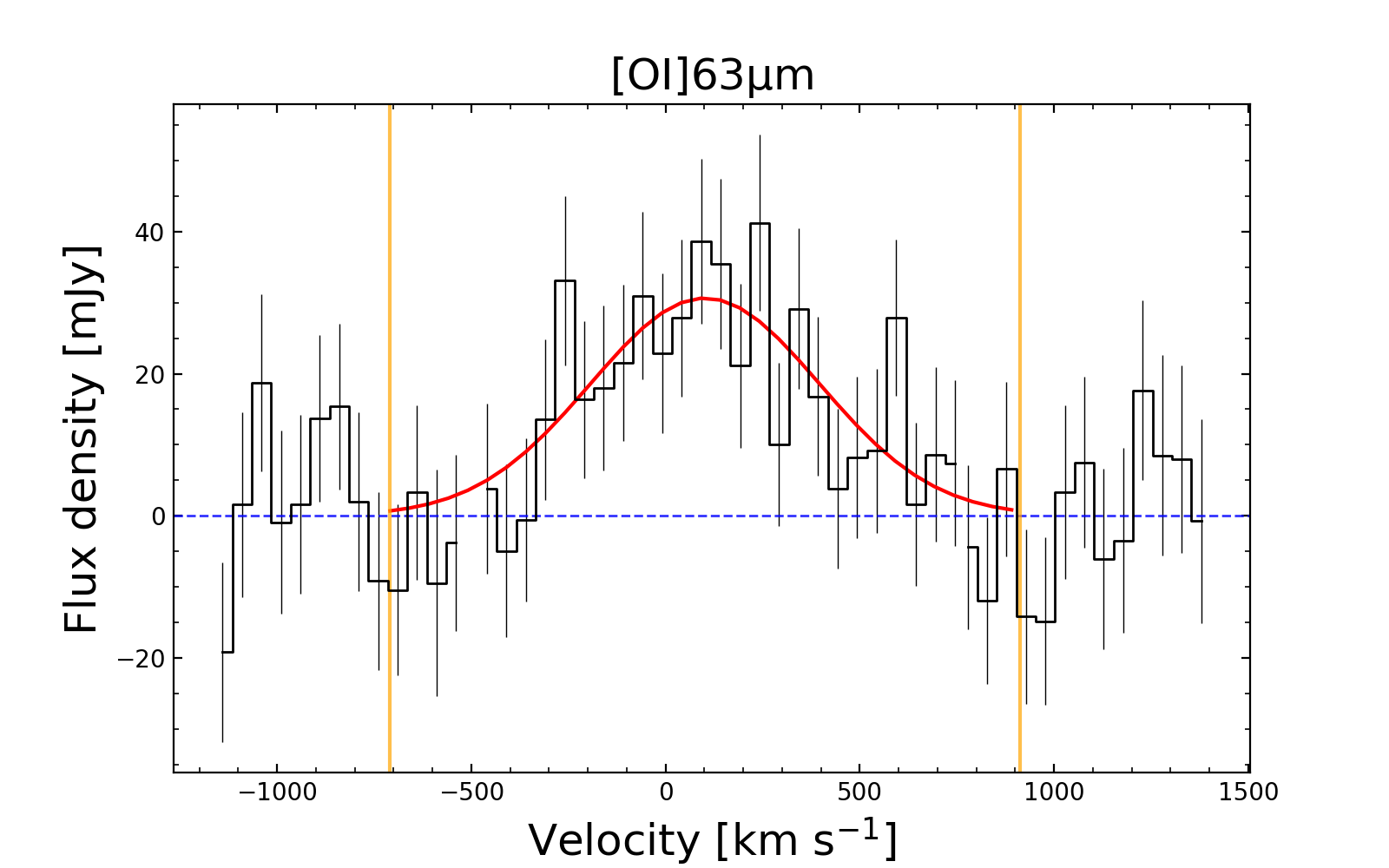}}
        \subfloat{\includegraphics[width=84mm]{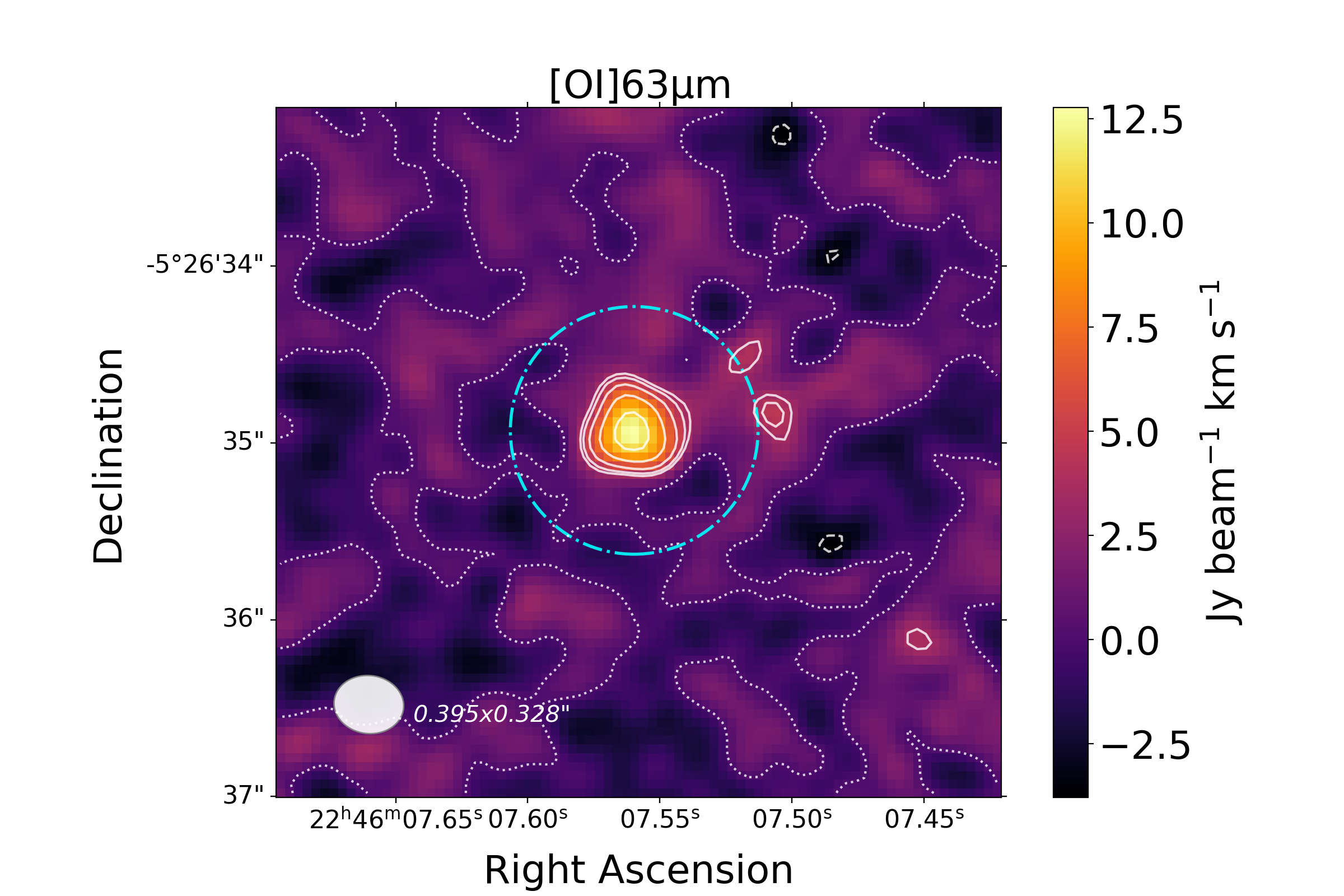}}
        \par
        \subfloat{\includegraphics[width=84mm]{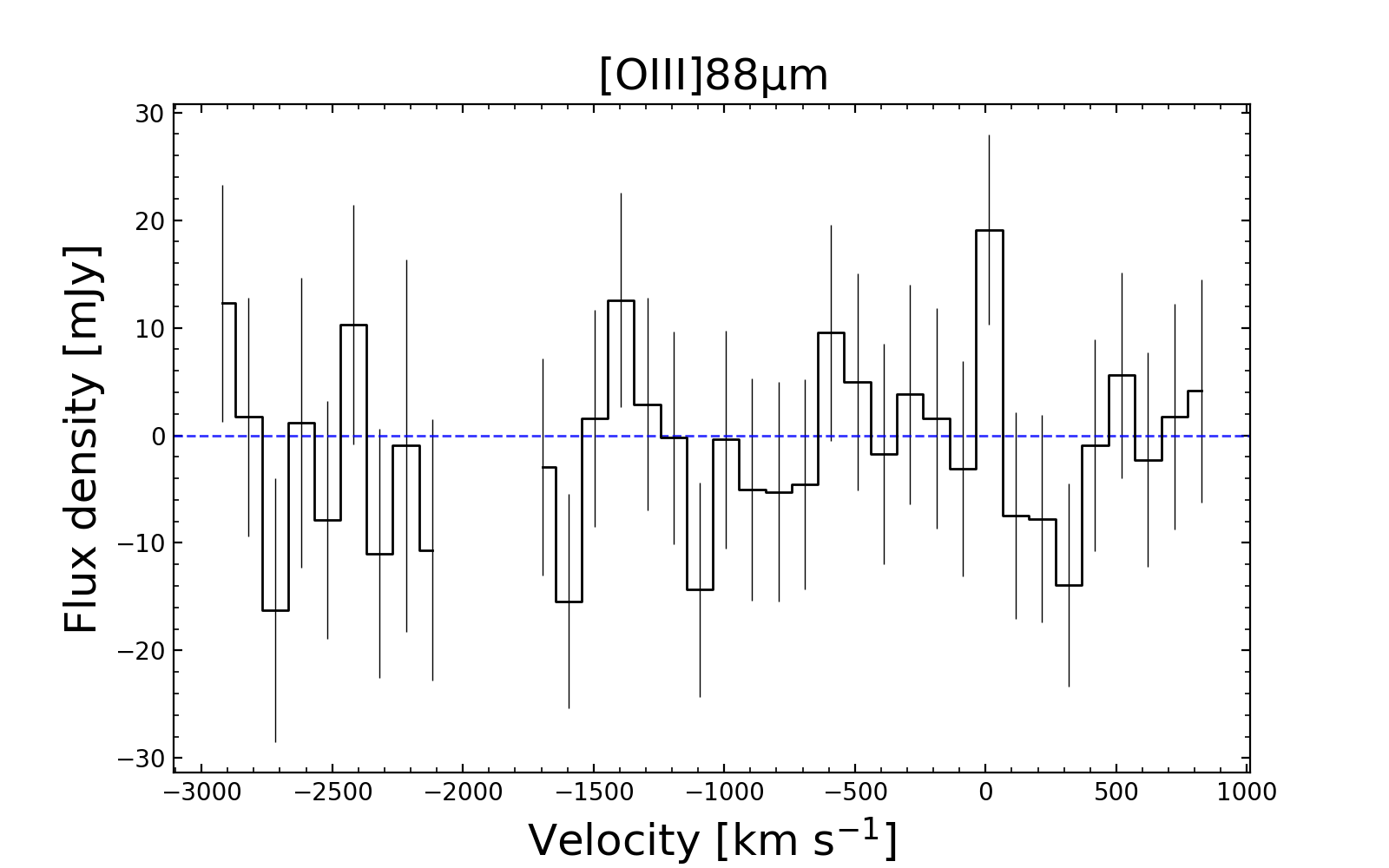}}
        \subfloat{\includegraphics[width=84mm]{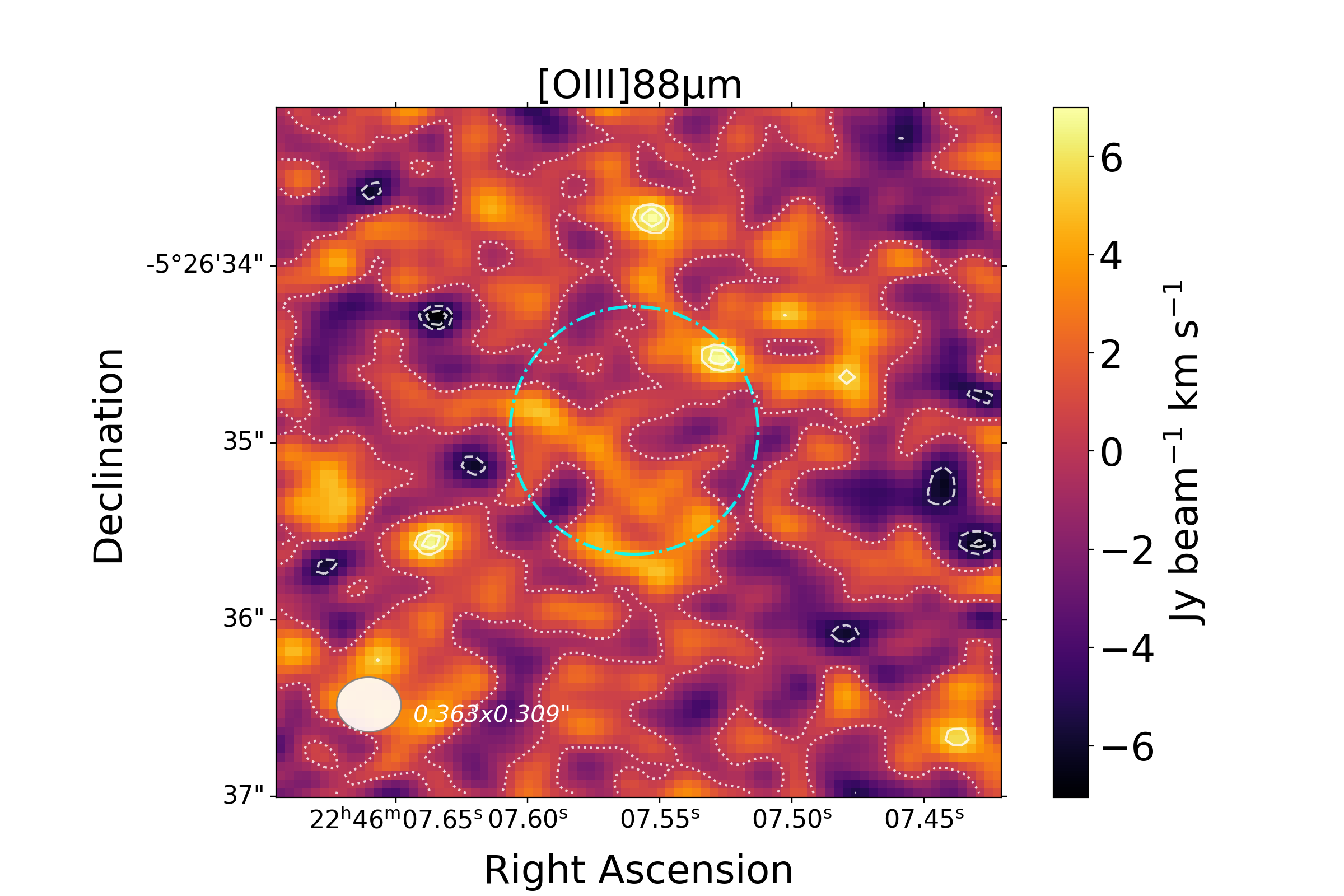}}
        \par
        \subfloat{\includegraphics[width=84mm]{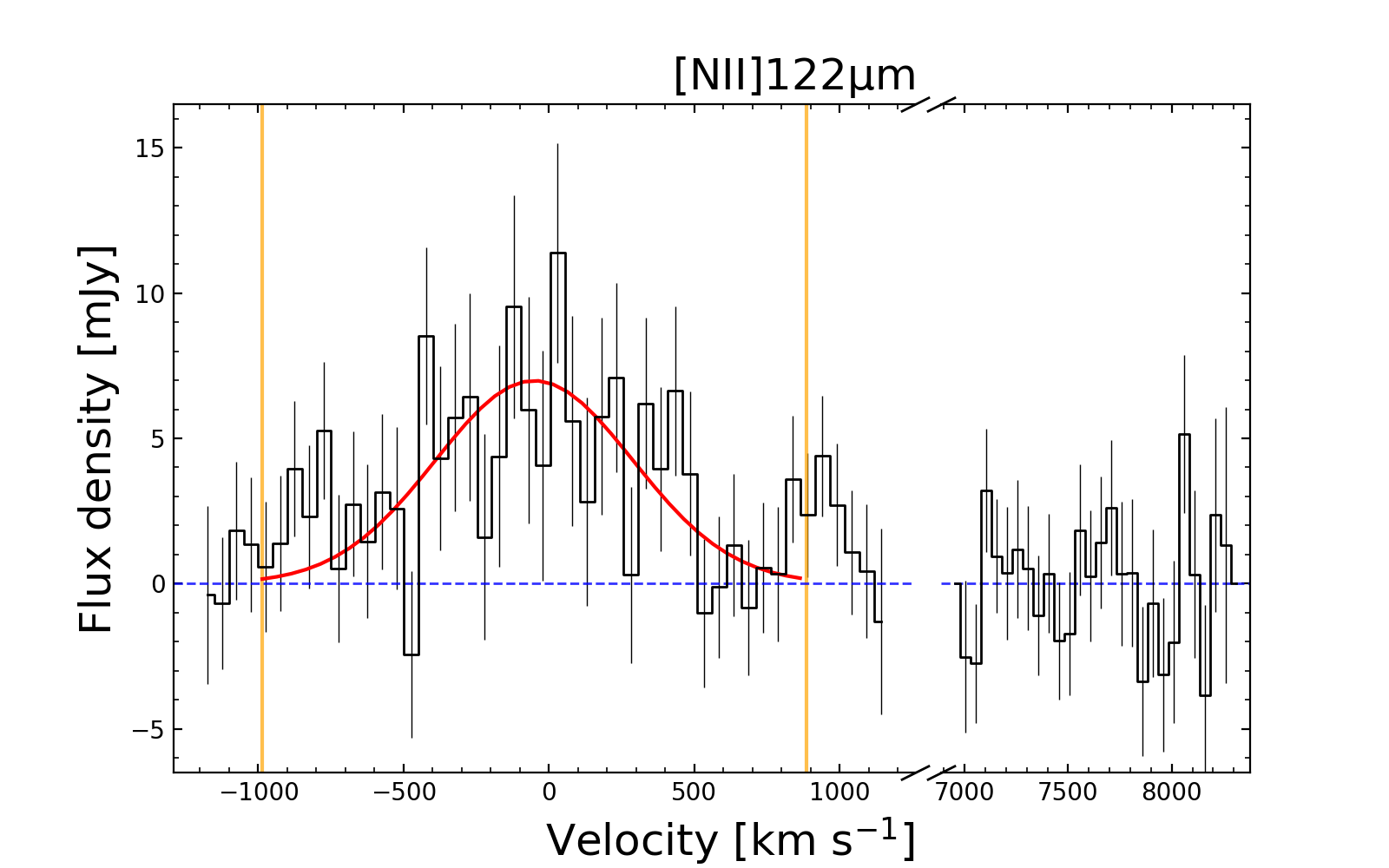}}
        \subfloat{\includegraphics[width=84mm]{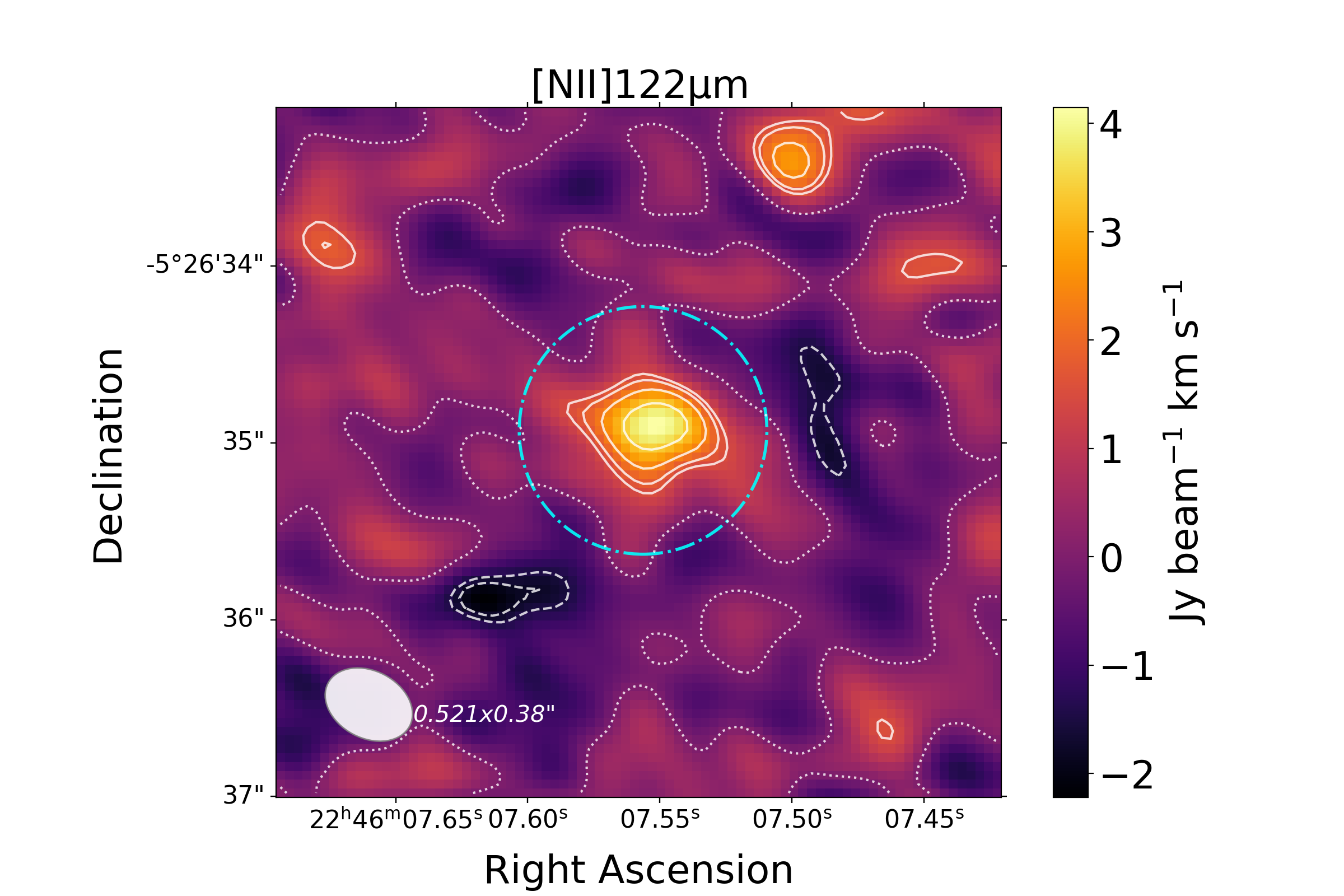}}
        \par
        \subfloat{\includegraphics[width=84mm]{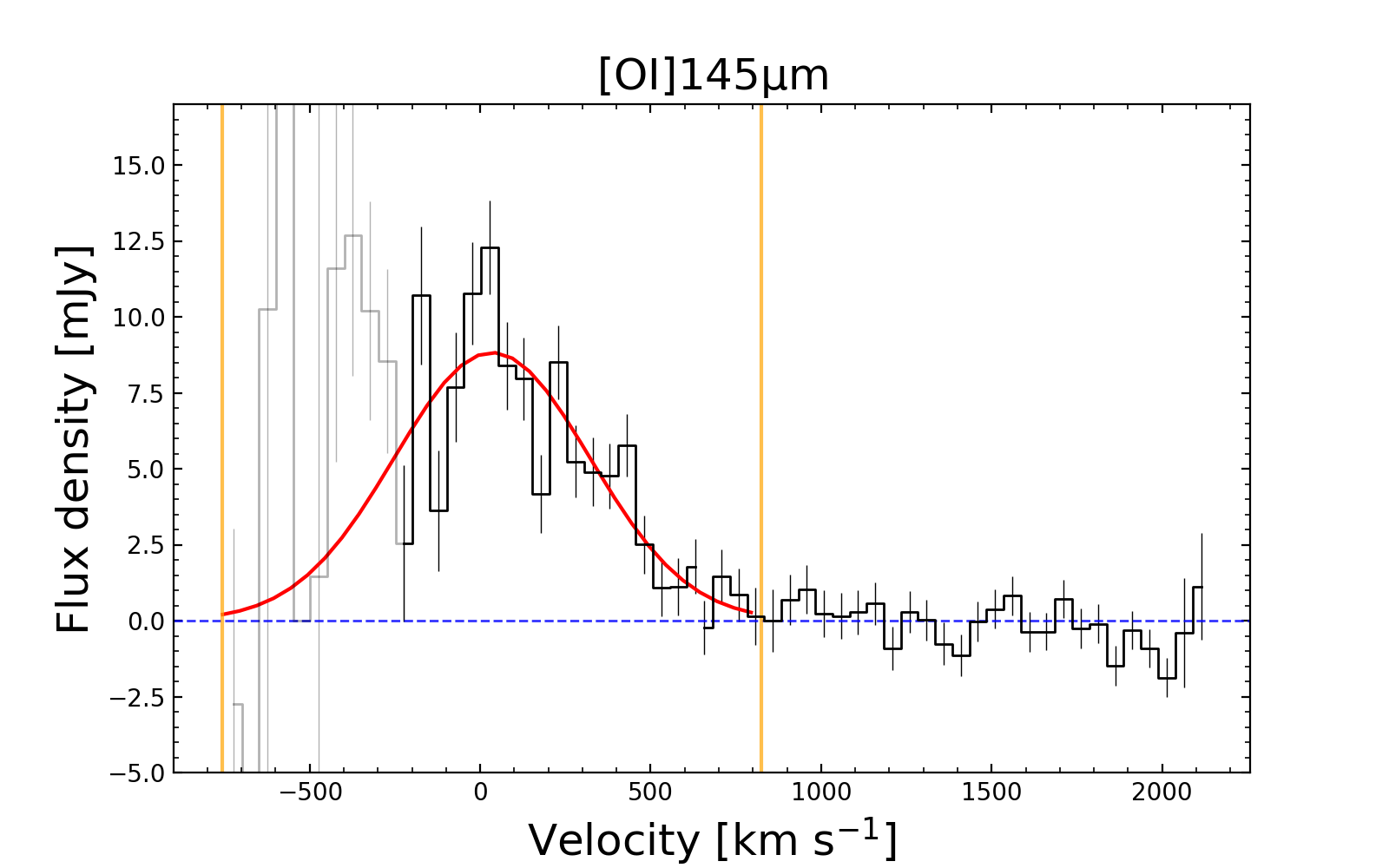}}
        \subfloat{\includegraphics[width=84mm]{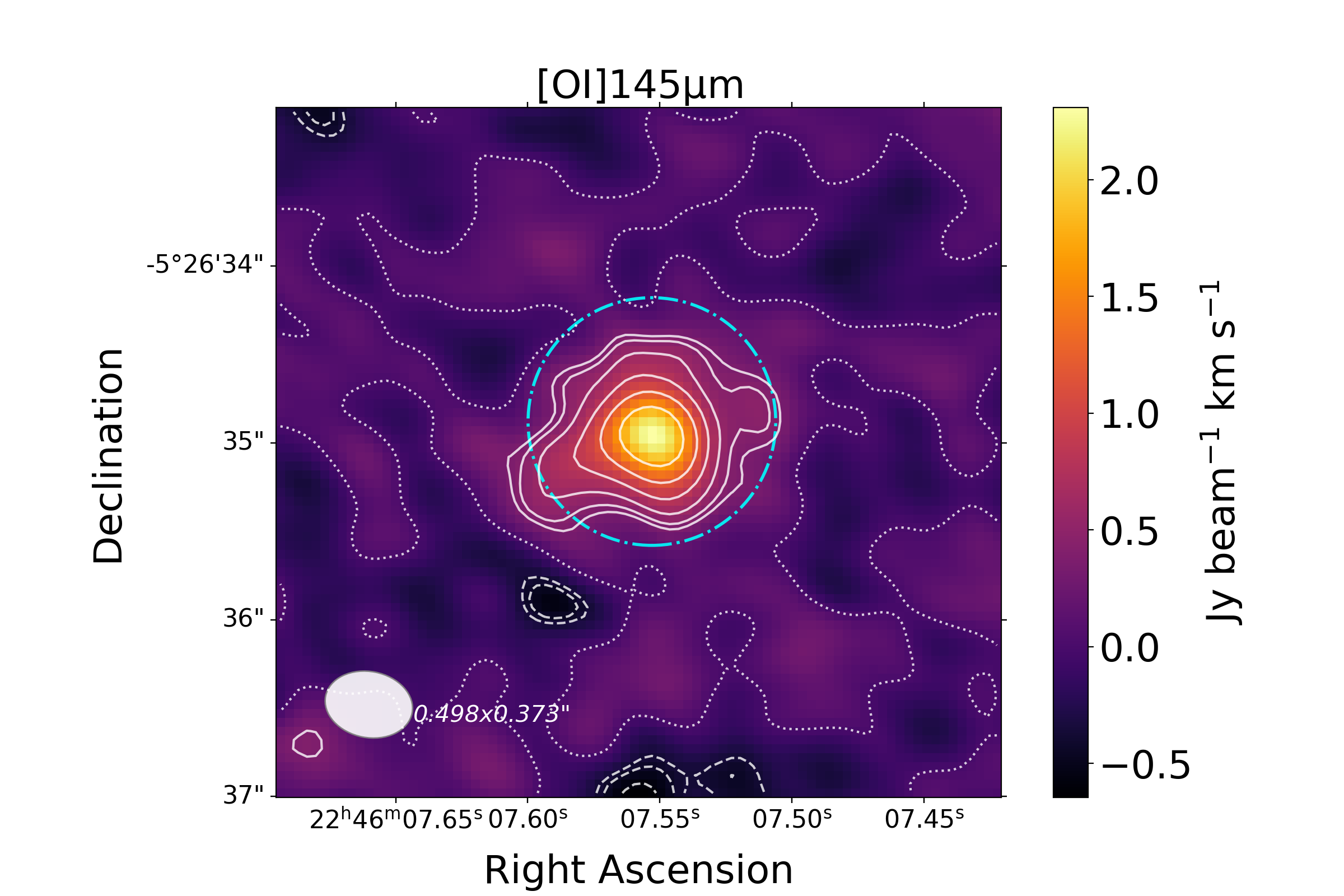}}      
        \par

    \caption{Spectra (left) and intensity map (right) for each observed line in W2246--0526. The blue dashed line indicates the zero flux density in each spectrum, the red line is the Gaussian fit for each emission line, and the two orange lines indicate the 2.75$\upsigma$ of each Gaussian fit that is used to integrate the flux of the emission line directly on the spectra. In the intensity maps, the blue dot-dashed line shows the aperture used to extract each spectrum, and the clean beam is shown at the bottom left. The zero-flux intensity level is shown as a white, dotted contour. Solid contours indicate [2.5, 3, 2$^{(4+n)/2}$]$\upsigma$ levels (with $\upsigma$ being the rms of the map, and $n=[0,1,2,...]$). Dashed contours show negative flux at the same absolute levels.}
    \label{fig:observations}
\end{figure*}

\begin{figure*}[!htp] 
    \ContinuedFloat 
    \centering
        \subfloat{\includegraphics[width=84mm]{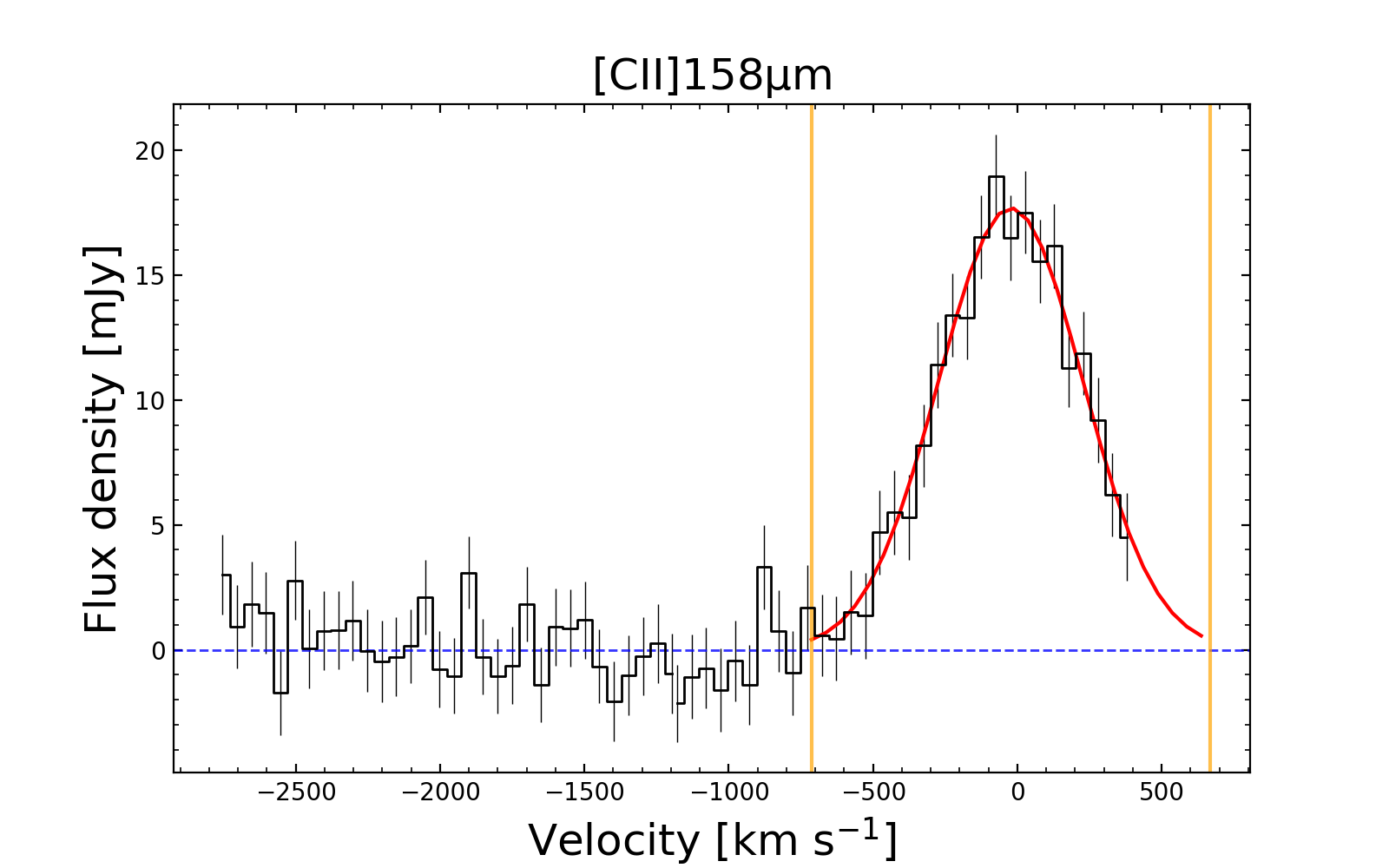}}
        \subfloat{\includegraphics[width=84mm]{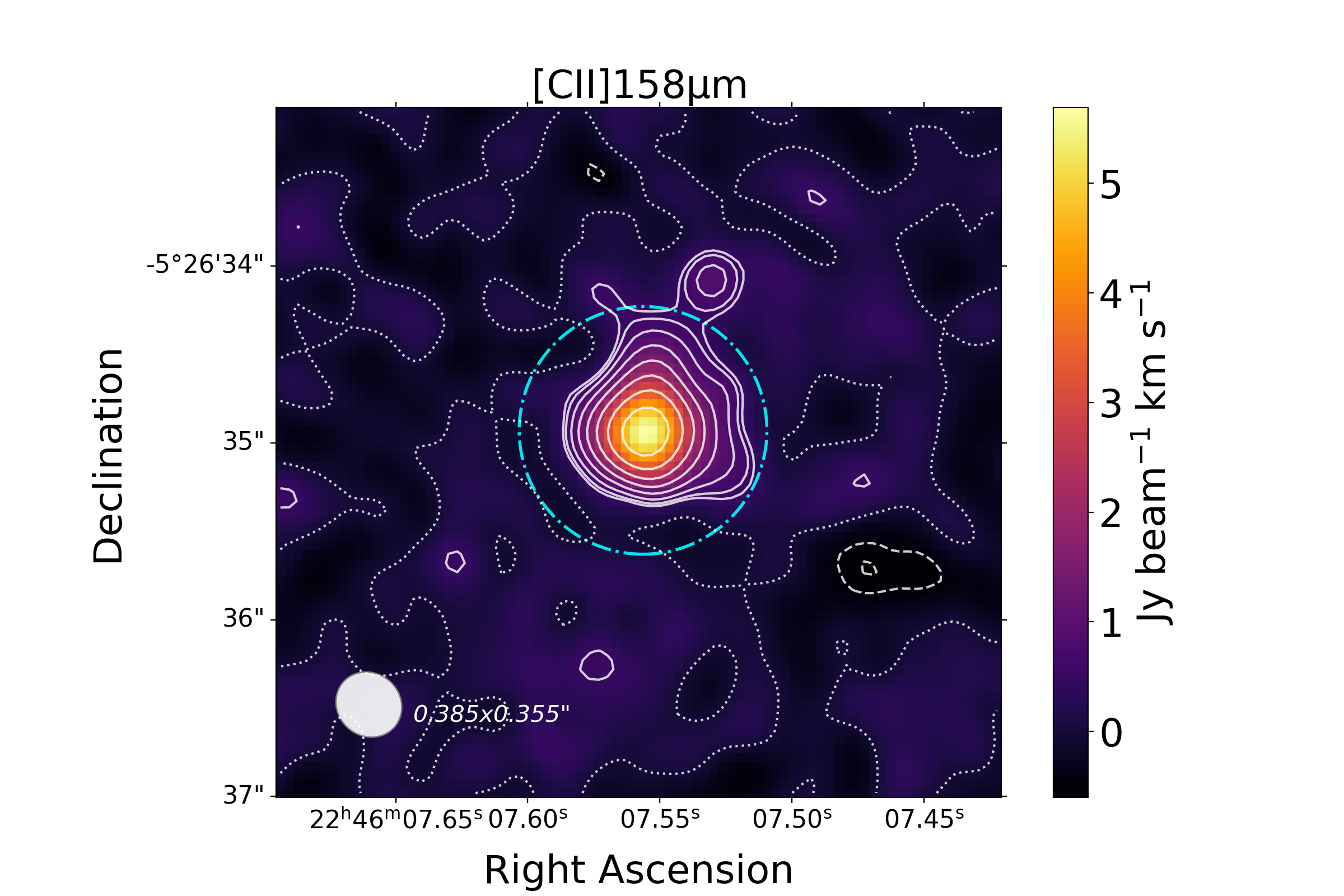}}   
        \par
        \subfloat{\includegraphics[width=84mm]{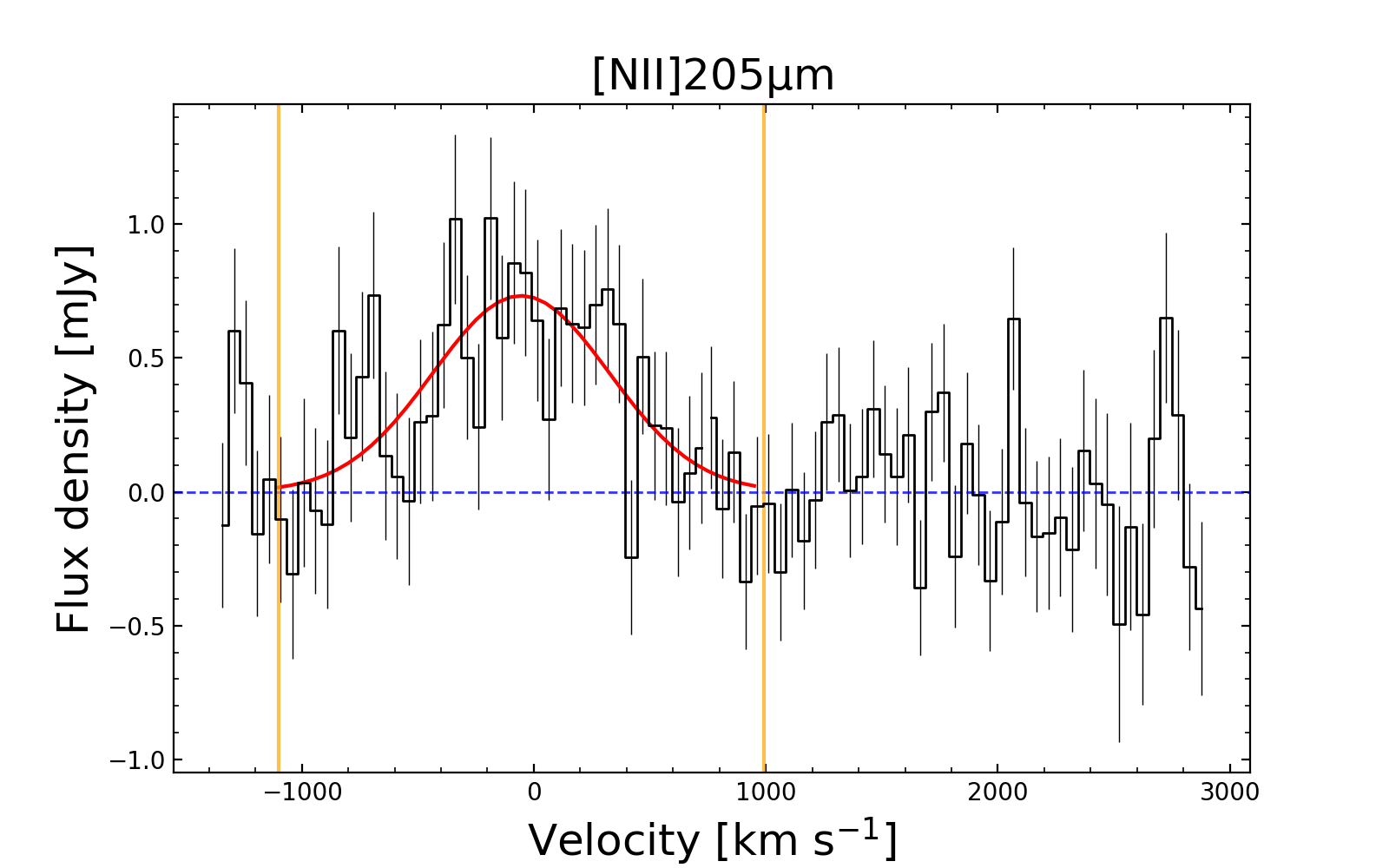}}
        \subfloat{\includegraphics[width=84mm]{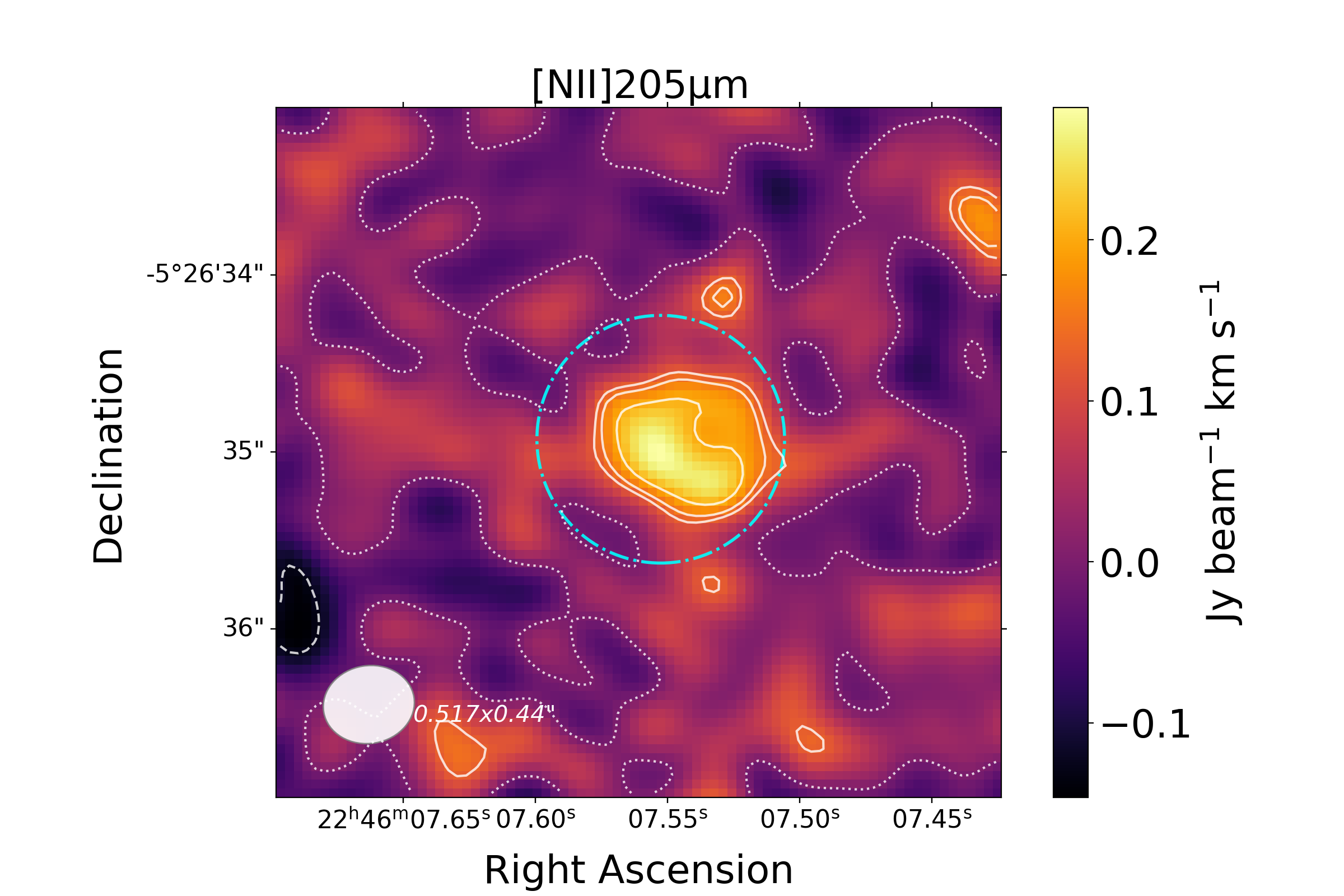}}
        \par
        \subfloat{\includegraphics[width=84mm]{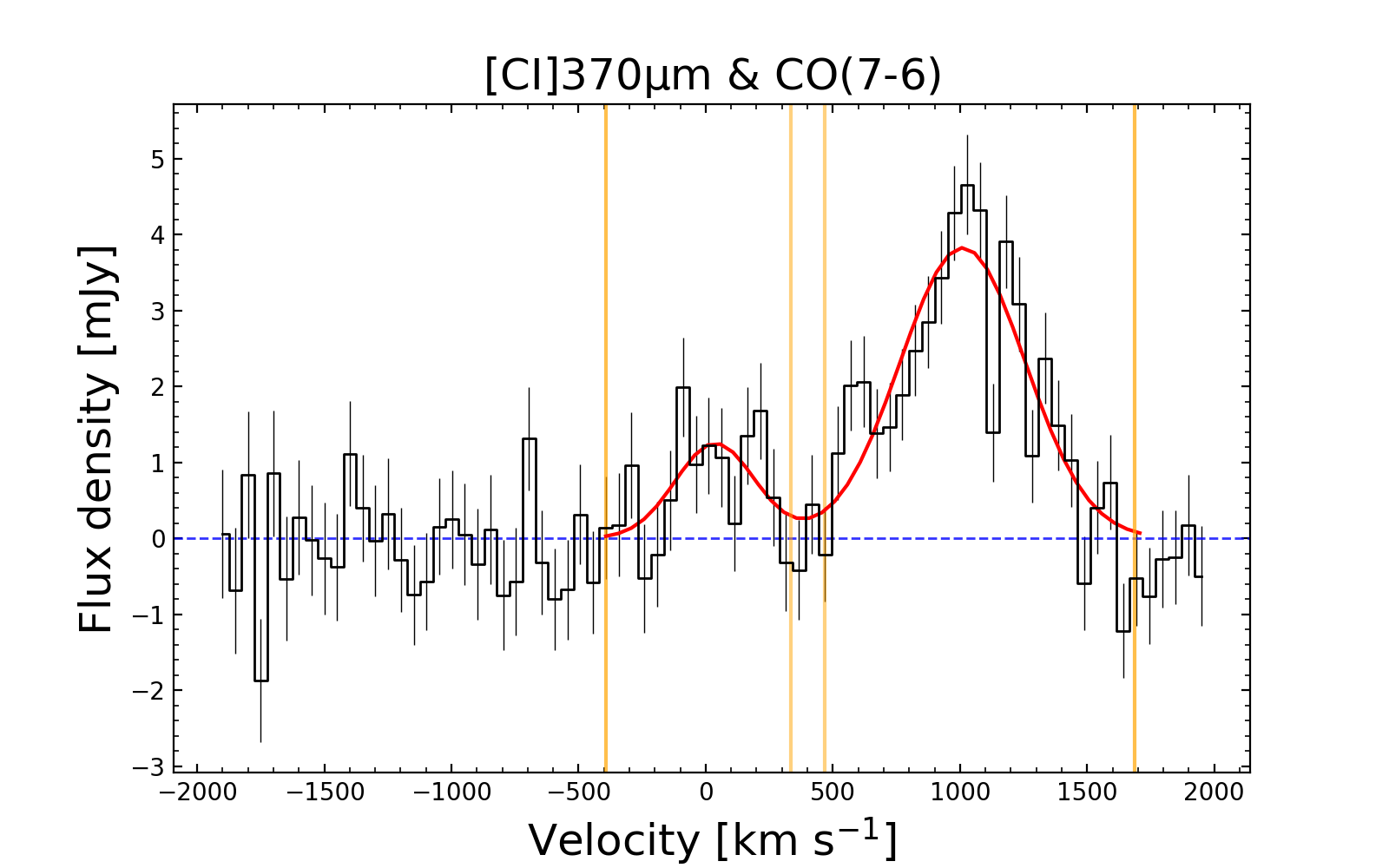}}
        \subfloat{\includegraphics[width=84mm]{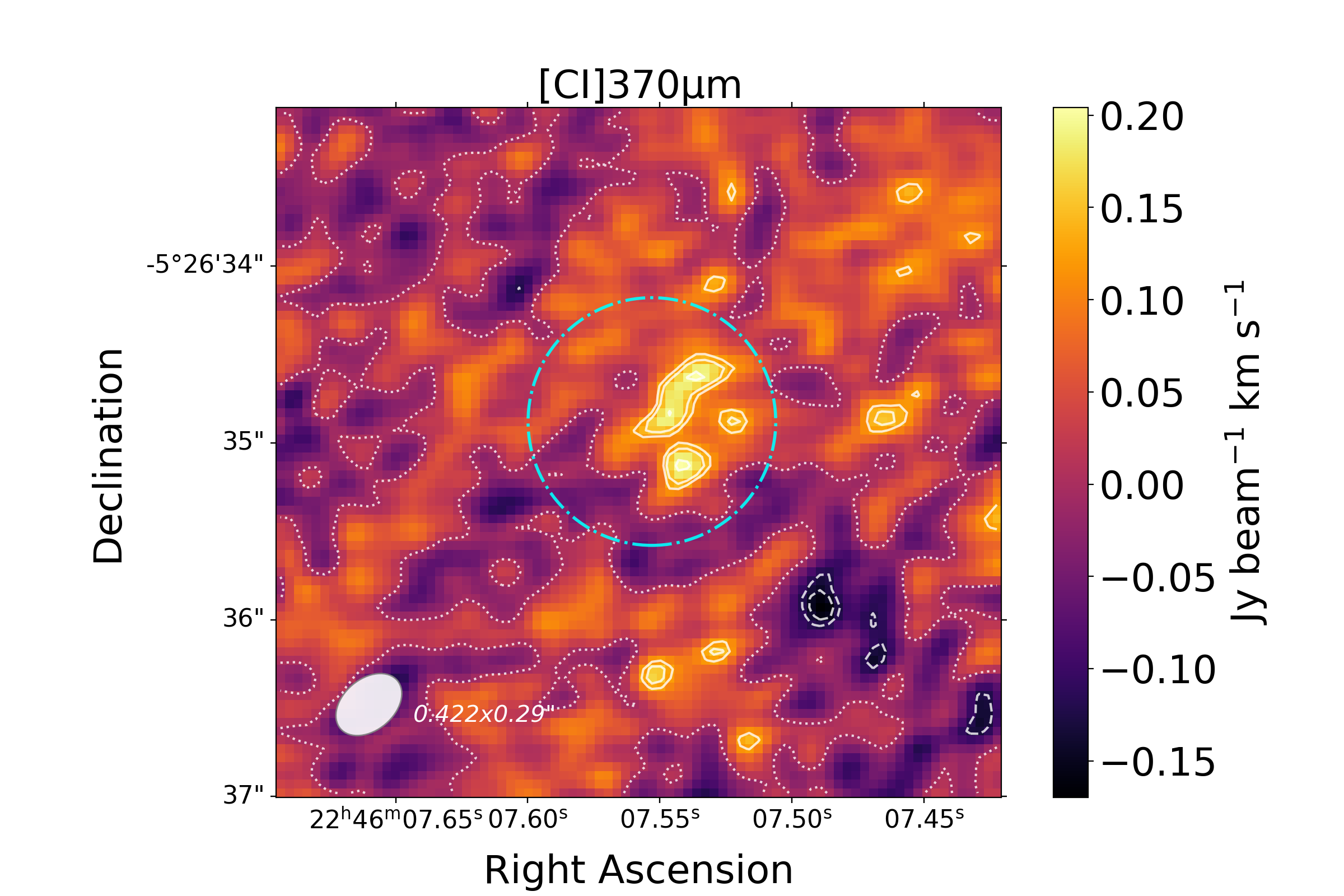}}
        \par
        \subfloat{\includegraphics[width=84mm]{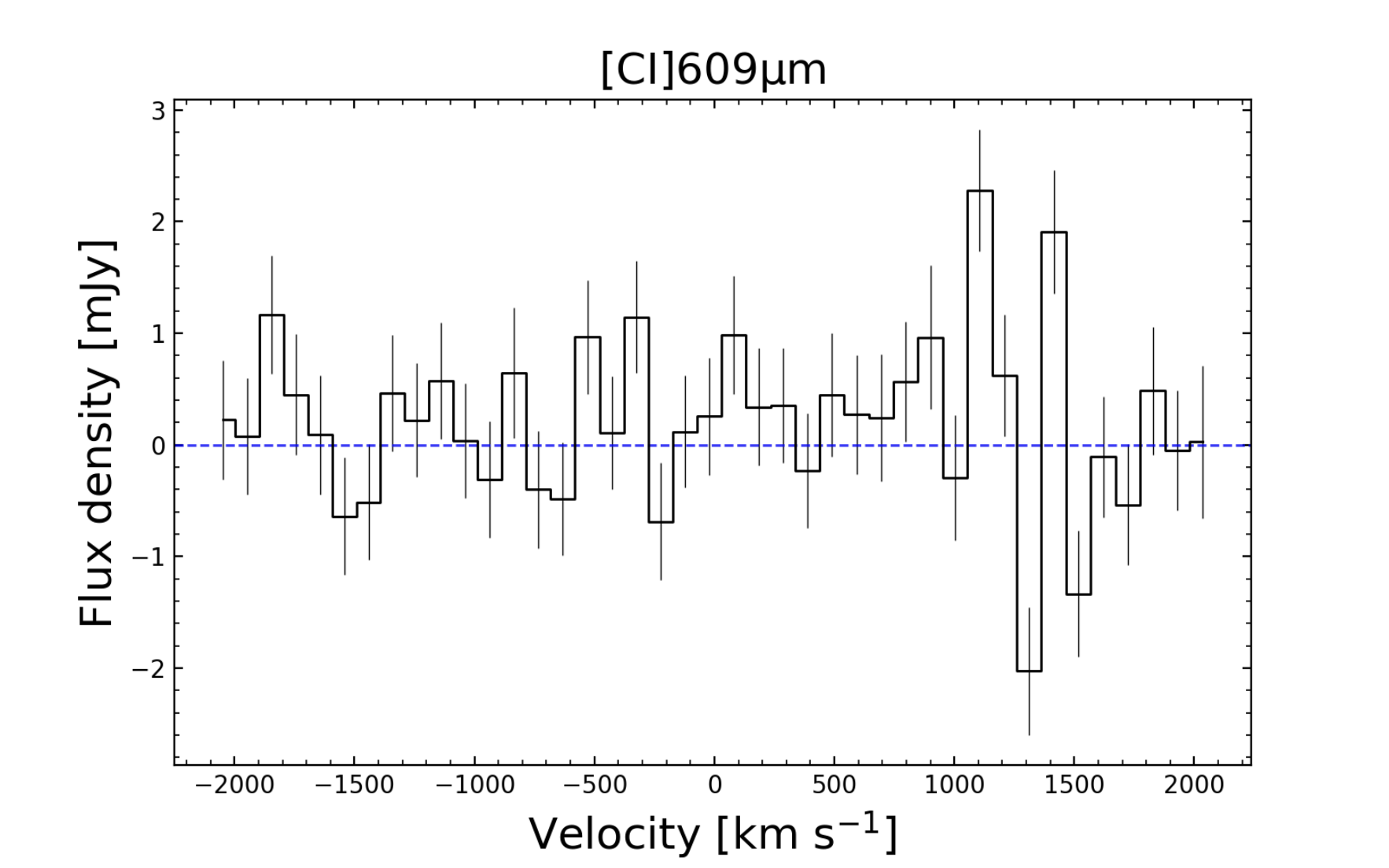}}
        \subfloat{\includegraphics[width=84mm]{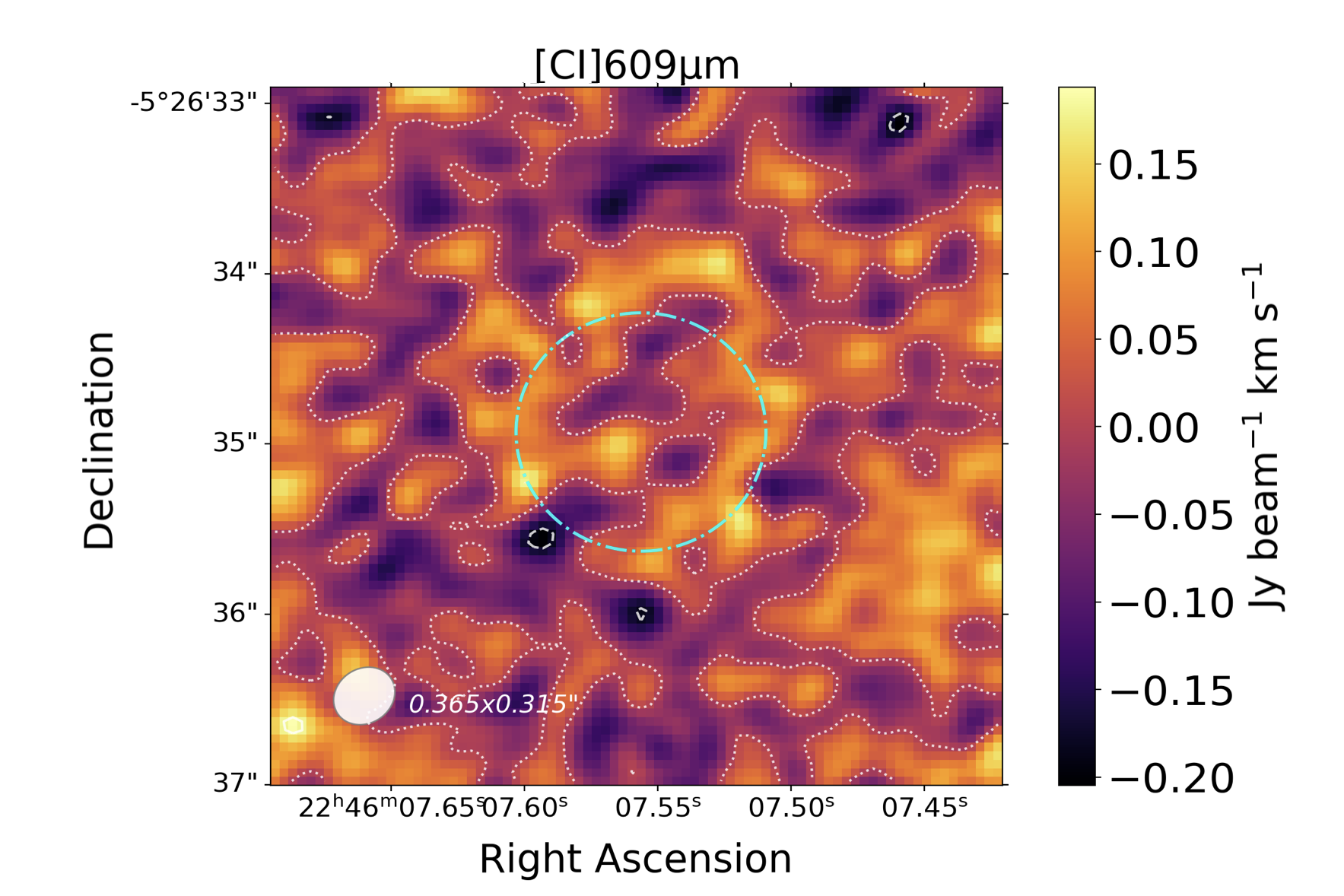}}
        \par
    
    \caption{Continuation.
    \vspace{4mm}}
\end{figure*}

W2246--0526 is surrounded by a complex structure of dust \citep{diaz2018}, close companion galaxies \citep{2016ApJ...816L...6D}, and Lyman-break galaxies at larger scales \citep{zewdie2023}. We note in Fig.~\ref{fig:observations} the resolved companion galaxy northwest of W2246--0526 in the [CII]$_{158\upmu \mathrm{m}}$ emission map \citep[labeled as C1 in][]{diaz2018}. The remaining lines are barely resolved or not resolved at all. W2246--0526 itself extends over $\sim 1\arcsec$ ($\sim 7$ kpc). The smallest maximum recoverable scale in our observations is $\sim2.7\arcsec$ for ALMA Bands 9 and 10, and therefore we can be sure that all the flux of the host galaxy is being recovered.
To perform the photometry and extract the spectra we used an aperture of diameter $0.7\arcsec$ centered at the peak emission pixel, enough to encircle the Hot DOG but not any of the known companion galaxies. This aperture size ensures that 99\% of the line flux is included assuming a Gaussian distribution the size of the largest beam in our observations ($0.517\arcsec\times0.440\arcsec$), meaning that effectively no aperture correction is needed. The continuum-subtracted spectra were then fitted with a single Gaussian (enough to describe all the line profiles) with the central wavelength of the line, peak intensity, and full width at half maximum (FWHM) as free parameters. To obtain the moment 0 map of the lines, the emission line cubes were collapsed using the channels within $\pm 2.75 \upsigma$ of the line peak (where $\upsigma$ is the standard deviation of the fitted Gaussian), a compromise between including most flux of the Gaussian to a level of $\sim99.5\%$, and avoiding including noise beyond the line.

As shown in Table~\ref{tab3}, all lines have a similar FWHM around 600--700 km s$^{-1}$, with the exception of [NII]$_{122\upmu \mathrm{m}}$, [NII]$_{205\upmu \mathrm{m}}$, and [CI]$_{370\upmu \mathrm{m}}$, with $\sim800$ km s$^{-1}$, $\sim900$ km s$^{-1}$, and $\sim350$ km s$^{-1}$ respectively. The range of line FWHM is close to the value previously found for [CII]$_{158\upmu \mathrm{m}}$ in W2246--0526 and other Hot DOGs \citep{2021A&A...654A..37D}. Interestingly, it is broader than the $\sim200$ km s$^{-1}$ typically observed for CO$(1-0)$ in Hot DOGs \citep[][]{2020MNRAS.496.1565P}. In general, the line FWHM in W2246--0526 are higher than high-redshift star-forming galaxies \citep[$\sim250$ km s$^{-1}$ in][]{2020A&A...643A...2B}, optical quasars \citep[$\sim385$ km s$^{-1}$ in][]{2018ApJ...854...97D}, or infrared quasars \citep[from $\sim200$ km s$^{-1}$ to $\sim500$ km s$^{-1}$ in][]{2017ApJ...836....8T}. We do not detect any evidence of underlying, broad emission line components (which may indicate outflows), suggesting that outflows may only be clearly identified in optical spectra of highly ionized gas \citep{2020ApJ...905...16F}, or that W2246--0526 does not have significant outflow activity at the moment.

The line luminosities are calculated following \cite{solomon1992}:
\begin{equation}
    L_{line}[\mathrm{L_{\odot}}] = 1.04\times10^{-3}\times S_{line}\Delta\upsilon \ \nu_{rest}(1+z)^{-1}D_{\mathrm{L}}^{2} \ ,
\end{equation}

\noindent
where $S_{line}\Delta\upsilon$ is the velocity integrated flux of the line over a range of $\pm2.75\upsigma$ from the line peak, in Jy km s$^{-1}$; $\nu_{rest}$ is the rest frequency in GHz; and $D_{\mathrm{L}}$ is the luminosity distance in Mpc. We also calculate the luminosity using the analytic expression of the Gaussian fit integral, especially for the cases in which the whole line profile is not covered by the observations. The fluxes and luminosities for all lines are presented, among other properties, in Table~\ref{tab3}.

The [CI]$_{609\upmu \mathrm{m}}$ and [OIII]$_{88\upmu \mathrm{m}}$ lines are not detected and we use them as upper limits in our analyses, calculated as the integrated luminosity of a Gaussian assuming a FWHM of 600 km s$^{-1}$ and a peak of $1\upsigma$ detection. To make sure the non-detections are not a product of the aperture choice or channel binning, we extracted the spectra using different aperture sizes and channel averaging. The lines were not detected in any case. We further discuss the non-detection in Sect.~\ref{sec:4.4}.

\begin{table*}
    \centering
    \caption{FIR fine-structure emission line properties}
    \begin{tabular}{*{10}{c}}
    \hline
     Line & $f_{obs}$ & $f_{fit}$ & $L_{obs}$ & $L_{fit}$ & FWHM & Beam & PA & S/N & r.m.s. \\
      & & & & & & & & & [mJy \\
      & [Jy km s$^{-1}$] & [Jy km s$^{-1}$] & [$10^9\mathrm{L_{\odot}}$] & [$10^9$L$_{\odot}$] & [km s$^{-1}$] & [$\arcsec$] & [$^{\circ}$] & & beam$^{-1}$] \\
     \hline 
     {[OI]$_{63\upmu \mathrm{m}}$} & $20.6 \pm 2.9$ & $20.8 \pm 4.4$ & $32.4 \pm 4.6$ & $32.7 \pm 6.9$ & $679 \pm 111$ & $0.395\times0.328$ & 83.0 & 7.4 & 4.04\\     
     {[OIII]$_{88\upmu \mathrm{m}}$} & ... & $<1.38$ & ... & $<1.61$ & ... & $0.363\times0.309$ & 88.7 & ... & 1.96$^{*}$ \\ 
     {[NII]$_{122\upmu \mathrm{m}}$} & $6.27 \pm 0.89$ & $5.73 \pm 1.26$ & $5.12 \pm 0.73$ & $4.68 \pm 1.03$ & $801 \pm 121$ & $0.521\times0.380$ & 62.2 & 6.7 & 1.10 \\
     {[OI]$_{145\upmu \mathrm{m}}$} & $5.07 \pm 0.44$ & $6.64 \pm 1.33$ & $3.47 \pm 0.30$ & $4.54 \pm 0.91$ & $677 \pm 117$ & $0.410\times0.330$ & -83.6 & 11.6 & 1.47 \\   
     {[CII]$_{158\upmu \mathrm{m}}$} & $10.5 \pm 0.4$ & $11.1 \pm 0.5$ & $6.60 \pm 0.26$ & $6.98 \pm 0.34$ & $590 \pm 23$ & $0.385\times0.355$ & 48.5 & 26.3 & 0.58 \\ 
     {[NII]$_{205\upmu \mathrm{m}}$} & $0.68 \pm 0.08$ & $0.70 \pm 0.14$ & $0.33 \pm 0.04$ & $0.34 \pm 0.07$ & $894 \pm 140$ & $0.517\times0.440$ & -80.5 & 7.2 & 0.12 \\  
     {[CI]$_{370\upmu \mathrm{m}}$} & $0.41 \pm 0.15$ & $0.45 \pm 0.22$ & $0.11 \pm 0.04$ & $0.12 \pm 0.06$ & $368 \pm 128$ & $0.422\times0.290$ & -49.8 & 3.3 & 0.20 \\
      {[CI]$_{609\upmu \mathrm{m}}$} & ... & $<0.08$ & ... & $<0.01$ & ... & $0.365\times0.315$ & -48.4 & ... &  0.16$^{*}$ \Tstrut \\
     \hline
     \\
    \end{tabular}
    \caption*{Note: Columns correspond to the line, flux and luminosity of both the data and the Gaussian fit, full width at half maximum (FWHM) of the Gaussian fit, size, and angle of the beam, signal-to-noise ratio (S/N) of the line, and depth (root mean square; r.m.s.) per channel. The channel width for all but [OIII]$_{88\upmu \mathrm{m}}$ and [CI]$_{609\upmu \mathrm{m}}$ FIR lines is 50 km s$^{-1}$ while the channel width for those two lines (denoted by $^{*}$) is 100 km s$^{-1}$, in order to enhance the S/N and possible detection of the lines. The upper limits correspond to the integrated luminosity of a Gaussian assuming a FWHM of 600 km s$^{-1}$ and a peak of 1$\upsigma$ detection.}
    \label{tab3}
\end{table*}

\section{Results}
\label{sec:results}

The targeted lines are among the brightest FIR fine-structure lines that can be observed in star-forming and AGN-dominated galaxies, and they trace different gas phases and their physical properties. When modeling and studying the ISM phases, a common division is made in the literature between PDRs and XDRs, depending on the dominant source of radiation (optical-UV photons or X-rays, respectively; see Fig.~\ref{fig:PDR}). X-ray photons can penetrate deeper into the ISM than UV photons, leading to a change in the ionization structure, chemistry, and physical conditions within the cloud. The difference between PDRs and XDRs, in terms of their effect on the gas phases and FIR lines emission, has been extensively discussed in the literature. For further details, we refer the reader to the review of \cite{2022ARA&A..60..247W}. Our \textsc{Cloudy} modeling indicates that the observed line ratios in W2246--0526 are reproducible with a single XDR component, which dominates the observed emission. We begin this section by discussing individual line ratios. Next we describe the setup of the CLOUDY models, followed by an explanation of the insights the models provide on the line ratio diagnostics, and conclude with a description of the best-fit model.

\subsection{Individual line ratio diagnostics}
\label{sec:3.1}

\begin{figure*}[h!]
    \centering
        \subfloat{\includegraphics[width=80mm]{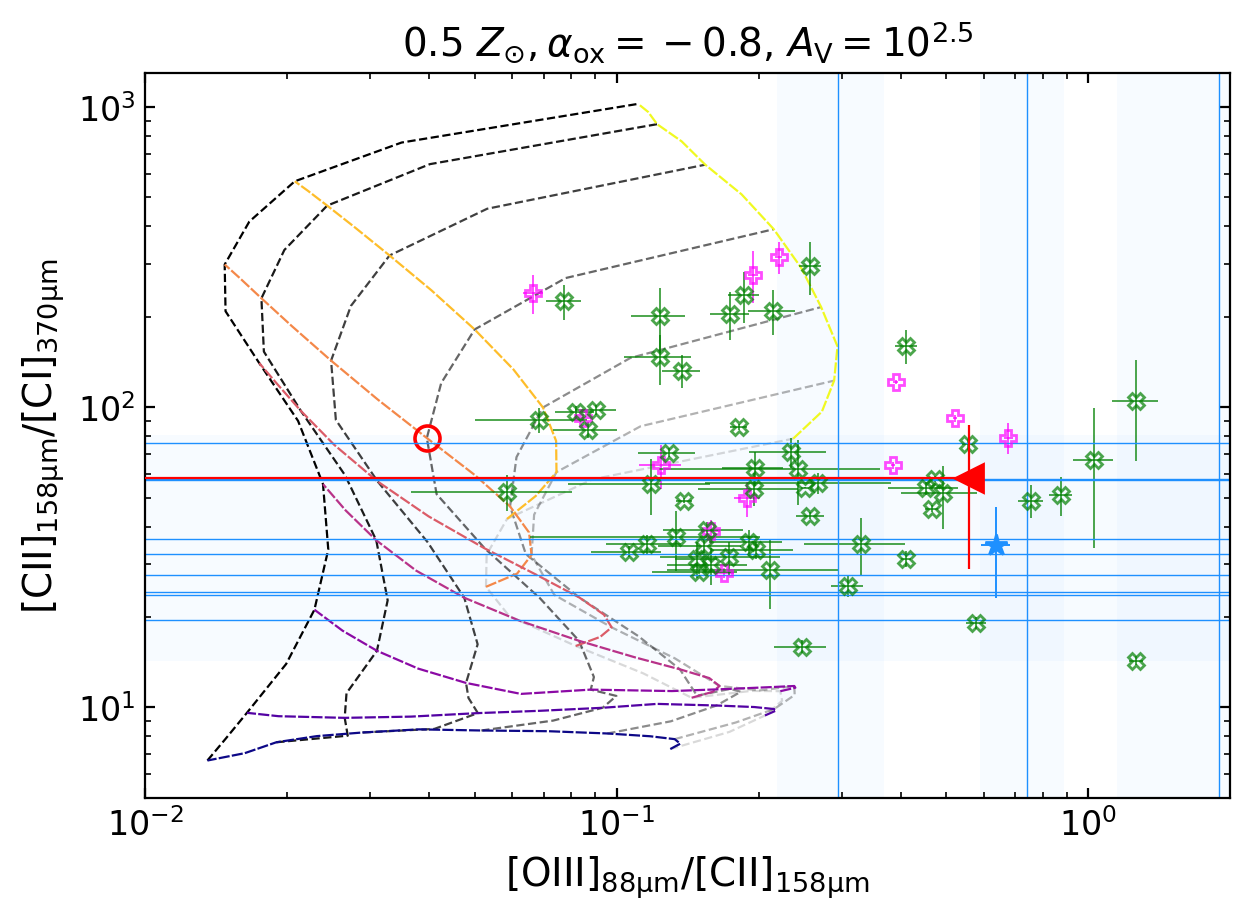}}
        \subfloat{\vspace{-0.5mm}\includegraphics[width=102mm,height=60.5mm] {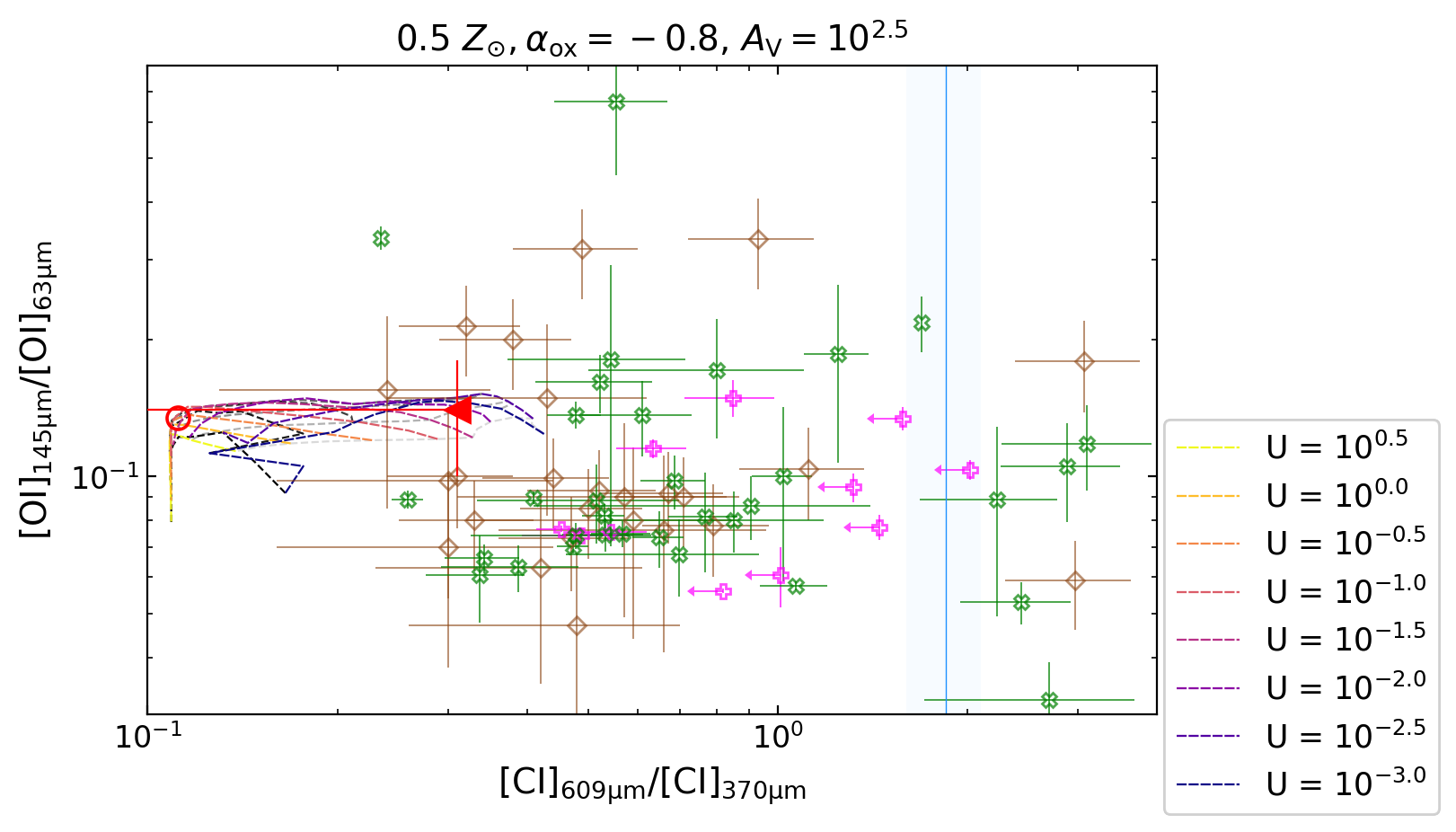}}\hspace{1.9cm}    
        \subfloat{\includegraphics[width=80mm]{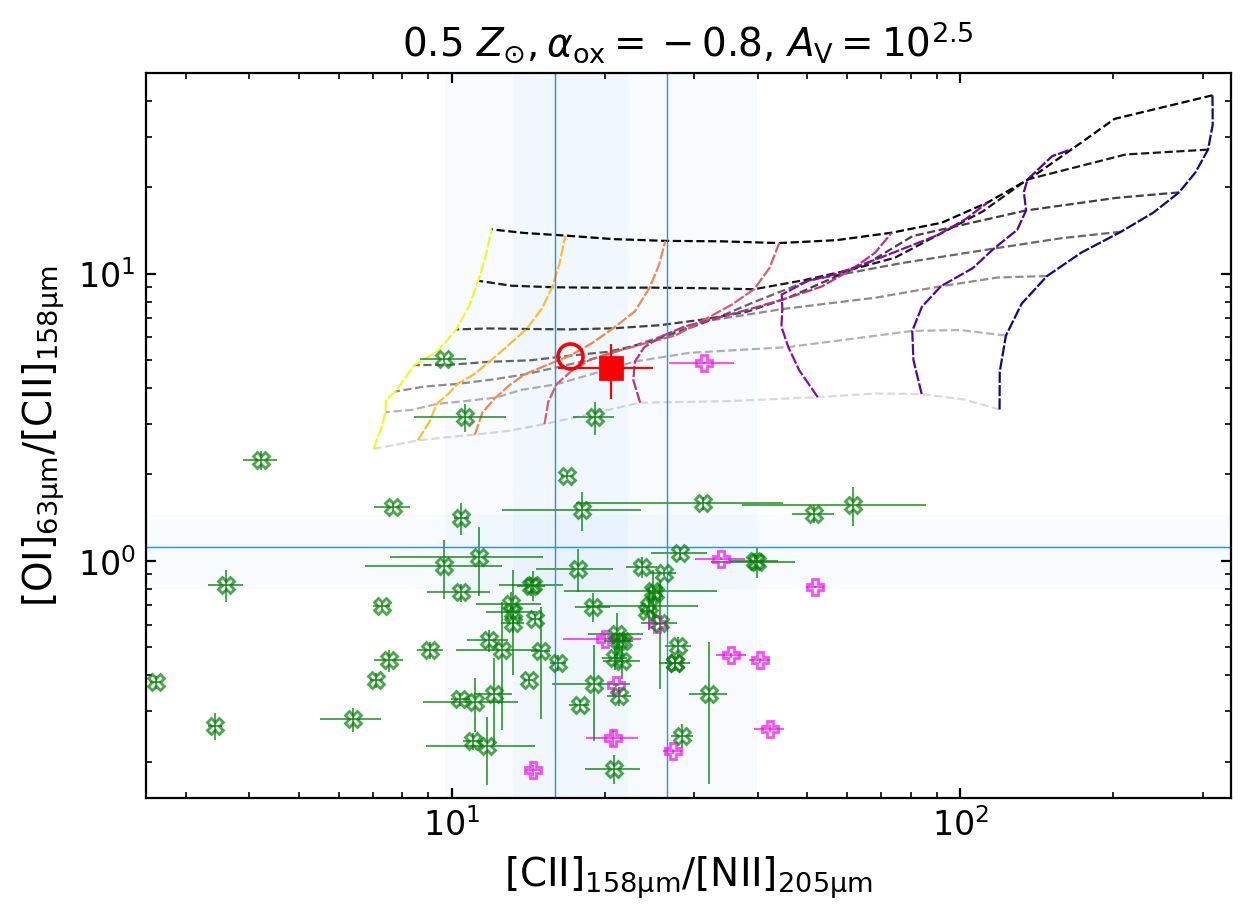}}
        \subfloat{\includegraphics[width=100mm,height=60mm]{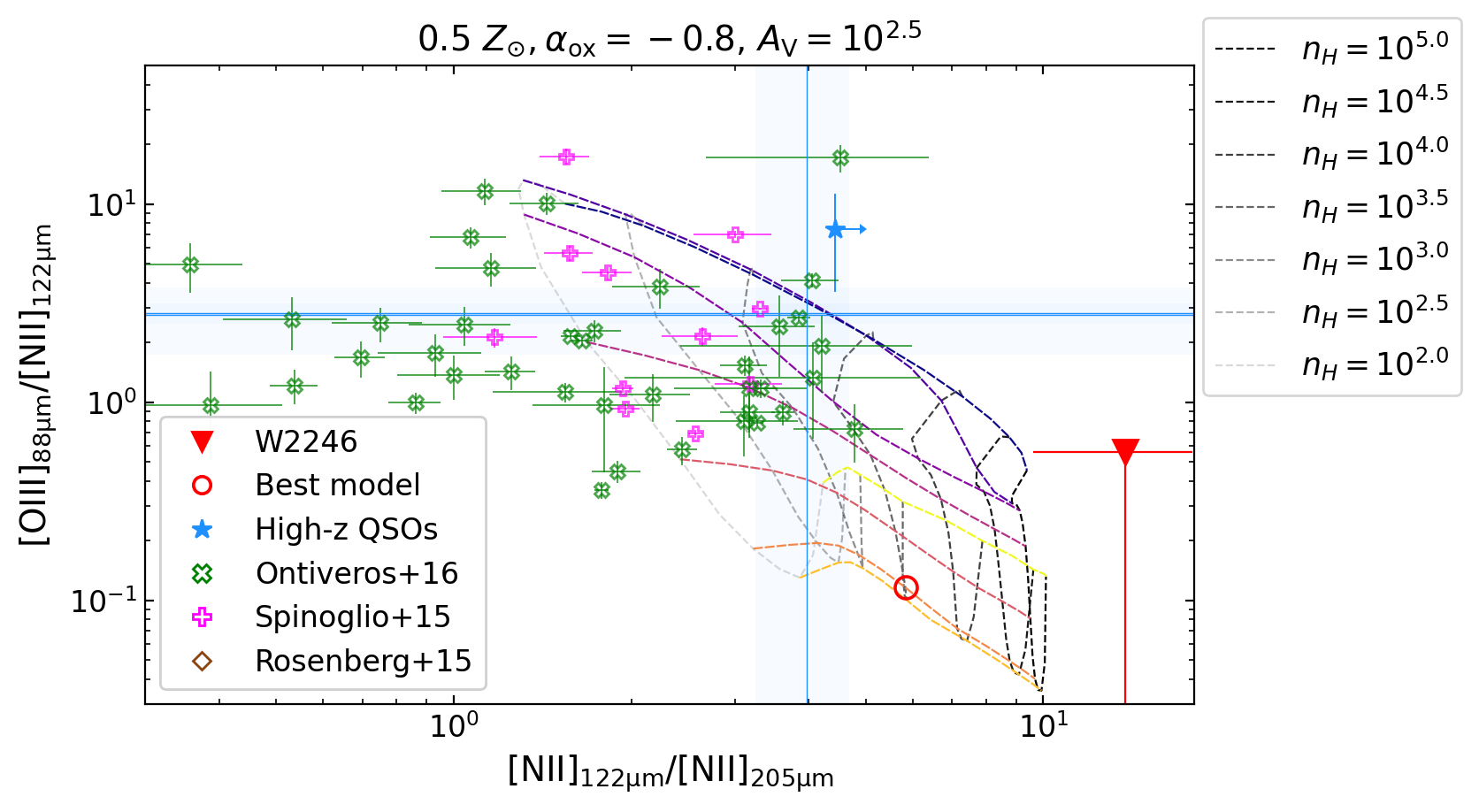}}
        \caption{Line ratio diagrams from the \textsc{Cloudy} models for the metallicity, $\alpha_{\mathrm{ox}}$ and $A_{\mathrm{V}}$ of the model with the lowest $\chi_{\nu}^2$ described in Sect.~\ref{sec:3.4}. The observed line ratios for W2246--0526 are indicated with a red square, and a red triangle for upper limits. The color grid spans the full range of values explored in this work for the ionization parameter and the density, as shown in the legend. The red open circle indicates the best-fit value for our source (i.e., lowest $\chi_{\nu}^2$ model). Literature values for high-z quasars \citep{weiss2003,uzgil2016,venemans2017,walter2018,lee2019,novak2019,2019ApJ...880..153Y,li2020,pensabene2021,meyer2022,2022A&A...662A..60D,decarli2023} are indicated with blue stars, or solid blue lines with shadow uncertainties if only one ratio is available. Literature values for local AGNs are indicated with open green crosses for \cite{2016ApJS..226...19F} and open magenta plus signs for \cite{2015ApJ...799...21S}, while local ULIRGs from \cite{rossenberg2015} are indicated with open brown diamonds.}
        \label{fig:grids}
\end{figure*}

Before interpreting the observations using the \textsc{Cloudy} models, we analyze key individual line luminosity ratios that are very useful to infer specific properties of the ISM with limited information. These line ratios are presented in Table~\ref{tab4}, with the Gaussian fits to the observed data, and the best-fit model explained in Sect.~\ref{sec:3.4}. We also report the observed line ratios when the FWHM is fixed to 600 km s$^{-1}$ for all the lines, to discard a major influence in the ratios due to line width differences. The same line ratios are shown in Fig.~\ref{fig:grids}.

\begin{table*}
    \centering
    \caption{Line ratios used as diagnostics}
    \begin{tabular}{*{10}{c}}
    \hline
     Line ratio & Data & Gaussian fit & Best model & Fixed FWHM
     \\
     \hline
     {[OIII]$_{88\upmu \mathrm{m}}$/[NII]$_{122\upmu \mathrm{m}}$} & ... & <0.34 & 0.12 & <0.39 \\
     {[NII]$_{122\upmu \mathrm{m}}$/[NII]$_{205\upmu \mathrm{m}}$} & $15.5\pm2.9$ & $13.8\pm4.2$ & 5.8 & $15.6\pm5.4$\\
     {[OI]$_{145\upmu \mathrm{m}}$/[OI]$_{63\upmu \mathrm{m}}$} & $0.11\pm0.02$ & $0.14\pm0.04$ & 0.13 & $0.12\pm0.03$ \\
     {[CI]$_{609\upmu \mathrm{m}}$/[CI]$_{370\upmu \mathrm{m}}$} & ... & <0.10 & 0.11 & <0.08 \\
     {[OI]$_{63\upmu \mathrm{m}}$/[CII]$_{158\upmu \mathrm{m}}$} & $4.91\pm0.72$ & $4.70\pm1.0$ & 5.2 & $4.66\pm0.89$\\
     {[CII]$_{158\upmu \mathrm{m}}$/[NII]$_{205\upmu \mathrm{m}}$} & $20.0\pm2.6$ & $20.5\pm4.3$ &  17 & $26.8\pm6.4$\\
     {[CII]$_{158\upmu \mathrm{m}}$/[CI]$_{370\upmu \mathrm{m}}$} & $60\pm22$ & $58\pm29$ & 79 & $45\pm24$\\
     {[OIII]$_{88\upmu \mathrm{m}}$/[CII]$_{158\upmu \mathrm{m}}$} & ... & <0.23 & 0.040 & <0.23\\
     \hline
     \\
    \end{tabular}
    \caption*{Note: Columns correspond to the line ratios measured directly from the data, the Gaussian fit of the lines, the ratio of the best-fit model described in Sect.~\ref{sec:3.4}, and also the best-fit model ratio but when assuming a Gaussian with a fixed FWHM of 600 km s$^{-1}$ for all the lines. For non-detections we report 1$\upsigma$ upper limits.}
    \label{tab4}
\end{table*}

\subsubsection{Ionized carbon and nitrogen}
\label{sec:3.1.1}
Owing to their ionization potentials, the emission from the [CII]$_{158\upmu \mathrm{m}}$ line in a PDR arises from both the neutral and the ionized gas medium, while [NII]$_{205\upmu \mathrm{m}}$ is produced only in the ionized phase. Because of this, the ratio between [CII]$_{158\upmu \mathrm{m}}$ and [NII]$_{205\upmu \mathrm{m}}$ has been extensively used to disentangle the fraction of [CII]$_{158\upmu \mathrm{m}}$ produced in PDRs \citep[e.g.,][]{2006MNRAS.368.1949A,oberst2006,2015ApJ...814..133G,2017ApJ...845...96C,diaz2017}. Within this context, the observed [CII]$_{158\upmu \mathrm{m}}$/[NII]$_{205\upmu \mathrm{m}}$ ratio of $20.5\pm4.3$ would correspond to an $\sim80\%$ of the [CII]$_{158\upmu \mathrm{m}}$ line emission produced in the neutral ISM (following the model of \cite{oberst2006} for $n_e\geq10^3$ cm$^{-3}$). However, as mentioned before, the presence of an XDR changes the chemistry of the ISM, and as described by \cite{pereira2017}, X-ray radiation from an AGN can cause nitrogen to be single ionized in regions where hydrogen is mainly neutral. Both [CII]$_{158\upmu \mathrm{m}}$ and [NII]$_{205\upmu \mathrm{m}}$ lines will arise from similar regions of the ISM, with their ratio strongly depending on the parameters of the model. This ratio is shown in the x-axis of the bottom left panel of Fig.~\ref{fig:grids}. 

\subsubsection{Optically thick [OI]$_{63 \upmu \mathrm{m}}$}
\label{sec:3.1.2}

The neutral oxygen ratio [OI]$_{145\upmu \mathrm{m}}$/[OI]$_{63 \upmu \mathrm{m}}$ provides information about the density and temperature of the dense neutral gas if both lines are optically thin, but the 63 $\upmu \mathrm{m}$ line is usually optically thick in most of ISM conditions \citep{1985ApJ...291..722T}. For W2246--0526, we obtain [OI]$_{145\upmu \mathrm{m}}$/[OI]$_{63 \upmu \mathrm{m}}$ = $0.14\pm0.04$. The [OI]$_{63 \upmu \mathrm{m}}$ line can become optically thick at $N_{\mathrm{H}} > 2\times10^{21}$ cm$^{-2}$ \citep{1985ApJ...291..747T}, implying that a value of the [OI]$_{145\upmu \mathrm{m}}$/[OI]$_{63 \upmu \mathrm{m}}$ ratio of $\sim$0.10 or higher is indicative of optically thick  [OI]$_{63 \upmu \mathrm{m}}$ emission. 

An alternative scenario for high values of this ratio that does not require [OI]$_{63 \upmu \mathrm{m}}$ to be optically thick is that the line emission is absorbed by less excited (cooler or more dense) neutral oxygen along the line of sight \citep{1998ApJ...503..785K}, where even small amounts of cold foreground neutral oxygen in the ground state could cause this effect \citep{2006A&A...446..561L}. However, we do not observe narrow absorption components in the [OI]$_{63 \upmu \mathrm{m}}$ spectral profile of W2246--0526, and therefore the optically thick scenario remains as the most likely cause of the observed high ratio.

An additional effect that could increase the oxygen ratio is the masering of [OI]$_{145 \upmu \mathrm{m}}$, which may occur via population inversion within warm giant molecular clouds \citep{2006MNRAS.365..779E,2006A&A...446..561L}, although it is not expected to affect the line emission significantly. Finally, dust obscuration can also act as a source of opacity at lower wavelengths, dimming more effectively the [OI]$_{63 \upmu \mathrm{m}}$ emission line compared to [OI]$_{145 \upmu \mathrm{m}}$. However, we can discard this effect as a potential mechanism for faint [OI]$_{63 \upmu \mathrm{m}}$ emission since we see that the ratio of [OI]$_{63 \upmu \mathrm{m}}$ to emission lines at longer wavelengths (e.g., [CII]$_{158 \upmu \mathrm{m}}$) is not particularly low when compared to other galaxies (see bottom left panel of Fig.~\ref{fig:grids}). The neutral oxygen ratio is shown in the y-axis of the top right panel of Fig.~\ref{fig:grids}.

\subsubsection{XDR diagnostics}
\label{sec:3.1.3}

The neutral carbon [CI]$_{609\upmu \mathrm{m}}$/[CI]$_{370 \upmu \mathrm{m}}$ ratio primarily depends on the temperature of the cold neutral gas, with higher temperatures lowering the value \citep{meijerink2007,papadopoulos2022}. The non-detection of [CI]$_{609\upmu \mathrm{m}}$ in W2246--0526 sets a 3$\upsigma$ upper limit of [CI]$_{609\upmu \mathrm{m}}$/[CI]$_{370 \upmu \mathrm{m}}<0.3$. The presence of intense X-ray emission is expected for this range of ratios, $\lesssim 0.19$ \citep{2016ApJS..226...19F}, as X-rays penetrate deeper in the cloud than UV radiation and can be an important source of heating in the neutral medium. The neutral carbon ratio is shown in the x-axis of the top right panel of Fig.~\ref{fig:grids}.

Together with the [CI]$_{370 \upmu \mathrm{m}}$ observations, the CO J=7--6 transition is also detected in W2246--0526 (Fig. \ref{fig:observations}), with a luminosity of $0.59\pm0.04\times10^9$ L$_{\odot}$. CO(7--6) is a warm dense gas tracer \citep[e.g.,][]{carilli2013}. Mid-$J$ CO transitions such as $J=7-6$ are more intense in XDRs \citep[][e.g.,]{meijerink2007,vallini2019}, and indicative of AGN heating. Comparing with the compiled galaxies and models of \cite{valentino2020} and \cite{hagimoto2023}, the very low ratio of [CI]$_{370 \upmu \mathrm{m}}$/CO$(7-6)=0.19\pm0.07$ and low neutral carbon luminosity for W2246--0526 points toward a high-density and highly ionized ISM, compatible with the existence of XDRs. 

Since in XDRs neutral carbon emission is not limited to a thin layer (as in PDRs) but distributed more smoothly through the ISM, the [CII]$_{158\upmu \mathrm{m}}$/[CI]$_{370 \upmu \mathrm{m}}$ ratio is also used to discern between PDR and XDR models \citep{venemans2017,novak2019,pensabene2021,2022A&A...662A..60D,decarli2023}. Ratios of [CII]$_{158\upmu \mathrm{m}}$/[CI]$_{370 \upmu \mathrm{m}} > 20$ discard XDR models in these previous studies. Our observation of [CII]$_{158\upmu \mathrm{m}}$/[CI]$_{370 \upmu \mathrm{m}} = 58\pm29$ for W2246--0526 favors PDR models, which appears to contradict the interpretation of the low [CI]$_{609\upmu \mathrm{m}}$/[CI]$_{370 \upmu \mathrm{m}}$ and [CI]$_{370 \upmu \mathrm{m}}$/CO$(7-6)$ ratios. Given that these models cover a range of parameters that may not be applicable to our source, we address this apparent tension together with the results of the \textsc{Cloudy} modeling in Sect.~\ref{sec:3.3}. The [CII]$_{158\upmu \mathrm{m}}$/[CI]$_{370 \upmu \mathrm{m}}$ ratio is shown in the y-axis of the top left panel of Fig.~\ref{fig:grids}.

\subsubsection{Extreme electron density}
\label{sec:3.1.4}

The [NII]$_{122\upmu \mathrm{m}}$ and [NII]$_{205 \upmu \mathrm{m}}$ emission lines originate from fine-structure transitions within the same ionization species. As such, the ratio between the two is insensitive to the metallicity of the gas and the intensity or hardness of the incident radiation field. Instead, it primarily depends on the electron density and weakly on the electron temperature of the low density ionized gas. Above a density of $n_e \sim 300$ cm$^{-3}$ both population levels are beyond their critical densities (shown in Table~\ref{tab1}) and become saturated. This places an upper limit to the [NII]$_{122\upmu \mathrm{m}}$/[NII]$_{205 \upmu \mathrm{m}}$ ratio of $\sim10$ \citep{oberst2006,2015ApJ...799...21S,pereira2017}. 

Our result of [NII]$_{122\upmu \mathrm{m}}$/[NII]$_{205 \upmu \mathrm{m}}=13.8 \pm 4.2$ indicates that the warm ionized gas in W2246--0526 is most likely in the high density limit. This ratio is shown in the x-axis of the bottom right panel of Fig.~\ref{fig:grids}.

\subsubsection{High gas temperatures}
\label{sec:3.1.5}

The [OI]$_{63\upmu \mathrm{m}}$/[CII]$_{158\upmu \mathrm{m}}$ ratio depends mainly on the density of the gas due to the different critical densities of both species. In addition, it is also sensitive to the gas temperature, as discussed in \cite{2015A&A...574A..32G}. A comparison of their ISM models with our observed ratio of $4.7\pm1.0$ in W2246--0526 suggests gas temperatures of a few hundred K, which is also expected in XDRs \citep{2022ARA&A..60..247W}. High [OI]$_{63\upmu \mathrm{m}}$/[CII]$_{158\upmu \mathrm{m}}$ ratios can also be found in high density and temperature PDRs exposed to the high radiation fields from OB stars \citep{stacey1983,stacey1993}. As discussed in Sect.~\ref{sec:3.1.2}, [OI]$_{63\upmu \mathrm{m}}$ is likely optically thick, so the ratio with [CII]$_{158\upmu \mathrm{m}}$ and therefore the implied gas temperatures may be even higher. This ratio is shown in the y-axis of the bottom left panel of Fig.~\ref{fig:grids}.

\subsubsection{Radiation field intensity}
\label{sec:3.1.6}

Recently, \cite{harikane2020} and \cite{algera2024} reported low [OIII]$_{88\upmu \mathrm{m}}$/[CII]$_{158\upmu \mathrm{m}}$ ratios in galaxies with low dust temperatures, associating the low ratios to low ionization parameters. The [OIII]$_{88\upmu \mathrm{m}}$/[NII]$_{122\upmu \mathrm{m}}$ ratio has also been proven useful as a radiation field intensity indicator \citep{ferkinhoff2011} and as a gas-phase metallicity diagnostic for high-redshift starburst galaxies \citep{2018MNRAS.473...20R,harikane2020}, with the value increasing with increasing ionization parameter and decreasing with increasing metallicity. Consequently, our 3$\upsigma$ upper limits of [OIII]$_{88\upmu \mathrm{m}}$/[CII]$_{158\upmu \mathrm{m}}<0.69$ and [OIII]$_{88\upmu \mathrm{m}}$/[NII]$_{122\upmu \mathrm{m}}<1.02$ for W2246--0526 may indicate low ionization parameters, but the models used in these previous works have a more restricted parameter space than our \textsc{Cloudy} models. We discuss these observed ratios thoroughly in Sect.~\ref{sec:3.3}. The [OIII]$_{88\upmu \mathrm{m}}$/[CII]$_{158\upmu \mathrm{m}}$ and [OIII]$_{88\upmu \mathrm{m}}$/[NII]$_{122\upmu \mathrm{m}}$ ratios are shown, respectively, in the x-axis of the top left panel and in the y-axis of the bottom right panel of Fig.~\ref{fig:grids}.

\subsubsection{FIR line deficits}
\label{sec:3.1.7}

\begin{figure*}[h!]
    \centering
        \subfloat{\includegraphics[width=180mm]{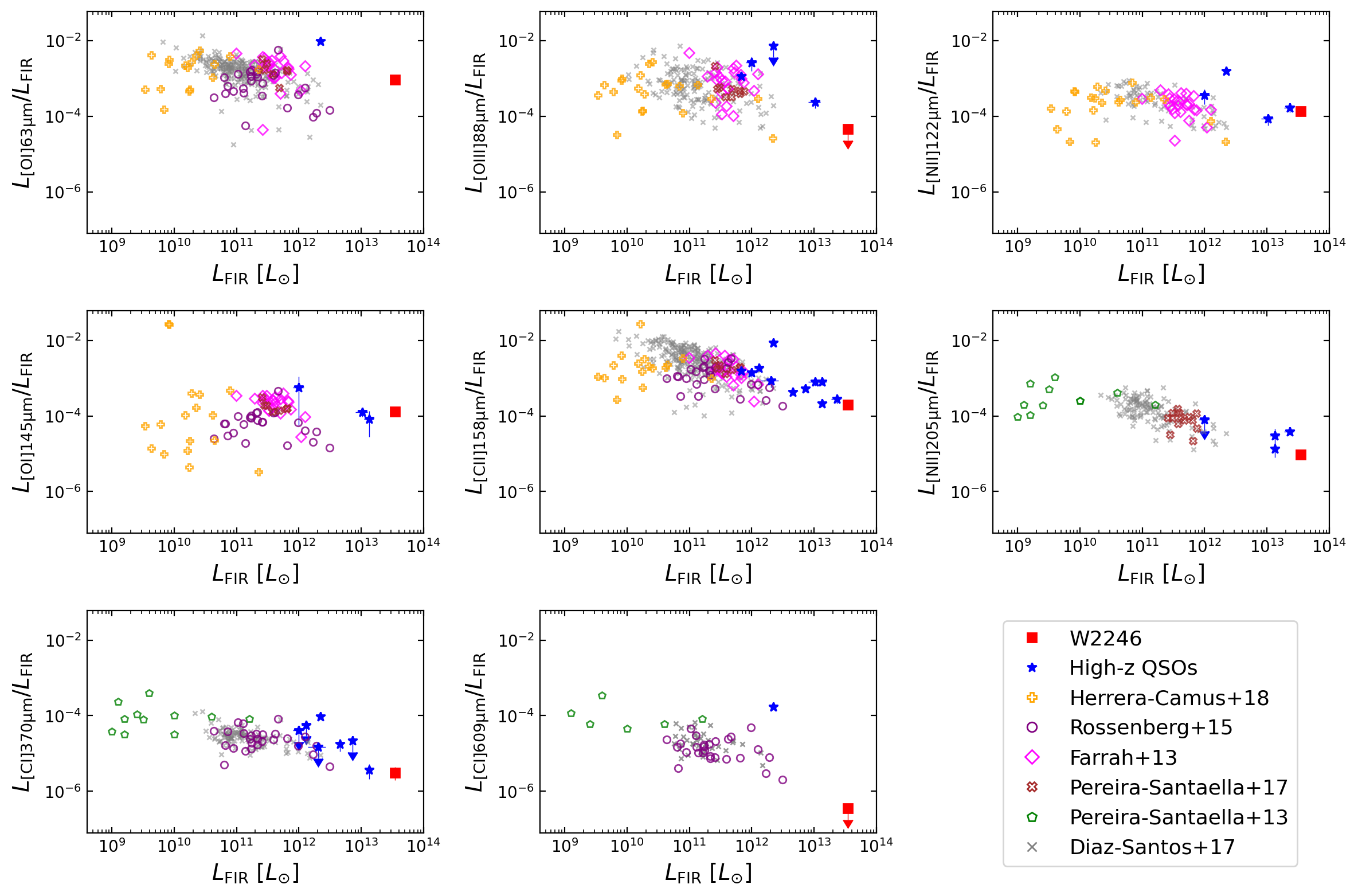}}
        \caption{Line-to-FIR$_{[42-122 \upmu\mathrm{m}]}$ luminosity ratio as a function of FIR$_{[42-122 \upmu\mathrm{m}]}$ luminosity for [OI]$_{63\upmu \mathrm{m}}$, [OIII]$_{88\upmu \mathrm{m}}$, [NII]$_{122\upmu \mathrm{m}}$ (top row), [OI]$_{145\upmu \mathrm{m}}$, [CII]$_{158\upmu \mathrm{m}}$, [NII]$_{205\upmu \mathrm{m}}$ (middle row), [CI]$_{370\upmu \mathrm{m}}$, and [CI]$_{609\upmu \mathrm{m}}$ (bottom row). W2246--0526 is indicated with a red square. The high-z quasars are taken from \cite{weiss2003,uzgil2016,lu2017,venemans2017,walter2018,lee2019,novak2019,li2020,pensabene2021,meyer2022}. Upper limits for high-z quasars and W2246--0526 are shown with downward arrows. We note that the only high-z quasar shown for [OI]$_{63\upmu \mathrm{m}}$ and [CI]$_{609\upmu \mathrm{m}}$ corresponds to the Cloverleaf quasar \citep{weiss2003,uzgil2016}.
        }
        \label{fig:FIR}
\end{figure*}

In Fig.~\ref{fig:FIR} we present the line-to-FIR$_{[42-122 \upmu\mathrm{m}]}$ ratio as a function of the FIR$_{[42-122 \upmu\mathrm{m}]}$ luminosity ($L_{\mathrm{FIR}}$) for all the targeted fine-structure emission lines in W2246--0526. In addition, for comparison we show a sample of high-z quasars gathered from the literature \citep{weiss2003,uzgil2016,lu2017,venemans2017,walter2018,lee2019,novak2019,li2020,pensabene2021,meyer2022}, local AGNs and ULIRGs from \cite{farrah2013,pereira2013,rossenberg2015,pereira2017,herrera2018a}, and local LIRGs from \cite{diaz2017}. We note that for galaxies without a measurement of the FIR luminosity \citep{farrah2013,pereira2013,pereira2017}, we assumed $L_{\mathrm{IR[8-1000\upmu m]}}=6.3\times L_{\mathrm{FIR[42-122\upmu m]}}$, the scaling used for W2246--0526 in \cite{2016ApJ...816L...6D}.

From previous studies, a pronounced deficit (reduced line-to-FIR ratio with increasing $L_{\mathrm{FIR}}$) is expected for [NII], [CI] and [CII] emission lines, while a less significant deficit is found for [OI] and [OIII] emission lines \citep[e.g.,][]{rossenberg2015,diaz2017,walter2018,li2020,pensabene2021}. Figure~\ref{fig:FIR} shows a clear tendency of lower line-to-FIR ratio with increasing $L_{\mathrm{FIR}}$ for [CII]$_{158\upmu \mathrm{m}}$, [NII]$_{205\upmu \mathrm{m}}$, [CI]$_{370\upmu \mathrm{m}}$, and [CI]$_{609\upmu \mathrm{m}}$, with some ratios in W2246--0526 at least two orders of magnitude smaller than lower luminosity galaxies, but well in line with high-z quasars (blue points in Fig.~\ref{fig:FIR}). The decreasing tendencies are weaker or not significant in [OI]$_{63\upmu \mathrm{m}}$, [OIII]$_{88\upmu \mathrm{m}}$, [NII]$_{122\upmu \mathrm{m}}$, and [OI]$_{145\upmu \mathrm{m}}$. The origin of such FIR line deficits is still under study, and we do not address them in the current work. 

\subsection{ISM models}
\label{sec:3.2}

In order to determine the physical conditions of the ISM, we simultaneously compare all observed FIR line ratios of W2246--0526 with grids of photoionization models built using the spectral synthesis code \textsc{Cloudy} v17.00 \citep{2017RMxAA..53..385F}. To simulate astrophysical environments, \textsc{Cloudy} requires as an input the SED and intensity of a radiation field, as well as the chemical and dust compositions, and the geometry of a cloud of gas. \textsc{Cloudy} solves the equations of statistical and thermal equilibrium of the gas cloud, obtaining the physical conditions (density, temperature, ionization) across the cloud and its resulting spectrum. 

With \textsc{Cloudy}, we model five physical parameters of the cloud and incident radiation, namely: the extinction, $A_{\mathrm{V}}$; the metallicity of the gas, Z; its volume gas density, $n_{\mathrm{H}}$; the X-ray to UV ratio of the input SED, $\alpha_{\mathrm{ox}}$; and the ionization parameter, $U$. The range and step of the model grid are presented in Table~\ref{tab5}. Due to the high obscuration present in W2246--0526, we adopted a spherical model with a covering factor of unity, which corresponds to a semi-infinite plane-parallel slab with a closed geometry. 

\begin{table*}
    \centering
    \caption{Details of the \textsc{Cloudy} model grids}
    \begin{tabular}{*{4}{c}}
    \hline
     Parameter & Range & Best model & $m\pm68\%$ $CI$ \\
     \hline
     $A_{\mathrm{V}}$ [mag] & $10^{1.0}\rightarrow10^{4.0}$, steps of $10^{0.5}$ & $10^{2.5}$ & $222_{-158}^{+546}$ \Tstrut \\ 
     $Z$ [Z$_{\odot}$] & (0.05, 0.10, 0.25, 0.5, 1, 2) & 0.5 & $0.49_{-0.29}^{+0.67}$ \Tstrut \\
     $n_{\mathrm{H}}$ [cm$^{-3}$] & $10^{2.0}\rightarrow10^{5.0}$, steps of $10^{0.25}$ & $10^{3.5}$ & $2754_{-1579}^{+3703}$ \Tstrut \\
     $\alpha_{\mathrm{ox}}$ & $-1.4\rightarrow-0.4$, steps of $0.2$ & $-0.8$ & $-0.78\pm0.29$ \Tstrut \\
     log(U) & $-3.0\rightarrow0.5$, steps of $0.25$ & $-0.5$ & $-0.47\pm0.62$ \Tstrut \\     
     \hline
     \\
    \end{tabular}
    \caption*{Note: Columns correspond to the parameters and range of values used to compute the grid of \textsc{Cloudy} models, value of the best-fit model, and marginalized posterior mean (m) of each parameter and its 68\% confidence interval ($CI$).}
    \label{tab5}
\end{table*}

Hot DOGs are expected to be powered by the central AGN, with little contribution from their star-forming host \citep{jones2014,tsai2015,2020ApJ...905...16F}. Specifically for W2246--0526, the best-fit SED longward of $\sim1\mathrm{\upmu \mathrm{m}}$ and the total bolometric luminosity is dominated by the central quasar \citep{diaz2018,2018ApJ...854..157F,2018ApJ...868...15T}. Exhaustively exploring the vast landscape of \textsc{Cloudy} setups with fine grids is computationally expensive and beyond the scope of this paper. Given the amount of ancillary data and literature supporting that W2246--0526 is an extremely obscured quasar, we thus selected a pure AGN as the source of radiation for the \textsc{Cloudy} models. This corresponds to a continuum SED consisting of a sum of two components: the "Big Bump" component, a rising power law peaking at a temperature $T_{\mathrm{BB}}$, with a low-energy slope parameterized by $\alpha_{\mathrm{uv}}$, and an infrared and a high-energy exponential cutoffs; and the X-ray component, parameterized by a slope $\alpha_\mathrm{x}$ between 1.36 eV and 100 keV, and decreasing with increasing frequency as $f_{\nu}\propto\nu^{-2}$. The two components are normalized by the X-ray to UV ratio $\alpha_{\mathrm{ox}}$, defined as follows: 

\begin{equation}
    \alpha_{\mathrm{ox}}=\frac{log(L_{\mathrm{2keV}} /L_{2500\AA})}{2.605} \ .
\end{equation}

\begin{figure}[h!]
    \centering
        \subfloat{\includegraphics[width=90mm]{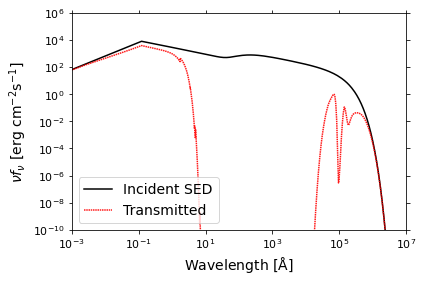}}
        \caption{Input AGN SED (black solid line)\, and reprocessed radiation by the cloud of gas and dust (red dotted line), for the best-fit model for W2246--0526 described in \ref{sec:3.4} with $log(U)=-0.5$ and $\alpha_{\mathrm{ox}}=-0.8$.}
        \label{fig:SED}
\end{figure}

\noindent
This multicomponent continuum of the AGN template in \textsc{Cloudy} can be shaped by varying $T_{\mathrm{BB}}$, $\alpha_{\mathrm{uv}}$, $\alpha_\mathrm{x}$, and $\alpha_{\mathrm{ox}}$. To introduce potential enhanced X-ray emission from the AGN as a parameter in our models we decided to vary $\alpha_{\mathrm{ox}}$ between $-1.4$ and $-0.4$ in steps of 0.2. We note that $\alpha_{\mathrm{ox}}$ becomes less negative as the ratio of X-ray to UV emission increases. The other parameters are degenerate with $\alpha_{\mathrm{ox}}$ and/or the ionization parameter, as they vary the X-ray to UV ratio and the total amount of ionization photons. Therefore, we decided to keep their default values ($T_{\mathrm{BB}} = 10^{6}$ K, $\alpha_{\mathrm{uv}} = -0.5$, and $\alpha_\mathrm{x} = -1$) in order to not increase the complexity and degeneracies of the fitting. An example of the AGN SED used by \textsc{Cloudy} is shown in Fig.~\ref{fig:SED}.

We run models with different values of the ionization parameter, which is defined as follows: 

\begin{equation}
    U\equiv \frac{Q(H)}{4\pi r_0^2n(H)c}\equiv \frac{\Phi(H)}{n(H)c} \ ,
\end{equation}

\noindent where $Q(H)$ is the number of hydrogen-ionizing photons, $r_0$ is the distance between the ionizing source and the inner part of the cloud, $n(H)$ is the hydrogen density of the cloud, $c$ is the speed of light, and $\Phi (H)$ is the flux of hydrogen-ionizing photons at $r_0$. Increasing $U$, therefore, increases the ratio between hydrogen-ionizing photons and total hydrogen density. In our grid of models, we vary $U$ between $10^{-3.0}$ and $10^{0.5}$ in steps of $10^{0.25}$, a range similar to other models in the literature \citep[e.g.,][]{moy2002,groves2004,nagao2006}. 

In addition to the intensity and SED of the radiation field, \textsc{Cloudy} requires the  chemical composition and the density profile of the cloud of gas and dust to be specified. We assume a constant hydrogen density through the cloud, ranging between $10^{2.0}$ and $10^{5.0}$ cm$^{-3}$ in steps of 0.25 dex, similar to other works in the literature \citep[e.g.,][]{meijerink2007,feltre2016,pensabene2021}. We let both the gas and dust metallicity vary and calculate models for 0.05, 0.10, 0.25, 0.5, 1 and 2 $Z_{\odot}$. We do not vary, however, relative elemental abundances, and assume the default gas-phase values provided by \textsc{Cloudy} for the ISM ($log(\mathrm{N/O}) = -0.6$), with the default grains and polycyclic aromatic hydrocarbons (PAH) size distribution from \cite{2008ApJ...686.1125A}. 
We note that some relative elemental abundances, specifically the N/O ratio, due to the secondary nature of nitrogen production in stellar evolution, depend on the overall, absolute metallicity (e.g., O/H) of the gas, especially at values below 0.25 $Z_{\odot}$ \citep{nicholls2017,chartab2022}. However, as we subsequently see, the best-fit metallicity for W2246--0526 is 0.5 $Z_{\odot}$, still sufficiently high for relative abundances not to be a dominant factor in the models. Finally, we also vary the depth of the cloud in the form of different $A_{\mathrm{V}}$ values: $10^{1.0}$, $10^{1.5}$, $10^{2.0}$, $10^{2.5}$, $10^{3.0}$, $10^{3.5}$, $10^{4.0}$ mag. The code stops when the different $A_{\mathrm{V}}$ values are reached. A cosmic ray background for the redshift of W2246--0526 is also included as it plays an important role in deep-heating the molecular cloud, as well as at low temperatures \citep{2022A&A...664A.150G}. Additional modeling options were investigated and are detailed in the Appendix \ref{sec:AppendixB}, supporting our choice of parameters and model grid.

The temperature of the coldest dust component in W2246--0526 is $\sim70$ K \citep[][Tsai et al. in prep.]{2016ApJ...816L...6D}, and the cosmic microwave background (CMB) temperature is around $\sim15$ K at $z\sim 4.6$. Therefore, following \cite{2013ApJ...766...13D}, we can neglect the CMB heating in the models. 

We also note that, in order to avoid introducing additional complexity into the models, we do not attempt to fit the continuum emission. Modeling the CO SLED of W2246--0526 is beyond the scope of this paper as well, and a separate analysis of the available CO line measurements will be addressed in a future publication.

\subsection{Line ratio diagnostics within \textsc{Cloudy}}
\label{sec:3.3}

In Sect.~\ref{sec:3.1.2} we proposed optically thick [OI]$_{63 \upmu \mathrm{m}}$ as the most likely cause of the observed [OI]$_{145\upmu \mathrm{m}}$/[OI]$_{63 \upmu \mathrm{m}}$ ratio. The comparison with our \textsc{Cloudy} models reinforce this scenario, as we can only reproduce the observed value with column densities above $10^{23}$ cm$^{-2}$. 

As mentioned in Sect.~\ref{sec:3.1.3} ratios of [CII]$_{158\upmu \mathrm{m}}$/[CI]$_{370 \upmu \mathrm{m}} > 20$ discard XDR models in previous studies, but we find in our \textsc{Cloudy} models that this applies only to $log(U)<-1.0$, while higher ratios are reproducible with higher ionization parameters. Our observation of [CII]$_{158\upmu \mathrm{m}}$/[CI]$_{370 \upmu \mathrm{m}} = 58\pm29$ for W2246--0526 points toward an extreme ionization of the ISM in XDR. Another mechanism that can enhance the [CII]$_{158\upmu \mathrm{m}}$/[CI]$_{370 \upmu \mathrm{m}}$ ratio is [CII] excitation by shocks and turbulence driven by jets \citep{2018ApJ...869...61A}, which can boost the ratio up to $\sim40\%$. The observed [CII]$_{158\upmu \mathrm{m}}$ kinematics in W2246--0526 show high turbulence, suggesting the central quasar might be blowing isotropically its ISM in large-scale outflows \citep{2016ApJ...816L...6D}. We do not address this case as it falls beyond the scope of the current study. 

\begin{figure}[h!]
    \centering
        \subfloat{\includegraphics[width=80mm]{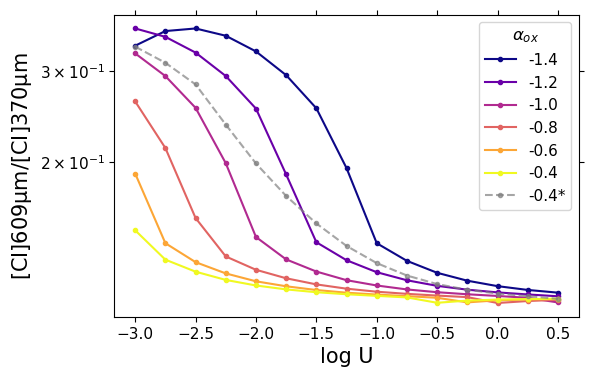}}
        \caption{Neutral carbon [CI]$_{609\upmu \mathrm{m}}$/[CI]$_{370 \upmu \mathrm{m}}$ ratio from our \textsc{Cloudy} models as a function of the ionization parameter. Each color represents varying X-ray to UV ratios. The metallicity is fixed to 1$Z_{\odot}$, the $A_{\mathrm{V}}$ to 10 mag, and the hydrogen density to 10$^3$ cm$^{-3}$. The gray dashed line (denoted by * in the legend) represents $\alpha_{\mathrm{ox}}=-0.4$ but with an $A_{\mathrm{V}}=100$ mag.}
        \label{fig:CI_aox}
\end{figure}

The neutral carbon [CI]$_{609\upmu \mathrm{m}}$/[CI]$_{370 \upmu \mathrm{m}}$ ratio supports the XDR scenario, as we see in our \textsc{Cloudy} models that increasing the X-ray to UV ratio decreases the neutral carbon line ratio (Fig.~\ref{fig:CI_aox}). We find that it also decreases with increasing density and $U$, and with decreasing $A_{\mathrm{V}}$, and it does not depend on the metallicity. Given our upper limit, the models favor higher $\alpha_{\mathrm{ox}}$, $U$ and densities, and lower $A_{\mathrm{V}}$.

Regarding radiation field intensity indicators involving the [OIII]$_{88\upmu \mathrm{m}}$ emission line (Sect.~\ref{sec:3.1.6}), previous studies find lower [OIII]$_{88\upmu \mathrm{m}}$/[CII]$_{158\upmu \mathrm{m}}$ and lower [OIII]$_{88\upmu \mathrm{m}}$/[NII]$_{122\upmu \mathrm{m}}$ for lower ionization parameters. In our \textsc{Cloudy} models, however, this dependence for [OIII]$_{88\upmu \mathrm{m}}$/[CII]$_{122\upmu \mathrm{m}}$ is seen only in the $log(U)$ range $\sim-3$ to $\sim-2$ that they cover in their models, while low ratios are also found in our models for much higher ionization intensities (and therefore much higher dust temperatures, as is the case for Hot DOGs). Lower [OIII]$_{88\upmu \mathrm{m}}$/[NII]$_{122\upmu \mathrm{m}}$ ratios for lower ionization parameters are seen in our models only in low density ($log(n_{\mathrm{H}})\le3$ cm$^{-3}$) and low $\alpha_{\mathrm{ox}}$ ($\le-1.0$). In other regimes, the ratio mainly decreases with increasing $\alpha_{\mathrm{ox}}$, increasing density, and increasing ionization parameter. Perhaps counter-intuitively, the emitting region where [OIII] is produced is actually smaller at high radiation field intensities and X-ray luminosities, as oxygen is ionized to even higher levels (see middle panel of Fig.~\ref{fig:CO}), while the emitting region of [NII] becomes larger with increasing X-ray ionization due to the change in the ISM structure. Our 3$\upsigma$ upper limit [OIII]$_{88\upmu \mathrm{m}}$/[NII]$_{122\upmu \mathrm{m}}<1.02$ for W2246--0526 strongly indicates again a high-density ISM being affected by high X-ray ionization. 

\subsection{Best-fit model}
\label{sec:3.4}

The line ratios described above are independent, powerful tools to derive specific properties of a particular phase of the ISM gas. Modeling various ISM phases together can be challenging, as they may not be, for instance, uniform in density, homogeneous or isotropic, and the emission lines can be produced in different regions in the galaxy nucleus or at different physical scales. Nonetheless, even when assumptions on the geometry and spatial distribution of the media are simplified (sometimes unavoidably due to a lack of observational information), models can still be used to provide insights into the average conditions of the multiphase gas that is exposed to the radiation field of the central quasar. 

\begin{figure*}[h!]
    \centering
        \subfloat{\includegraphics[width=140mm]{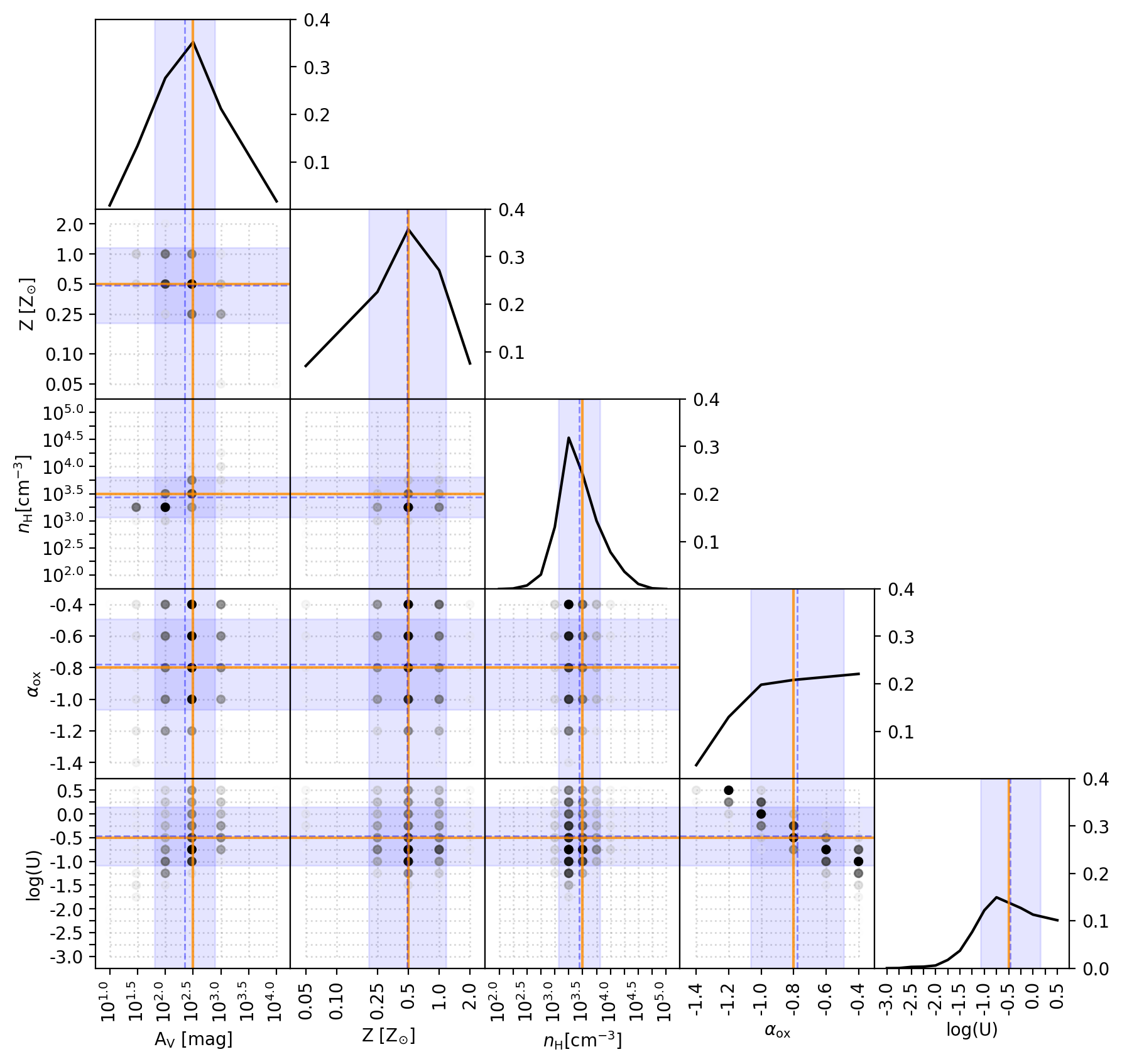}}
        \caption{Corner plot showing the marginalized 1D and 2D posterior probability distributions (PPDs) for the parameters in the \textsc{Cloudy} models used to fit the emission line ratios of W2246--0526. The orange solid lines indicate the parameter values of the best model (maximum likelihood estimation, MLE), the blue dashed line the median of each parameter marginalized over the rest, and the blue shaded region the 68\% confidence interval. The y-axes of the 1D PPDs show the probabilities with their sum normalized to one for each parameter.}
        \label{fig:posteriors}
\end{figure*}

To this end, we compare all the observed lines with our \textsc{Cloudy} model grid using a Bayesian framework from which we can estimate the marginalized 1D and 2D posteriors of the probability distribution functions ($P = e^{-\chi^2/2}$) for each of the considered parameters. An uncertainty of 15\% is assumed for the model ratios to account for the sparsity of the parameter grid. As we see in the corner plot shown in Fig.~\ref{fig:posteriors}, the most probable solution (and in parenthesis the posterior mean with 68\% confidence intervals) corresponds to a hydrogen density $log(n_{\mathrm{H}}) = 3.5 \ \mathrm{cm}^{-3}\ (3.44\pm0.37)$, an ionization parameter $log(U) = -0.5 \ (-0.47\pm0.62)$, an extinction $A_{\mathrm{V}} = 316 \ \mathrm{mag}\ (222_{-158}^{+546})$, an X-ray to UV ratio $\alpha_{\mathrm{ox}} = -0.8 \ (-0.78\pm0.29)$, and a metallicity $Z = 0.5 \ Z_{\odot}\ (0.49_{-0.29}^{+0.67})$. The best fit between model and observations has a reduced chi-square $\chi_{\nu}^{2}=3.34$, and it is shown in Fig.~\ref{fig:ratios}. For fixed FWHM, the ratios are very similar and the reduced chi-square is $\chi_{\nu}^{2}=4.12$.

The average gas density is well constrained and comparable to values obtained for high-z quasars \citep{2016ApJ...819...24W,venemans2017,novak2019,2019ApJ...880..153Y,pensabene2021,meyer2022}. A density of $log(n_{\mathrm{H}}) = 3.5 \ \mathrm{cm}^{-3}$ is above the critical densities of the lines arising from ionized gas (Table~\ref{tab1}), indicating they are all near or above the high density limit. Densities higher than $n_{\mathrm{cr}}$ saturate the emission, decreasing thus the efficiency of the lines as coolants of the ISM. The extinction has a very high value; an $A_{\mathrm{V}}$ of $10^{2.5}$ mag corresponds to a column density close to $10^{24}$ cm$^{-2}$, in agreement with the column densities expected for Hot DOGs \citep{stern2014,piconcelli2015,assef2015,assef2016,assef2020}, as well as for obscured high-z quasars \citep{2019A&A...630A.118V,pensabene2021,meyer2022}. A metallicity of 0.5 $Z_{\odot}$ suggests a fairly evolved and enriched ISM at $z\sim4.6$, and falls within the range of metallicities (0.2 to 2.0 $Z_{\odot}$) reported by \cite{2020MNRAS.492.5675C} for AGNs at both low and high redshifts ($z=0-7$). The ionization parameter is also well constrained and significantly higher than other high-z AGNs \citep[$-2.2<log(U)<-1.4$ in][]{nagao2006}. The X-ray to UV ratio $\alpha_{\mathrm{ox}}$ is again extreme among the population of quasars. \cite{2011ApJ...726...20M} finds values of $\alpha_{\mathrm{ox}}$ varying between $-2.0$ and $-0.8$. We can see in the corner plot that $U$ is degenerate with $\alpha_{\mathrm{ox}}$, in the sense that the model is able to reproduce the line ratios with almost the same probability by increasing $\alpha_{\mathrm{ox}}$ if $U$ is reduced. Because of this, $\alpha_{\mathrm{ox}}$ is unbound at the high end and we can only put a lower limit of $\alpha_{\mathrm{ox}} \geq -0.8$. 

The lower limit in $\alpha_{\mathrm{ox}}$ and the high value of $U$ correspond to intense X-ray emission from the central AGN. Observations with XMM-\textit{Newton} \citep{2018MNRAS.474.4528V} set an upper limit to the X-rays that is a factor of $\sim2$ lower than what is predicted by the best model. We address and discuss this result in Sect.~\ref{sec:4.2.2}.

\begin{figure}[h!]
    \centering
        \subfloat{\includegraphics[width=90mm]{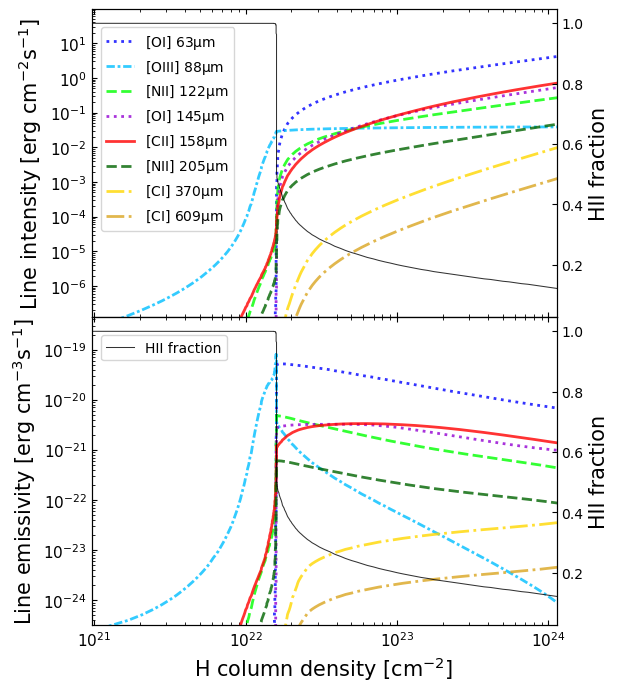}} 
        \par\vspace{1.5mm}

        \caption{Cumulative line intensity (top panel) and line emissivity (bottom panel) at each depth in the cloud, for the \textsc{Cloudy} model that best fits the W2246–0526 observations. Both panels show each modeled FIR fine-structure line as a function of the cloud depth, measured in hydrogen column density. The HII fraction is included to visualize the ionization front, where it drops sharply.}
        \label{fig:cloud_lines}
\end{figure}

\begin{figure}[h!]
    \centering        \subfloat{\includegraphics[width=80mm]{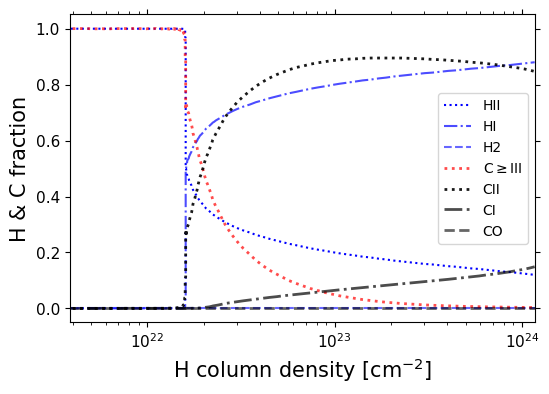}} \par\vspace{1.5mm}
    \subfloat{\includegraphics[width=80mm]{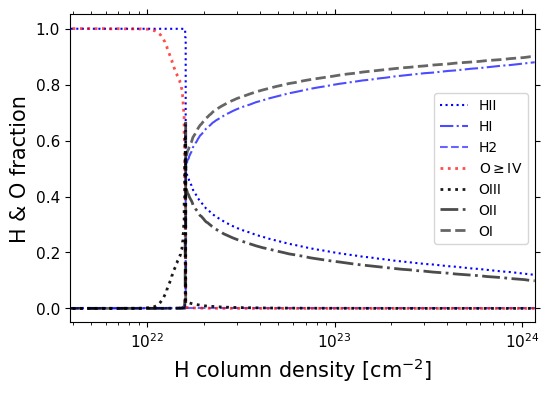}} \par\vspace{1.5mm}
    \subfloat{\includegraphics[width=80mm]{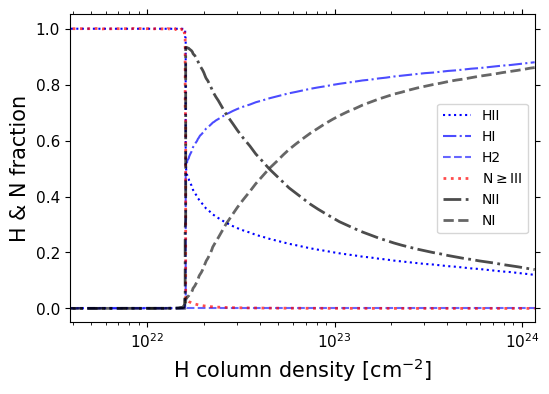}} \par\vspace{1.5mm}
    
    \caption{Ionized, neutral, and molecular hydrogen and carbon (top), oxygen (middle), and nitrogen (bottom) number density fractions as a function of the cloud depth, measured in hydrogen column density, for the \textsc{Cloudy} model that best fits the W2246--0526 observations. We note the negligible fraction of molecular gas and CO (top), even at depths as large as $10^{24} \ \mathrm{cm^{-2}}$.}
    \label{fig:CO}
\end{figure}

In Fig.~\ref{fig:cloud_lines}, we present the emission of each FIR fine-structure line at each layer of the cloud of gas for the best model. The structure of the gas cloud, also seen in Fig.~\ref{fig:CO} for the different ion abundances, corresponds to that of an XDR: the [NII] lines continue to arise even where hydrogen is mostly neutral, and the neutral carbon emission is distributed through the ISM and not limited only to a thin layer as expected in PDRs. We find that $\sim$90\% of [CII] emission is emitted after the ionization front (i.e., in the non-ionized gas). While this fraction depends on the parameters of the model, the difference with the estimation in Sect.~\ref{sec:3.1.1} (80\%) clearly shows that in sources with important contribution from XDRs, using a scaling based on PDR models leads to different results.


\section{Discussion}
\label{sec:discussion}

In Sect.~\ref{sec:3.4} we presented the best-fit model parameters and their posterior distributions and found that, even with the limitations imposed by the simple geometry used to model the ISM in the vicinity of W2246--0526's quasar, \textsc{Cloudy} is able to broadly reproduce the measured emission line ratios relatively well. However, we also find tensions between some of the predictions of the best model and  other independent observations available for W2246--0526. In this section, we discuss potential explanations for those discrepancies, address the implications for the non-detections and multiphase gas budget, and put our results in the context of other populations of quasars at high redshift.

\subsection{Comparison with other high-z quasars and local AGNs and ULIRGs}
\label{sec:4.1}

We compare the FIR emission line ratios measured for W2246--0526 to those found in local AGNs and ULIRGs (ultra-luminous infrared galaxies) as well as high-redshift quasars reported in the literature. This compilation of sources is shown in Fig.~\ref{fig:grids}, together with the values of W2246--0526 and the \textsc{Cloudy} model grids for varying $n_{\mathrm{H}}$ and $U$.

We note that the model grids shown in Fig.~\ref{fig:grids} are only a small part of the parameter space we covered by the \textsc{Cloudy} models. We only use the best-model values obtained for the metallicity, $\alpha_{\mathrm{ox}}$, and $A_{\mathrm{V}}$, which are representative only of models with high X-ray to UV emission and high extinctions. Figure~\ref{fig:grids2} in the Appendix illustrates the effect of varying these two parameters ($\alpha_{\mathrm{ox}}$ and $A_{\mathrm{V}}$). We also caution that our modeling considers an AGN as the only source of radiation, while in some quasar populations there may be an additional significant contribution to the ionization from star formation in the host galaxy. Therefore, the main goal of including these galaxies along with the grids of models is not to characterize them, but to put W2246--0526 in the context of other AGN-powered galaxies.

From a modeling point of view, W2246--0526 stands out in terms of the large X-ray and radiation field intensities needed to reproduce the observed line ratios when compared to other AGNs and quasars at both low and high redshift. This is reflected in some specific individual line diagnostics, where we find that W2246--0526 is at the edge of the distribution of other populations, showing extremely high values of the [OI]$_{63\upmu \mathrm{m}}$/[CII]$_{158\upmu \mathrm{m}}$ and [NII]$_{122\upmu \mathrm{m}}$/[NII]$_{205\upmu \mathrm{m}}$ ratios, and very low values of the [CI]$_{609\upmu \mathrm{m}}$/[CI]$_{370\upmu \mathrm{m}}$ ratio, all of them independently suggesting a very dense and highly ionized ISM with X-ray dominated heating. Therefore, the unusual ISM conditions in W2246--0526 can serve as a benchmark for comparisons with other quasar populations at cosmic noon and beyond that may be less extreme.

\subsection{X-ray emission}
\label{sec:4.2}

\subsubsection{$\alpha_{\mathrm{ox}}$ as a free parameter}
\label{sec:4.2.1}
 
With the default \textsc{Cloudy} value of $\alpha_{\mathrm{ox}}=-1.4$, the ionization parameter, $U$, for the best model is extremely high and unconstrained (Fig.~\ref{fig:Appendix_posterior}). According to \cite{2012ApJ...757..108Y}, high radiative pressure will create a pressure gradient in ionized regions, that can make $U$ to saturate, implying that values of $U$ higher than $10^{0.5}$ (the upper value in our grid of \textsc{Cloudy} models) are highly unlikely. In fact, such high (or higher) values have never been reported in the literature when modeling sources. 

Since hyper-luminous AGNs and quasars are expected to be powerful X-ray emitters \citep[e.g.,][]{2010A&A...512A..34L}, we introduced $\alpha_{\mathrm{ox}}$ as an additional parameter in order to search for a more physical solution. The result is that $\alpha_{\mathrm{ox}}$ and $U$ are degenerate, as shown in the 2D marginalized posterior distributions presented in Fig.~\ref{fig:posteriors}. In this case, however, the value of $\alpha_{\mathrm{ox}}$ is unbound, with a lower limit of $\sim-0.8$, but the value of $U$ is more constrained and physical, with the best fit peaking at $\sim10^{-0.5}$.

In order to quantitatively assess the model assumptions that better reproduce the data, that is, that with fixed $\alpha_{\mathrm{ox}}=-1.4$ (henceforth model $A$), or that with $\alpha_{\mathrm{ox}}$ as a free parameter (henceforth model $B$), we use the $\chi^2$ statistic defined as $\chi^2=\sum{\frac{(d_i-m_i)^2}{\Delta d_i^2+\Delta m_i^2}}$, where $d_i$ and $\Delta d_i$ are the observed ratios and their uncertainties, and $m_i$ and $\Delta m_i$ are the model ratios and their uncertainties, which account for the sparsity of the parameter grid. We obtain  $\chi^2(A)=15.2$, $\chi^2(B)=6.7$, $\chi^2_{\nu}(A)=5.1$, and $\chi^2_{\nu}(B)=3.3$, with $\chi^2_{\nu}$ defined as $\chi^2_{\nu}=\frac{\chi^2}{n-k}$, $n$ being the number of data points (number of ratios), and $k$ the number of free parameters of each model. Both $\chi^2$ and $\chi^2_{\nu}$ are smaller for model $B$. To investigate if the difference is significant, we use the Bayesian Information Criteria ($BIC$), defined as $BIC = log(n)k-2log(\hat{L})$, with $log(\hat{L})$ being the maximum log-likelihood of the model. The $BIC$ statistic takes into account the performance of a model penalizing its complexity. A lower $BIC$ means less information is lost, and hence the model is better. Following \cite{burnham2004}, we can compute the posterior model probability as $p_i=\frac{exp(-1/2\Delta BIC_i)}{\sum_i(exp(-1/2\Delta BIC_i))}$, with $\Delta BIC_i$ being the difference of the $BIC$ of model $i$ with the minimum $BIC$ of all models. The probability of model $A$ over $B$ is $p(A)=0.15$, and of model $B$ over $A$ is $p(B)=0.85$. Therefore, model $B$ is statistically better than model $A$, indicating that it is preferable to vary $\alpha_{\mathrm{ox}}$. 

\subsubsection{X-ray non-detection}
\label{sec:4.2.2}

\cite{2018MNRAS.474.4528V} reported a non-detection of X-rays in W2246--0526, as part of a study of X-ray properties in Hot DOGs. To compare with our best-fit model we extract, from its associated input SED (Fig.~\ref{fig:SED}), the X-ray flux between 2--10 keV transmitted through the cloud and the bolometric incident flux, with a ratio of $\frac{f_{\mathrm{(2-10\ keV)}}}{f_{\mathrm{bol}}}\simeq10^{-2}$. The 3$\upsigma$ upper limit of X-ray emission between 2 -- 10 keV reported by \cite{2018MNRAS.474.4528V} is $<6.9\times10^{45}$ erg s$^{-1}$. The bolometric luminosity of W2246--0526 is $3.6\times10^{14}$ L$_{\odot} = 1.4\times10^{48}$ erg s$^{-1}$, which translates into a ratio of $\frac{\mathrm{L_{(2-10\ keV})}}{\mathrm{L_{bol}}} \simeq 0.5\times10^{-2}$. Comparing the two ratios, the best model predicts a factor of two more X-rays compared to the upper limit of the observations. Given the conversion factor $A_{\mathrm{V}}/N_{\mathrm{H}}$ provided by \textsc{Cloudy} and the best fit value for $A_{\mathrm{V}}$ we estimated a hydrogen column density $N_{\mathrm{H}}=(1.2_{-0.8}^{+3.1})\times10^{24}$ cm$^{-2}$. To be able to obscure the model X-rays below the observed limit (a factor of two less), we would need an extra $N_{\mathrm{H}}=4.6\times10^{23}$ cm$^{-2}$ along the line of sight, which is within $0.5\upsigma$ from the best-fit model given the uncertainties in $N_{\mathrm{H}}$. Therefore, the predictions of our models indicate that we may be close to detecting the X-ray emission from W2246--0526, highlighting the importance of future, deeper X-ray observations. 

A non-detection of hard X-rays in heavily obscured quasars is not surprising \citep[see, e.g., ][]{2022MNRAS.517.2577A}. In \cite{piconcelli2015}, the Hot DOG W1835+4355 is also extremely obscured in X-rays and in the Compton thick regime. \cite{2015ApJ...807..129S} presented a mid-infrared (MIR) X-ray relation that broadly agrees with our best model predictions estimate. Namely, given the MIR luminosity of W2246--0526, $\mathrm{\nu L_{\nu}} (6\upmu \mathrm{m}) \simeq 2\times10^{14} \ \mathrm{L_{\odot}}$ \citep[see Fig.~S4 in][]{diaz2018}, the relation predicts an X-ray luminosity of $\mathrm{L(2-10\ keV)=6.4\times10^{45}\ erg\ s^{-1}}$, compatible with the observational upper value. Interestingly, the Hot DOG WISE J1036+0449 analyzed by \cite{2017ApJ...835..105R} is predicted to have around $\sim 3$ times more X-ray emission than observed, when using the \cite{2015ApJ...807..129S} relation. To reconcile the difference, they suggest either X-ray weakness or significantly more obscuration than estimated. Given our model prediction, and the fact that a large fraction of MIR-selected AGNs are Compton thick \citep{carroll2023}, we favor the second scenario for W2246--0526 and propose that the reason for a non-detection of W2246--0526 in X-rays from the literature is likely due to additional obscuring columns of gas along the line of sight, not associated with the cloud we are modeling. 

\subsection{Multiphase gas budgets}
\label{sec:4.3}

\subsubsection{Cold molecular gas}
\label{4.3.1}

The amount of CO predicted by the models is extremely low (Fig.~\ref{fig:CO}), as expected from the high ionization and especially the intense X-ray emission implied by the best-fit model. X-ray photons cause the CO to dissociate much deeper in the gas cloud than in PDRs where photo-dissociation is dominated by far--UV photons. AGN X-ray dissociation of CO has been observed, as reported in \cite{2020ApJ...895..135K}. The amount of single and double ionized species is also very high in the presence of intense ionization and X-rays, and ions can also destroy CO molecules via charge transfer \citep{2013RMxAA..49..137F}.

Interestingly, \cite{diaz2018} used JVLA to observe the CO(2--1) emission line transition and reported the existence of large amounts of molecular gas in W2246--0526, which is in contrast with our model predictions. In particular, the observational CO(2--1)/[CII]$_{158\upmu\mathrm{m}}$ luminosity ratio in W2246--0526 is $\sim5\times10^{-3}$, versus $\sim3\times10^{-12}$ from the best-fit model. Moreover, the measured CO(7--6)/CO(2--1) luminosity ratio is $\sim17$, but the best-fit model yields a value of $\sim100$. This agrees with the findings of Lee et al. (in prep.), who studied a sample of ten Hot DOGs and identified that molecular gas is, in general, remarkably excited and abundant. However, it is key to note that the angular resolution of the JVLA CO(2--1) observations of W2246--0526 is much lower, $2.47\arcsec\times2.01\arcsec$, than the aperture we use to extract the fluxes of the fine-structure lines and CO(7--6), suggesting that the cold molecular gas component detected in W2246--0526 may be located and spread over a region at much larger scales than those being radiatively affected by the central quasar. Critically, this would be also in agreement with our findings regarding the non-detection of X-ray emission, as we discussed in the previous section. We note, however, that the presence of multiple gas phases and the likely not very homogeneous ISM distribution are not accounted by the relative simplicity of the \textsc{Cloudy} models and could also affect the CO luminosity and X-ray absorption. A detailed discussion regarding the properties of the molecular gas component of the Hot DOG population will be addressed in a future work via CO SLED modeling.

\subsubsection{Physical size and total gas mass of the cloud}
\label{4.3.2}

\cite{diaz2018} calculated the mass of molecular hydrogen, $M(H_2)$, in W2246--0526 using the CO(2--1) emission line. They estimate a $M(H_2)=7\pm1.5\times10^{10}$ M$_{\odot}$, assuming an $\alpha_{\mathrm{CO}}$ conversion factor for ultra luminous infrared galaxies (ULIRGs; $L_{IR}\gtrsim10^{12} L_{\odot}$) of $\alpha_{\mathrm{CO,ULIRG}}=0.8$ M$_{\odot}$ [K km s$^{-1}$ pc$^2$]$^{-1}$. Assuming that the geometry we have selected for the \textsc{Cloudy} modeling is a reasonable approximation, we can estimate the hydrogen mass (ionized + neutral, since the fraction of H$_2$ is negligible; see Fig.~\ref{fig:CO}) contained in the cloud using its volume and the best-fit value of $n_{\mathrm{H}}$. To calculate the volume, we derive the inner radius of the cloud by comparing the observed bolometric luminosity of W2246--0526 with the intrinsic bolometric flux (black solid line in Fig.~\ref{fig:SED}) obtained from the best-fit model. We use the observed, tens-of-kiloparsecs integrated bolometric luminosity as an upper limit for our modeled SED, as the latest only considers the emission from the central quasar. This puts an upper limit on the inner radius of the cloud of $r_0 \lesssim 670$ pc. \textsc{Cloudy} also reports the size of the cloud, which is $\sim\,$120 pc. Using the volume of this shell and $n_{\mathrm{H}}$, we estimate an upper limit for the hydrogen mass of the cloud (again, ionized + neutral) of $M_{H}<0.6\times10^{11} M_{\odot}$, which is smaller than the amount of molecular gas derived from the CO(2--1) line.

\subsubsection{Neutral carbon mass}
\label{4.3.3}

Following \cite{weiss2003}, we can derive the neutral carbon mass from the [CI]$_{370 \upmu \mathrm{m}}$ line luminosity via
\begin{equation}
    M_{CI}=4.566\times10^{-4}Q(T_{ex})\frac{1}{5}e^{62.5/T_{ex}}L'_{[CI]370\upmu \mathrm{m}} \ ,
\end{equation}
where $Q(T_{ex})=1+3e^{-23.6/T_{ex}}+5e^{-62.5/T_{ex}}$ is the $CI$ partition function, and $L'=3.25\times10^7S_{line}\Delta\upsilon \ \nu_{obs}^{-2}(1+z)^{-3}D_{L}^{2}$ is the luminosity expressed in units of K km s$^{-1}$ pc$^2$ \citep{solomon1992}. With the ratio $R_{CI}=L'_{[CI]370\upmu \mathrm{m}}/L'_{[CI]609\upmu \mathrm{m}}$ we can calculate the excitation temperature $T_{ex}=38.8/ln(2.11/R_{CI})$, assuming both lines are optically thin. Given our observational 3$\upsigma$ upper limit on [CI]$_{609\upmu \mathrm{m}}$ emission, we estimate $T_{ex}>37$ K, and therefore $M_{CI}<1.3\times10^7 \ M_{\odot}$. This upper limit of neutral carbon mass is in the range of estimations for other high-z quasars \citep{walter2011,pensabene2021}.

\subsection{Implications for the [OIII]$_{88 \upmu \mathrm{m}}$ and [CI]$_{609 \upmu \mathrm{m}}$ non-detections}
\label{sec:4.4}

Intense [OIII]$_{88 \upmu \mathrm{m}}$ line emission has been successfully detected in high-z galaxies up to $z\sim9$ \citep[e.g.,][]{harikane2020,witstok2022}, which in principle would make the non-detection of the line in W2246--0526 surprising. However, [OIII]$_{88 \upmu \mathrm{m}}$ emission arises most effectively from mildly ionized, diffuse gas \citep{2012A&A...548A..91L,witstok2022} while instead, our best-fit model favors very high hydrogen densities, $U$, and $\alpha_{\mathrm{ox}}$. In these XDR conditions most of the oxygen in the ionized gas is in higher ionization states (see middle panel on Fig.~\ref{fig:CO}) and [OIII]$_{88 \upmu \mathrm{m}}$ emission becomes highly suppressed, thus likely explaining the non-detection. This may have critical implications for the detection of this line in sources at the epoch of reionization, where the average intensity of the radiation field in galaxies should be stronger than at later cosmic times \citep{hashimoto2019,algera2024}. 

Regarding the non-detection of the [CI]$_{609 \upmu \mathrm{m}}$ line, the \textsc{Cloudy} models clearly show that the combination of high ionization and X-rays leads to less [CI]$_{609 \upmu \mathrm{m}}$ emission, and an overall lower abundance of neutral carbon with respect to ionized carbon (see top panel of Fig.~\ref{fig:CO}). Following \cite{papadopoulos2004}, we can also estimate the expected flux of [CI]$_{609 \upmu \mathrm{m}}$ using the value of $M(H_2)$ derived from the CO(2--1) observations mentioned above. Based on this, we obtain a flux of $S([\mathrm{CI}]_{609 \upmu \mathrm{m}})\sim0.2$ Jy km s$^{-1}$, which is consistent with the $3\upsigma$ observational upper limit ($<0.24$ Jy km s$^{-1}$). Given these estimations, deeper observations with ALMA of [CI]$_{609 \upmu \mathrm{m}}$ should achieve this sensitivity and successfully detect the emission line.

\section{Summary and conclusions}
\label{sec:conclusions}

In this work we have presented multiline ALMA observations of the extremely luminous, hot, dust-obscured galaxy W2246--0526 at redshift $z=4.6$. We detected the FIR fine-structure lines [OI]$_{63\upmu \mathrm{m}}$ (one of the few band 10 detections in the literature to date), [NII]$_{122\upmu \mathrm{m}}$, [OI]$_{145\upmu \mathrm{m}}$, [CII]$_{158\upmu \mathrm{m}}$, [NII]$_{205\upmu \mathrm{m}}$ and [CI]$_{370\upmu \mathrm{m}}$, and we measure upper limits for [OIII]$_{88\upmu \mathrm{m}}$ and [CI]$_{609\upmu \mathrm{m}}$. 

With the goal of characterizing the average physical conditions of the ISM of W2246--0526, we created a large grid of models with the photoionization code \textsc{Cloudy}. The main results are summarized as follows:

\begin{itemize}
    \item The model that best fits the observations indicates that W2246--0526 stands out from other quasars at high and low redshift in terms of its very high X-ray to UV ratio ($\alpha_{\mathrm{ox}}\geq-0.8$) and ionization parameter ($U=10^{-0.5}$). The average density of the cloud is comparable to other high-z quasars ($\sim3\times10^3$ cm$^{-3}$), and with a metallicity close to $0.5\ Z_{\odot}$. The extinction is very high ($A_{\mathrm{V}}\sim300$ mag), with hydrogen column densities in the Compton thick regime.
    \item The observed emission line ratios require high gas temperature, intense ionization, and X-ray emission in the ISM, conditions typical of XDRs. This also explains the non-detections of [OIII]$_{88\upmu \mathrm{m}}$ and [CI]$_{609\upmu \mathrm{m}}$ emission lines, which may be critical in the observation of these lines in other high-z AGNs and galaxies in general.
    \item The models predict an X-ray flux from W2246--0526 larger (although compatible within the model uncertainties) than the upper limit set by a non-detection with XMM-\textit{Newton}. Also, the models predict almost no CO emission, which is in contrast with previous works reporting a strong JVLA detection of CO(2--1) on scales of a few tens of kiloparsecs. When combined, these two pieces of evidence suggest the presence of an additional molecular gas component located farther away from the central quasar, distributed over spatial scales much larger than the region we are modeling with \textsc{Cloudy}. 
\end{itemize}

This study demonstrates the need for multiline FIR observations to properly characterize the multiphase ISM properties of galaxies and quasars at high redshift. In particular, W2246--0526 serves as a benchmark for the extreme conditions that may be potentially present in dust-obscured quasar populations at and beyond cosmic noon.

\begin{acknowledgements}
      We thank the referee for their constructive comments and suggestions to improve the paper. R.F.A. thanks ESO for the hospitality during the period of the 2022/2023 studentship, and Anna Feltre, Luke Maud, and Dirk Petry for the advice during the ALMA data reduction and the \textsc{Cloudy} modeling. 
      R.F.A. acknowledges support from the Hellenic Foundation for Research and Innovation (H.F.R.I.) under the "First Call for H.F.R.I. Research Projects to support Faculty members and Researchers and the procurement of high-cost research equipment grant" (Project 1552 CIRCE). R.J.A. was supported by the ANID BASAL project FB210003 and by FONDECYT grant and 1231718. This research was carried out in part at the Jet Propulsion Laboratory, California Institute of Technology, under a contract with NASA. M.A. acknowledges support from FONDECYT grant 1211951 and ANID BASAL project FB210003. A.P. acknowledges support from Fondazione Cariplo grant no. 2020-0902. G.J.S. acknowledges support from the US National Science Foundation through grants AAG 1910107 and AAG 1716229. 
      
      This paper makes use of the following ALMA data: ADS/JAO.ALMA\#2013.1.00576.S, ADS/JAO.ALMA\#2015.1.00883.S, ADS/JAO.ALMA\#2016.1.00668.S, ADS/JAO.ALMA\#2017.1.00899.S, ADS/JAO.ALMA\#2018.1.00119.S,
      ADS/JAO.ALMA\#2021.1.00726.S. ALMA is a partnership of ESO (representing its member states), NSF (USA) and NINS (Japan), together with NRC (Canada), MOST and ASIAA (Taiwan), and KASI (Republic of Korea), in cooperation with the Republic of Chile. The Joint ALMA Observatory is operated by ESO, AUI/NRAO and NAOJ.
\end{acknowledgements}

\bibliographystyle{aa}
\bibliography{refs} 

\newpage

\begin{appendix}
\section{ALMA data reduction}

\begin{figure}[h!]
    \centering
        \subfloat{\includegraphics[width=90mm]{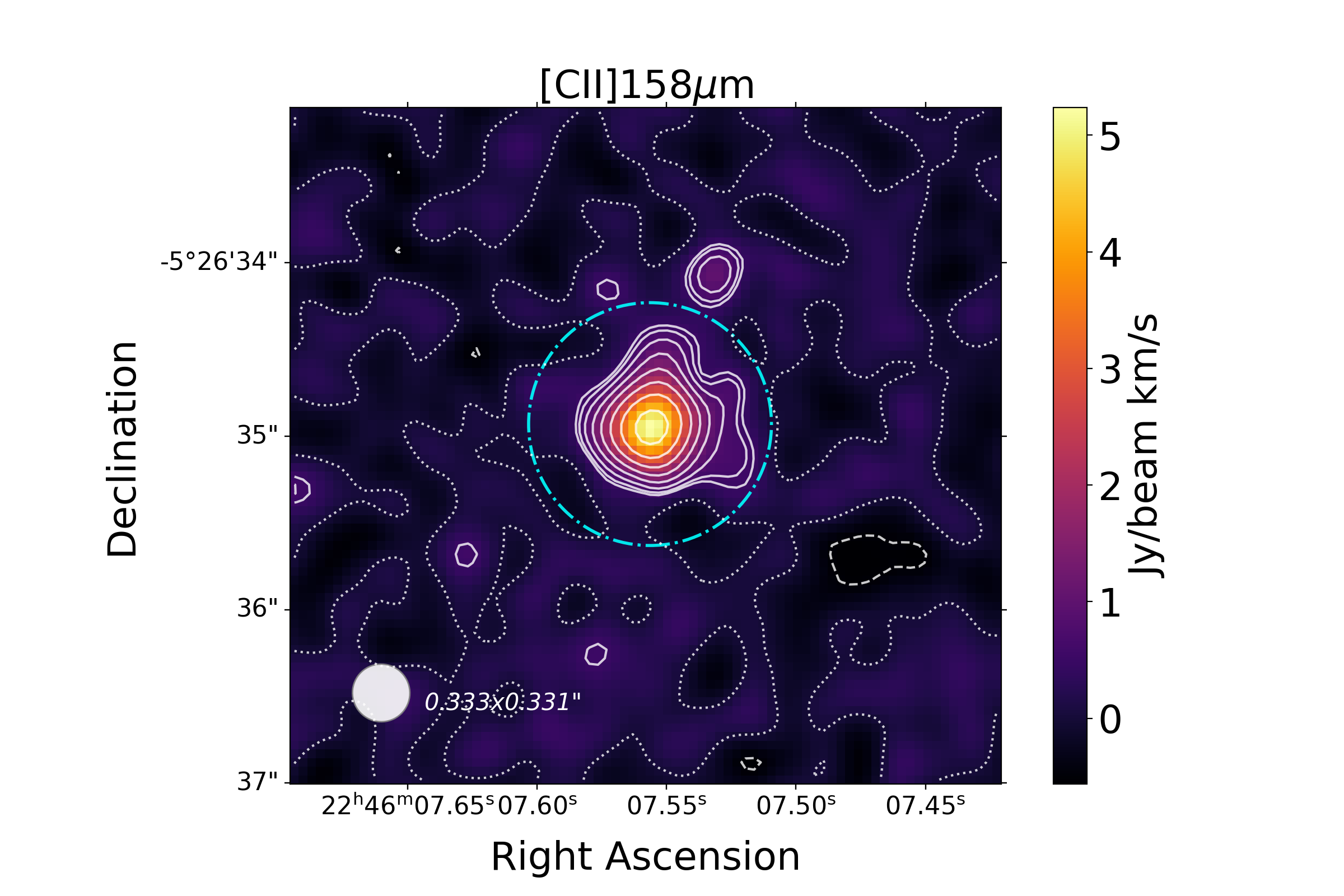}} 
        \par\vspace{1.5mm}
        \subfloat{\includegraphics[width=90mm]{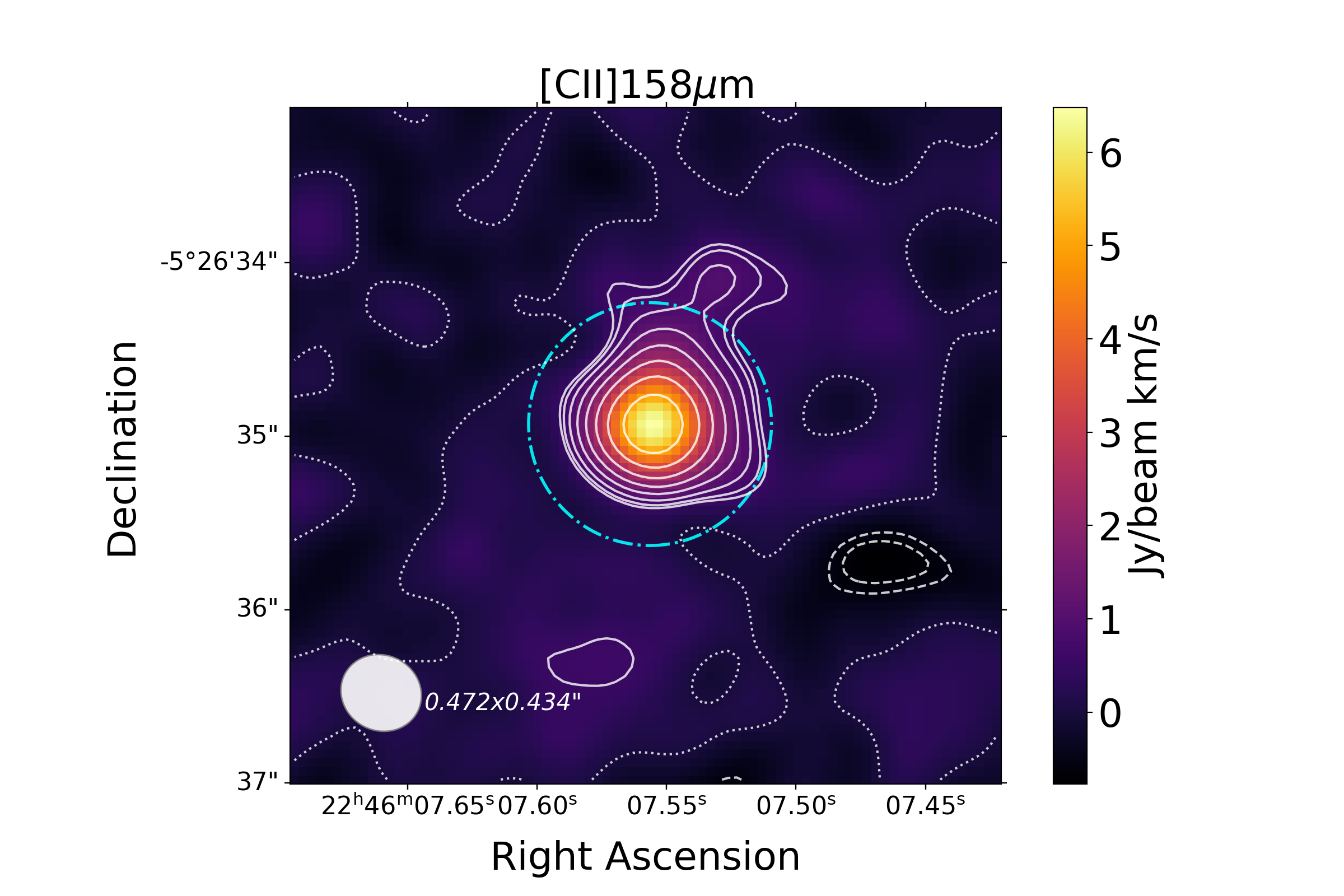}}

        \caption{[CII] line emission maps for W2246--0526 obtained using different CASA cleaning weightings (top, $r = 0.5$; bottom, $r = 2.0$ with a uvtaper = 0.3$\arcsec$).}
        \label{fig:Appendix_briggs}
\end{figure}

When using the CASA task \texttt{tclean}, a Briggs weighting was chosen, together with a robust parameter of 2.0 (close to natural weighting). Other options were also explored, shown in Fig.~\ref{fig:Appendix_briggs} for the [CII] line. For example, we can see that extended emission is missed when using a robust parameter of $r=0.5$ (between uniform and natural weighting). On the other hand, with a $r=2$ and applying a tapering of the visibilities with a value of $0.3\arcsec$ no extra emission is recovered, while the angular resolution is degraded.

Regarding the [OI]$_{145\upmu \mathrm{m}}$ line in W2246--0526, a telluric atmospheric feature can potentially affect the observations. To ensure that the detected line profile was not affected by an increase in the noise toward the edge of the spectral window where the atmospheric line lies, we investigated the transmission at the Atacama Pathfinder Experiment (APEX)  site\footnote{\url{http://www.apex-telescope.org/}; APEX and ALMA are both located in Llano de Chajnantor at the same altitude. APEX website provides tools to explore the atmospheric transmission.} and confirmed that the center of the telluric feature is offset by at least 600 km s$^{-1}$ with respect to the redshifted center of the [OI]$_{145\upmu \mathrm{m}}$ line. Hence, most of the line profile is not affected by the atmospheric feature. 

\section{\textsc{Cloudy} modeling}
\label{sec:AppendixB}

\subsection{Best-fit model $\chi_{\nu}^2$}

\begin{figure}[h!]
    \centering
        \subfloat{\includegraphics[width=90mm]{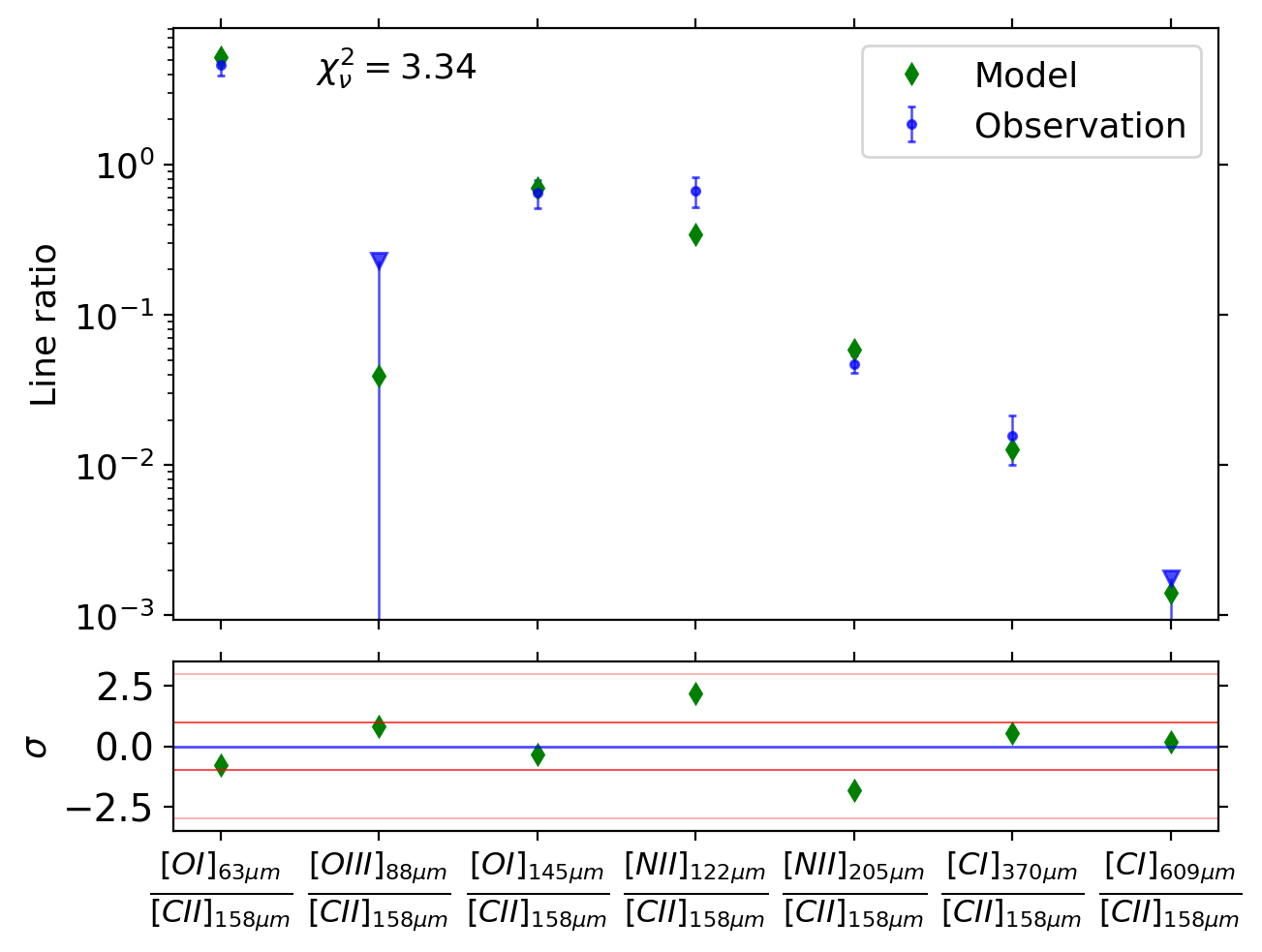}}
        \caption{Comparison between the W2246--0526 observed and best-fit model line ratios used for the minimization. The reduced $\chi^2$ is shown in the top left. The bottom panel shows the differences between the two in units of standard deviations.
        }
        \label{fig:ratios}
\end{figure}

To estimate the $\chi^2$ of the different \textsc{Cloudy} models we compare all the observed and modeled independent line ratios for W2246--0526 using [CII]$_{158\upmu \mathrm{m}}$ as a reference. In Fig.~\ref{fig:ratios} we present the reduced $\chi^2$ of the best-fit model presented in Sect. \ref{sec:3.4}. The observed upper limits for [OIII]$_{88\upmu \mathrm{m}}$ and [CI]$_{609\upmu \mathrm{m}}$ are consistent with the best-fit model, and we do not observe any systematics in the residuals.

\subsection{Fixed $\alpha_{\mathrm{ox}}$}

\begin{figure}[h]
    \centering
        \subfloat{\includegraphics[width=90mm]{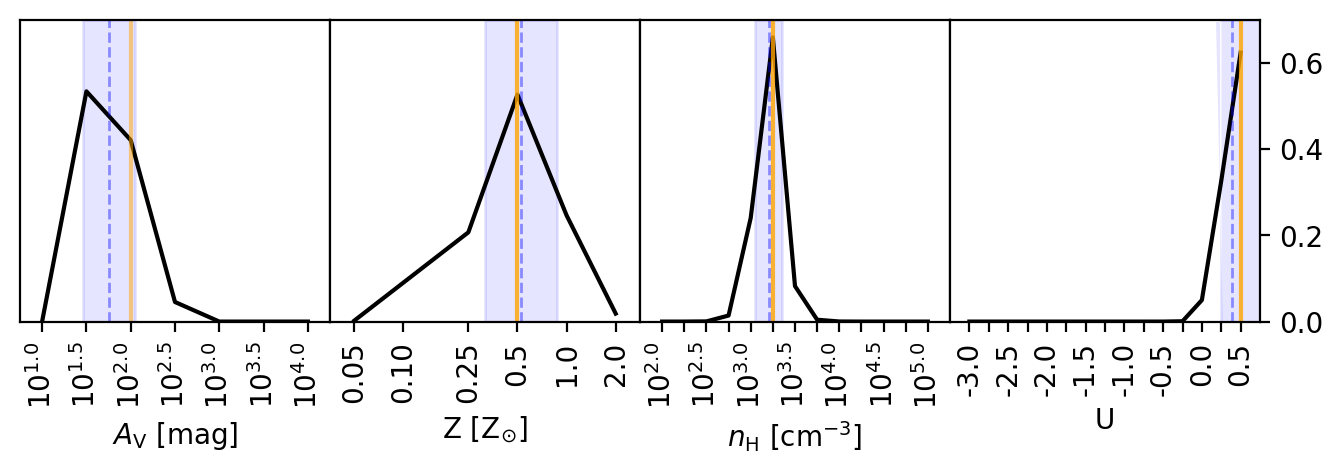}}
        \caption{Marginalized 1D posterior probability distributions for each parameter of the models without varying $\alpha_{\mathrm{ox}}$. The lines and shaded regions are as defined in Fig.~\ref{fig:posteriors}.}
        \label{fig:Appendix_posterior}
\end{figure}

Following the discussion in \ref{sec:4.2.1}, we show in Fig.~\ref{fig:Appendix_posterior} the posterior probability distributions for each of the \textsc{Cloudy} parameters with a fixed $\alpha_{\mathrm{ox}}=-1.4$. The ionization parameter is unconstrained and extremely high. 

\begin{figure*}[h!]
    \centering
        \subfloat{\includegraphics[width=90mm]{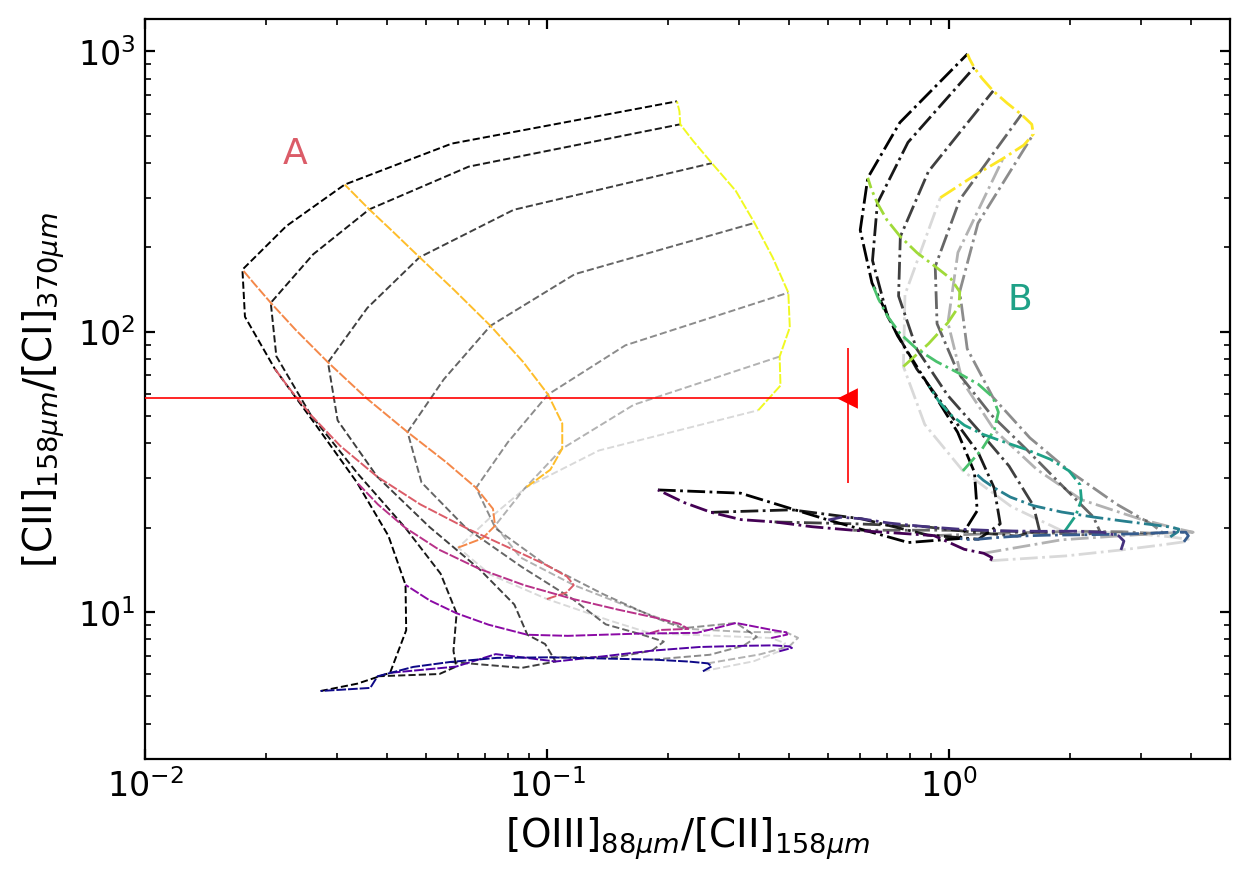}}
        \subfloat{\includegraphics[width=90mm]{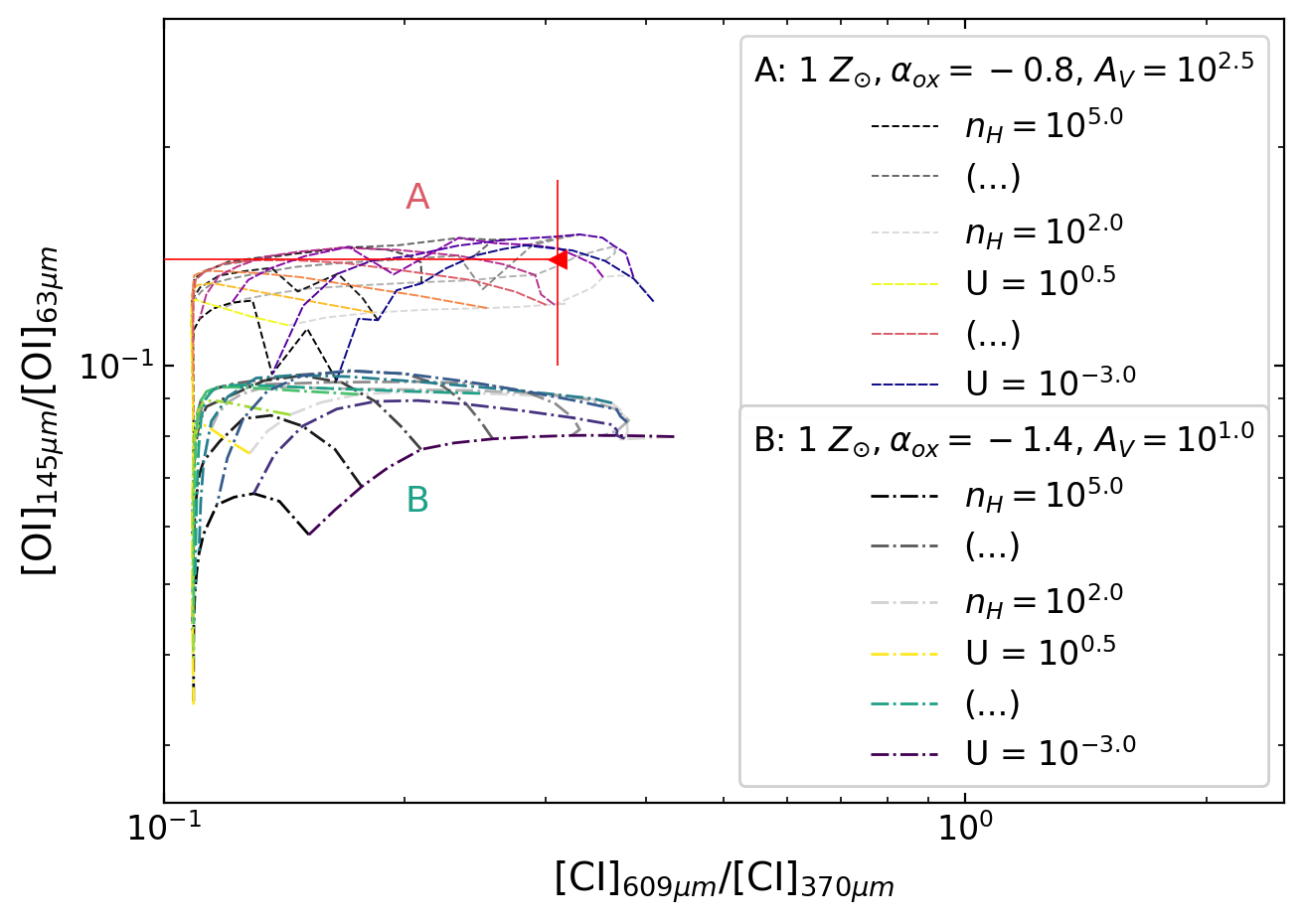}}  
        \\
        \subfloat{\includegraphics[width=90mm]{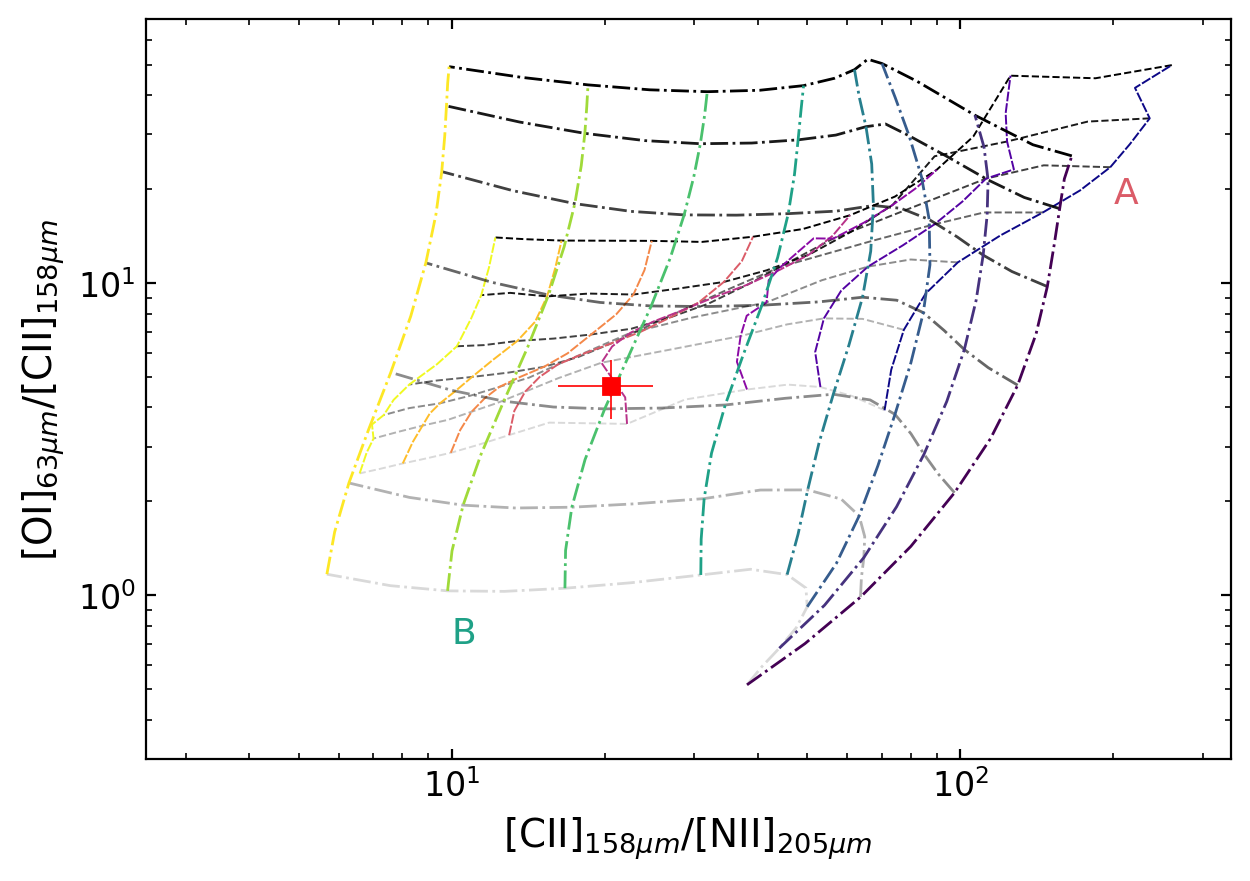}}
        \subfloat{\includegraphics[width=90mm, height=63mm]{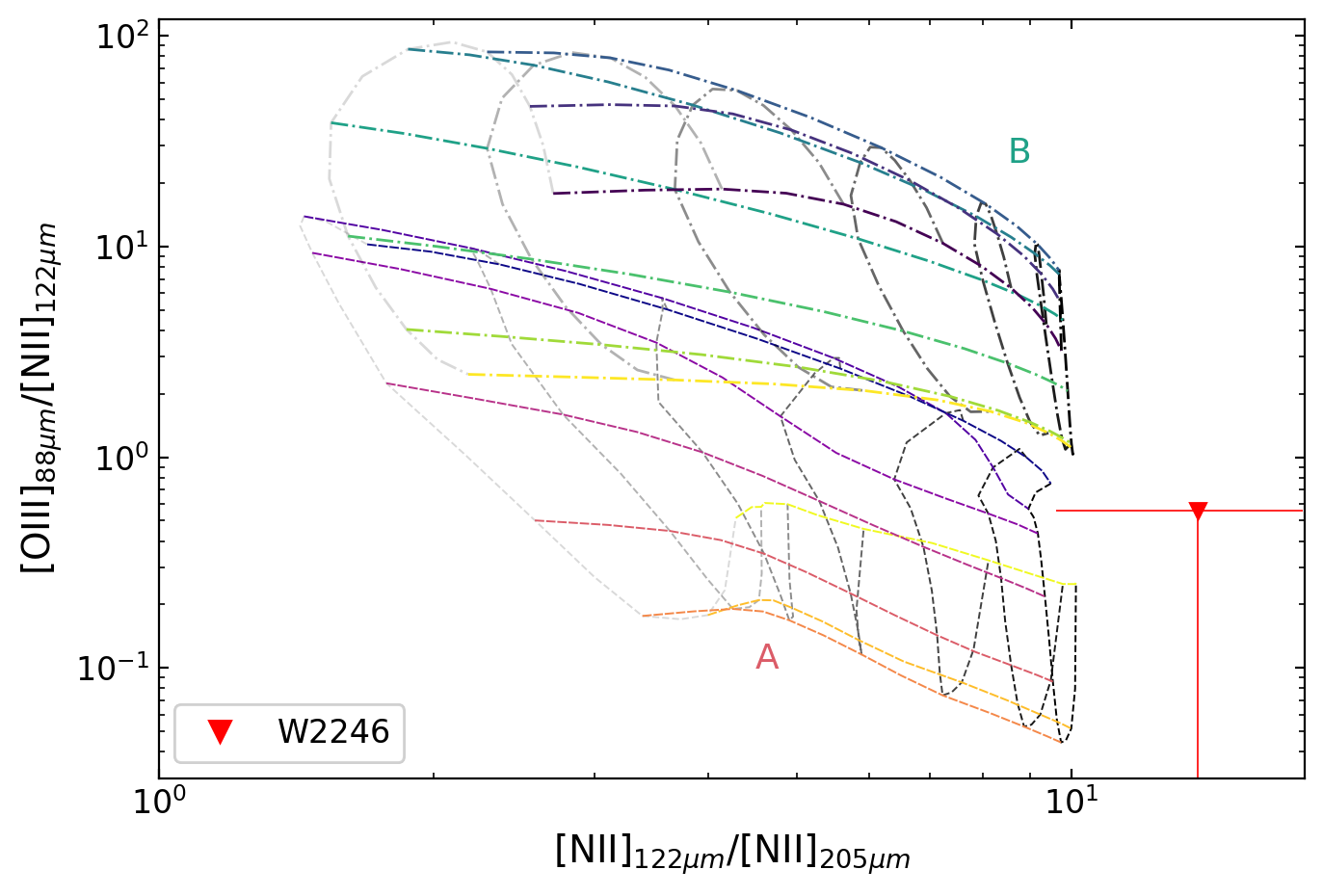}}
        \caption{Line ratio diagrams from \textsc{Cloudy} AGN models for solar metallicity, $\alpha_{\mathrm{ox}}=-1.4$ with $A_{\mathrm{V}}=10^{1}$ mag (labeled as A, grid with dashed lines), and $\alpha_{\mathrm{ox}}=-0.8$ with $A_{\mathrm{V}}=10^{2.5}$ mag (labeled as B, grid with dash-dotted lines). Both grids span the full range of values explored in this work for the ionization parameter and the density, as shown in the legend and Fig.~\ref{fig:grids}. The observed line ratios for W2246--0526 are indicated with a red square, or a red triangle for upper limits. }
        \label{fig:grids2}
\end{figure*}

Additionally, Fig.~\ref{fig:grids2} shows the same line ratios as in Fig.~\ref{fig:grids} for grids of \textsc{Cloudy} models with $\alpha_{\mathrm{ox}}=-1.4$ and $A_{\mathrm{V}}=10$ mag. This showcases how lower X-ray to UV ratios and lower extinctions are not able to reproduce our observed line ratios.

\subsection{Tests of additional parameters}

\begin{figure*}
    \centering
        \subfloat{\includegraphics[width=90mm]{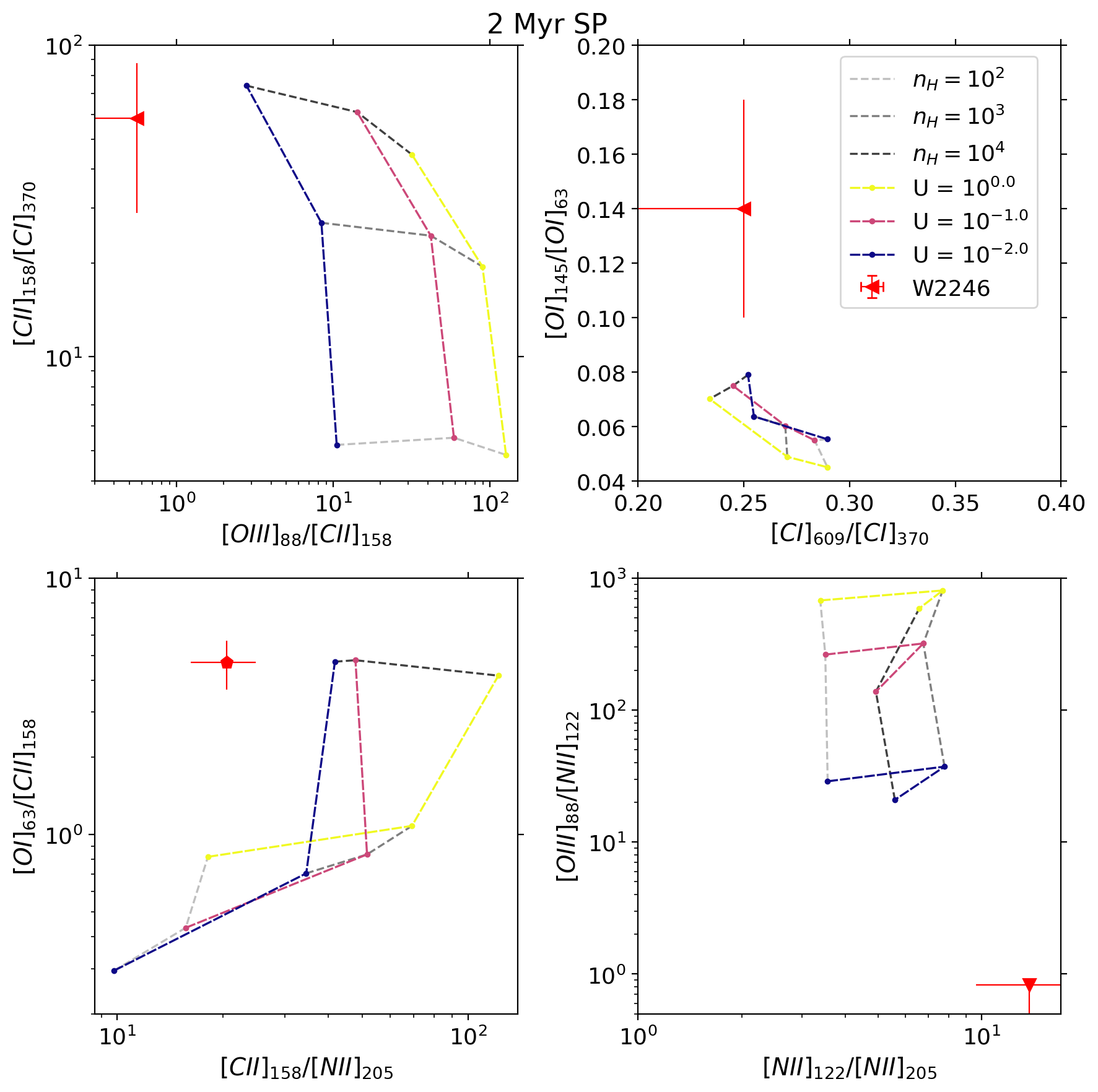}}        \subfloat{\includegraphics[width=90mm]{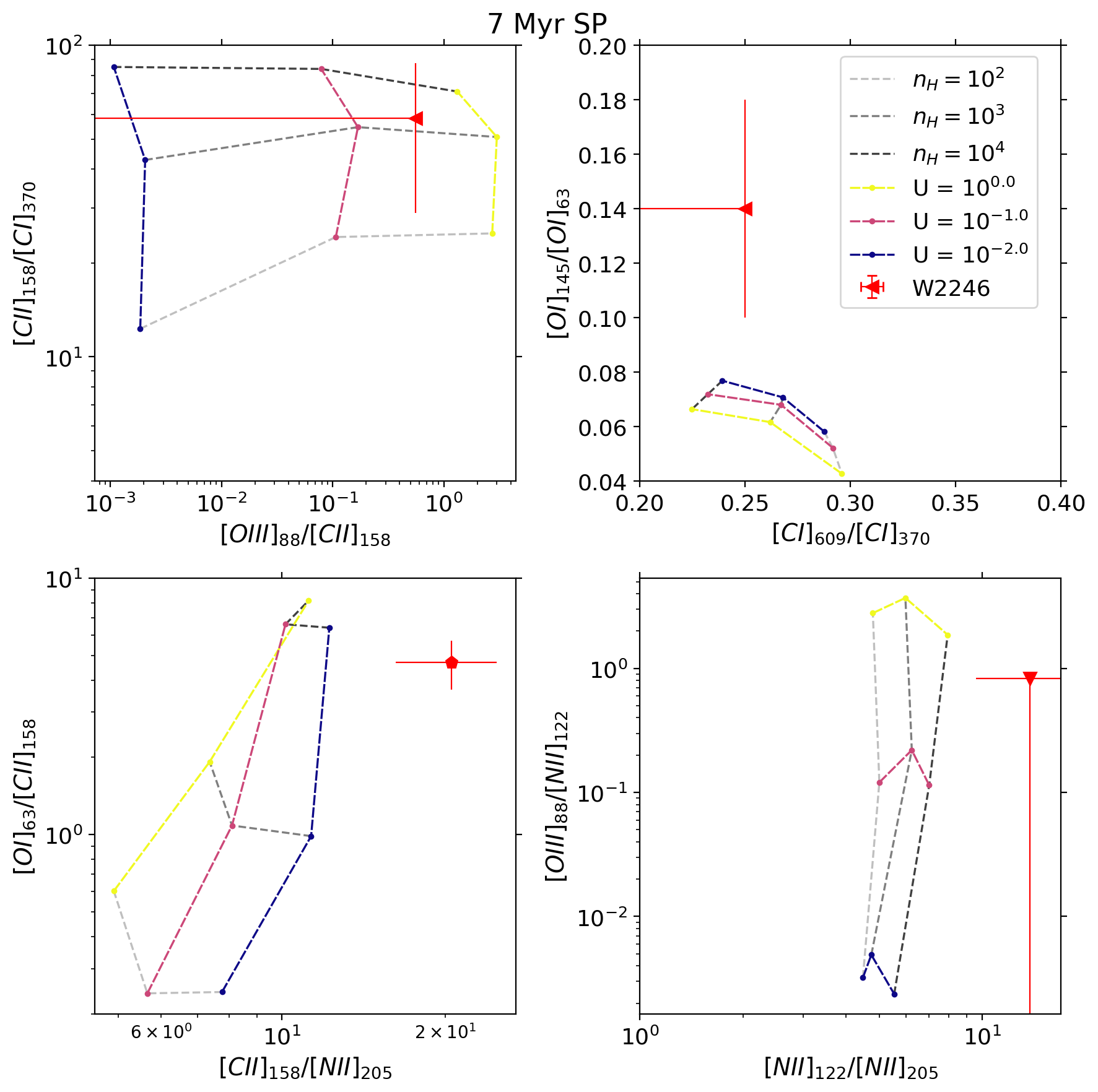}}
        \par\vspace{1.5mm}
        \subfloat{\includegraphics[width=90mm]{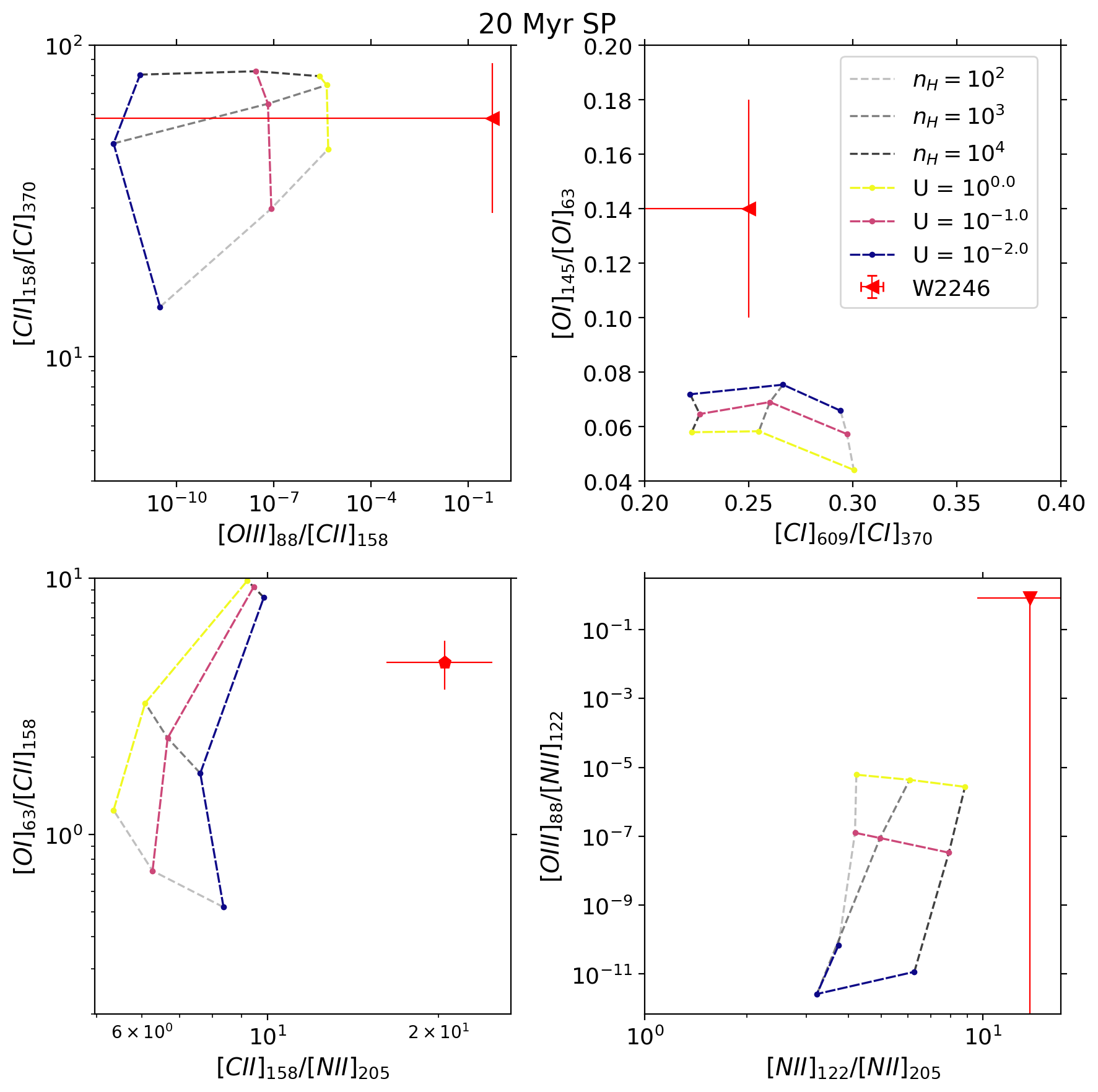}}
        \subfloat{\includegraphics[width=90mm]{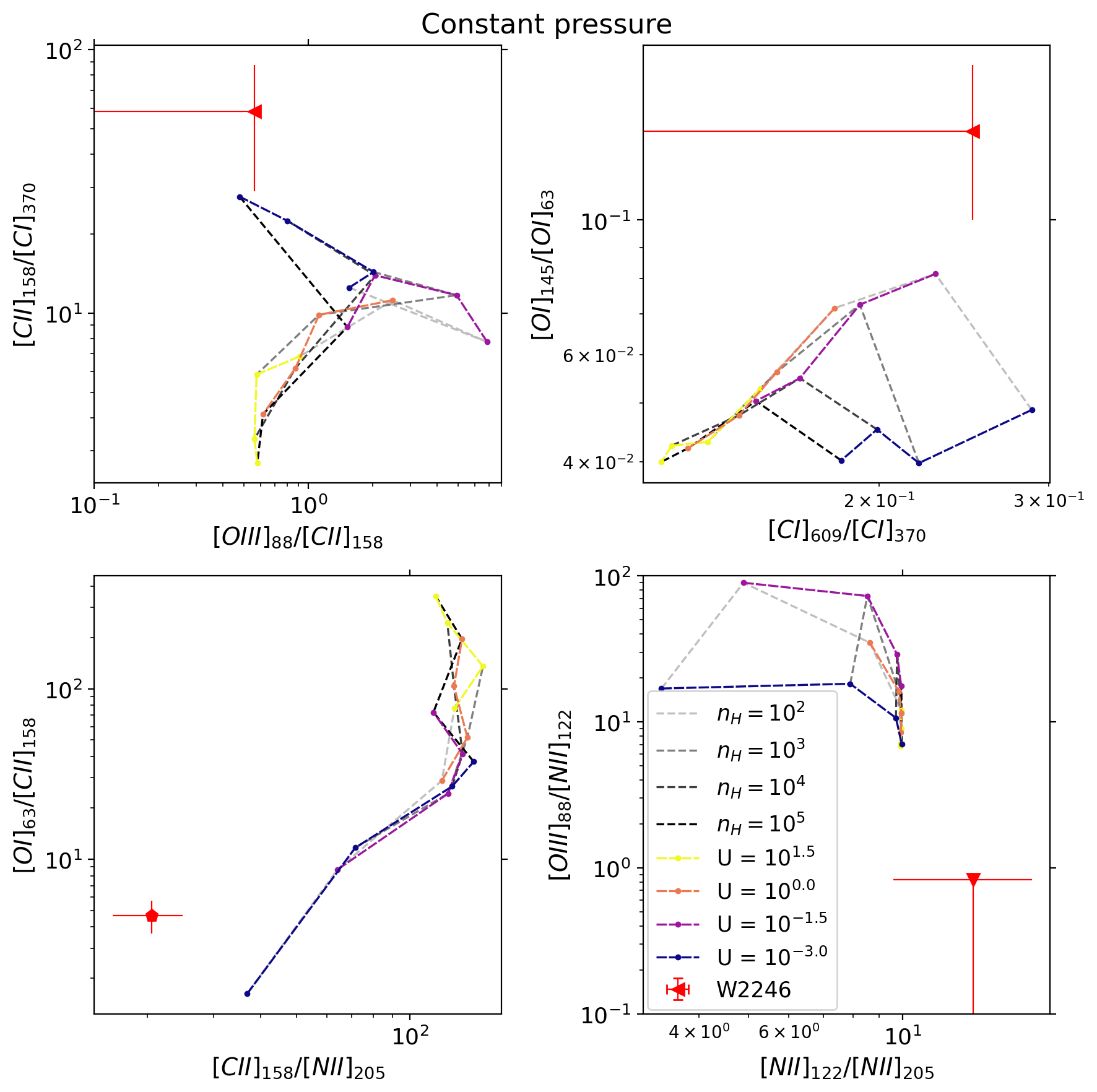}}

        \caption{Same line ratios as in Fig.~\ref{fig:grids}, but for smaller \textsc{Cloudy} grids varying the ionization parameter and the density of the cloud to test stellar populations (a single 2 Myr population in the top left, 7 Myr in the top right, and 20 Myr in the bottom left) as a source of radiation, instead of an AGN. In the bottom right we test a constant pressure density profile instead of the constant density one.}
        \label{fig:Appendix_smallgrids}
\end{figure*}

Figure~\ref{fig:Appendix_smallgrids} shows the same line ratios as in Fig.~\ref{fig:grids} for grids of \textsc{Cloudy} models testing additional physical parameters. For all of them, the metallicity is set to 1 $Z_{\odot}$, and the $A_{\mathrm{V}}$ is 100 mag. The choice of density profiles in ISM modeling can highly affect the molecular gas and CO emission predictions. We selected a constant density profile, but other options may be more appropriate for modeling the molecular phase as discussed in \cite{2019MNRAS.482.4906P}. We investigated the possibility of using a constant gas pressure through the cloud instead of a constant density, with an AGN with $\alpha_{\mathrm{ox}}=-1.4$ as the ionization source (bottom right in Fig.~\ref{fig:Appendix_smallgrids}). In this case, the density of the cloud $n_{\mathrm{H}}$ is the initial density at the inner part of the cloud. We also explored three different stellar SEDs as the input source instead of an AGN, using single stellar populations from \cite{2009MNRAS.398..451M} of ages 2 Myr, 7 Myr, and 20 Myr \citep[with initial mass function from][and solar metallicity.]{2003PASP..115..763C} The density for the stellar populations in these cases is constant through the cloud. 

As seen in Fig.~\ref{fig:Appendix_smallgrids}, none of the test grids fall in the region of the parameter space where the observed emission ratios lay, a finding that favors a constant density modeling with an AGN as a radiation source.

\end{appendix}

\end{document}